\documentclass[fleqn,usenatbib]{mnras}

\usepackage[T1]{fontenc}

\DeclareRobustCommand{\VAN}[3]{#2}
\let\VANthebibliography\thebibliography
\def\thebibliography{\DeclareRobustCommand{\VAN}[3]{##3}\VANthebibliography}

\usepackage{savesym}
\usepackage{amsmath}
\savesymbol{iint}
\usepackage{txfonts}
\restoresymbol{TXF}{iint}
\usepackage{graphics,graphicx}
\usepackage{amssymb}
\usepackage{color}
\usepackage[]{hyperref}
\usepackage{array}
\usepackage{rotating}
\usepackage{xcolor}
\usepackage{natbib}
\usepackage{float}
\usepackage{flafter}
\usepackage{booktabs}
\usepackage{afterpage}
\usepackage{lipsum}
\usepackage{longtable,pdflscape}

\newcommand{\xmm}{{\it XMM-Newton}}
\newcommand{\rosat}{{\it ROSAT}}

\newcommand{\spitzer}{{\it Spitzer}}

\newcommand{\vpshock}{{\tt vpshock}}

\newcommand{\vphabs}{{\tt vphabs}}
\newcommand{\vapec}{{\tt vapec}}

\newcommand{\pow}{{\tt powerlaw}}

\newcommand{\vnei}{{\tt vnei}}

\newcommand{\hb}{H$\beta$}
\newcommand{\ha}{H$\alpha$}
\newcommand{\halpha}{H$\alpha$}
\newcommand{\sii}{[$\ion{S}{ii}$]}
\newcommand{\nii}{[$\ion{N}{ii}$]}
\newcommand{\oiii}{[$\ion{O}{iii}$]}
\newcommand{\ratio}{[$\ion{S}{ii}$]/H$\alpha$}
\newcommand{\nh}{$N_{\rm{H}}$}

\newcommand\tablefoot[1]{\raggedright{\textbf{Notes:} #1}}

\DeclareUnicodeCharacter{2212}{-}

\begin{document}

\title[\xmm\ observations of faint, evolved SNRs in the LMC]{New \xmm\ observations of faint, evolved supernova remnants in the Large Magellanic Cloud\thanks{Based on observations obtained with \xmm, an ESA science mission with instruments and contributions directly funded by ESA Member States and NASA}}

\author[P. J. Kavanagh et al.]{P.~J.~Kavanagh,$^{1}$\thanks{E-mail: pkavanagh@cp.dias.ie}
M.~Sasaki,$^{2}$
M.~D.~Filipovi\'c,$^{3}$
S.~D.~Points,$^{4}$
L.~M.~Bozzetto,$^{3}$
F.~Haberl,$^{6}$
\newauthor
P.~Maggi,$^{5}$
C.~Maitra$^{6}$
\\
$^{1}$School of Cosmic Physics, Dublin Institute for Advanced Studies, 31 Fitzwilliam Place, Dublin 2, Ireland\\
$^{2}$Dr Karl Remeis Observatory and ECAP, Friedrich-Alexander-Universit\"{a}t Erlangen-N\"{u}rnberg, Sternwartstr. 7, 96049 Bamberg, Germany\\
$^{3}$Western Sydney University, Locked Bag 1797, Penrith South DC, NSW 2751, Australia\\
$^{4}$NSF's NOIRLab/CTIO, Casilla 603, La Serena, Chile\\
$^{5}$Universit\'{e} de Strasbourg, CNRS, Observatoire astronomique de Strasbourg, UMR 7550, F-67000 Strasbourg, France\\
$^{6}$Max-Planck-Institut f\"{u}r extraterrestrische Physik, Gie{\ss}enbachstra{\ss}e 1, D-85748 Garching, Germany\\
}
\date{Accepted 2022 March 13. Received 2022 February 25; in original form 2021 November 1}

\pubyear{2021}

\label{firstpage}
\pagerange{\pageref{firstpage}--\pageref{lastpage}}
\maketitle

\begin{abstract}
The Large Magellanic Cloud (LMC) hosts a rich population of supernova remnants (SNRs), our knowledge of which is the most complete of any galaxy. However, there remain many candidate SNRs, identified through optical and radio observations where additional X-ray data can confirm their SNR nature and provide details on their physical properties. In this paper, we present \xmm\ observations that provide the first deep X-ray coverage of ten objects, comprising eight candidates and two previously confirmed SNRs. We perform multi-frequency studies using additional data from the Magellanic Cloud Emission Line Survey (MCELS) to investigate their broadband emission and used \spitzer\ data to understand the environment in which the objects are evolving. We confirm seven of the eight candidates as \textit{bona-fide} SNRs. We used a multi-frequency morphological study to determine the position and size of the remnants. We identify two new members of the class of evolved Fe-rich remnants in the Magellanic Clouds (MCs), several SNRs well into their Sedov-phase, one SNR likely projected towards a HII region, and a faint, evolved SNR with a hard X-ray core which could indicate a pulsar wind nebula. Overall, the seven newly confirmed SNRs represent a $\sim$10\% increase in the number of LMC remnants, bringing the total number to 71, and provide further insight into the fainter population of X-ray SNRs.

\end{abstract}




\section{Introduction}
Supernova remnants (SNRs) are the objects that result from the interaction of ejecta from supernova (SN) explosions with the surrounding medium. These types of explosions are generally divided into two categories: the core-collapse of a massive star, or the explosion of a carbon-oxygen white dwarf in a binary system which has exceeded the Chandrasekhar limit due to accretion or merger. SNe and their SNRs are critical engines of the matter cycle in galaxies. They power the mechanical evolution of the interstellar medium (ISM), they seed the ISM with heavy elements which are available for future generations of star-formation, and they are prodigious producers of cosmic rays \citep[for a review see][]{Vink2012,2021pma..book.....F}. Therefore, by understanding their physics and evolution we will better understand the matter cycle in galaxies. Studies of evolved SNRs in the Galaxy are difficult because of uncertain distance estimates and high foreground absorption, which effectively shields the emission from their low temperature plasmas from our view. Therefore, to perform unbiased studies of evolved SNRs we must look to nearby galaxies. 

The Large Magellanic Cloud (LMC) is perfectly suited for the study of evolved SNRs. It is located at a distance of 50.0$\pm$1.3~kpc \citep{Pie2013}, is oriented almost face-on \citep[inclination angle of 30--40$^\circ$,][]{vanderMar2001,Nikolaev2004}, and its location outside of the Galactic disk means it is subject to relatively low line-of-sight absorption (average Galactic foreground $N_{\rm{H}} \approx 7 \times 10^{20}$~cm$^{-2}$) making evolved SNRs more easily observable. Prior to this work there were 64 confirmed SNRs in the LMC and many more candidates \citep[][and references therein]{Yew2021,Maitra2021,2022MNRAS.tmp..324F}, and we have performed an extensive study of the known population in X-rays
\citep{Maggi2016} and radio \citep{Bozzetto2017}.
 
Our SNR candidate sample in recent years was identified based on \rosat\ data, in particular from the study of \citet{Haberl1999}. Many of these objects were confirmed as evolved SNRs using follow-up \xmm\ observations or data from the Very Large Programme (VLP) survey of the LMC (PI: Frank Haberl), along with radio observations, e.g., \citet{Bozzetto2014}, \citet{Maggi2014}, \citet{Kavanagh2015}, \citet{Kavanagh2015b}, \citet{Kavanagh2016}. In this paper we present \xmm\ observations to SNR candidates identified with either \rosat\ in \citet{Haberl1999} or in the radio continuum in \citet{Bozzetto2017}, supplemented by optical emission line data from the Magellanic Cloud Emission Line Survey \citep[MCELS,][]{Smith2006}. 

\begin{figure*}
\begin{center}
\resizebox{\hsize}{!}{\includegraphics[trim= 0cm 0cm 0cm 0cm, clip=true, angle=0]{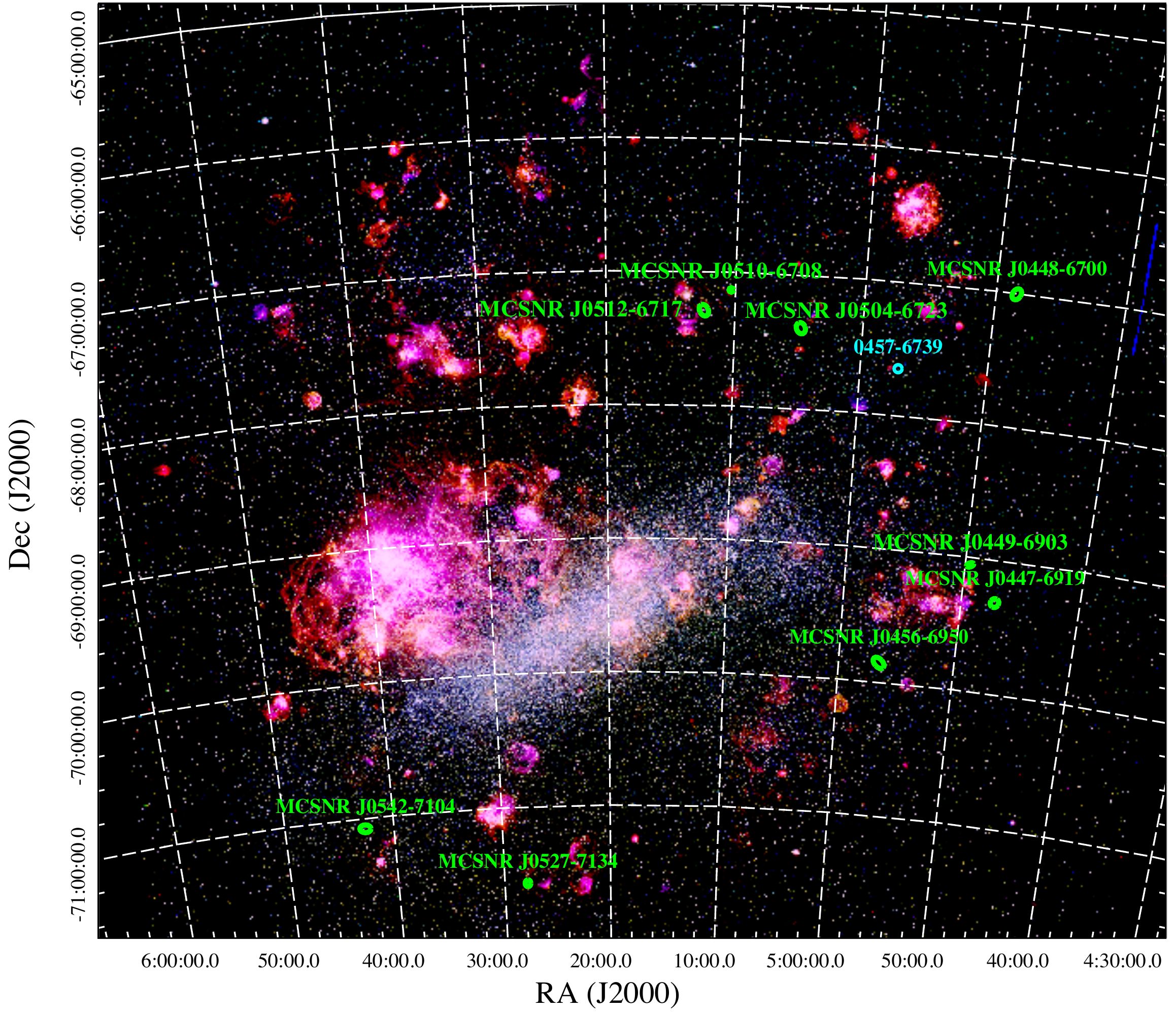}}
\caption{The location of our SNR sample plotted on a three-colour MCELS mosaic of the LMC (RGB = \ha, \sii, \oiii). All confirmed SNRs are shown in green. The unconfirmed candidate 0457--6739 is shown in cyan.}
\label{figures:snrs_loc}
\end{center}
\end{figure*}

In this and our previous works, we adopt the criteria of the Magellanic Cloud Supernova Remnant (MCSNR) Database for SNR classification, i.e. a candidate must meet at least two of the following three criteria to be confirmed as an SNR: significant \ha, \sii, and/or \oiii\ line emission with an \ratio$>$0.4 \citep{Mathewson1973,Fesen1985,1998A&AS..130..421F,book2}; extended non-thermal radio emission, typically with spectral index $\alpha \approx -0.5$\footnote{Defined by $S \propto \nu^{\alpha}$, where $S$ is flux density, $\nu$ is the frequency and $\alpha$ is the spectral index.} for an evolved remnant; and/or soft thermal X-ray emission. As shown in the forthcoming text, seven of our eight candidates were confirmed as SNRs based on their radio, optical, and/or X-ray properties conforming to the MCSNR classification scheme. Therefore, to avoid confusion, we will hereafter refer to these objects using their MCSNR identifier. These are given in Table~\ref{table:mcsnr} and their locations within the LMC are shown in Fig.~\ref{figures:snrs_loc}.

\begin{table}
\caption{MCSNR identifiers for the confirmed SNRs in our sample.
}
\begin{center}
\label{table:mcsnr}
\begin{tabular}{lcl}
\hline\hline\noalign{\smallskip}
Candidate name & Reference$^{\left(a\right)}$ & MCSNR Id. \\
\noalign{\smallskip}\hline
0447--6918     & B & MCSNR~J0447--6919  \\
$[$HP99$]$~460 & H, M$^{\left(b\right)}$ & MCSNR~J0448--6700 \\
0449--6903     & B & MCSNR~J0449--6903  \\
0456--6951     & B & MCSNR~J0456--6950  \\
$[$HP99$]$~529 & H & MCSNR~J0504--6723 \\
0510--6708     & B & MCSNR~J0510--6708 \\
0512--6717     & B & MCSNR~J0512--6717 \\
0527--7134     & B & MCSNR~J0527--7134 \\
0542--7104     & Y$^{\left(b\right)}$ & MCSNR~J0542--7104  \\
\hline
\end{tabular}
\end{center}
\tablefoot{$^{\left(a\right)}$ B: \citet{Bozzetto2017}, H:  \citet{Haberl1999}, M: \citet{Maggi2016}, Y:  \citet{Yew2021}.
$^{\left(b\right)}$ Previously confirmed as SNR.}
\end{table}%

\par Our paper is arranged as follows: the observations and data reduction are described in Sect.~\ref{sect:oadr},  data analysis is outlined in Sect.~\ref{sect:analysis}, results are given in Sect.~\ref{results}, we discuss our results in Sect.~\ref{discussion}, and finally we summarise our work in Sect.~\ref{sum}.

\section{Observations and data reduction}
\label{sect:oadr}
\subsection{X-ray}
\label{Xobs}
Each of our candidates was observed with \xmm\ \citep{Jansen2001} with the European Photon Imaging Camera (EPIC) as the primary instrument. The EPIC comprises a pn CCD \citep{Struder2001} and two MOS CCD \citep{Turner2001} imaging spectrometers. All the observational data were reduced using the standard reduction tasks of SAS\footnote{Science Analysis Software, see \url{http://xmm.esac.esa.int/sas/}} version 18.0.0. The details of each observation are given in Table~\ref{table:xray_observations} which also lists each of the exposure times after filtering for periods of high particle background. We note that MCSNR~J0512--6717 was observed serendipitously by \xmm\ on May~2014 (Obs.~ID~0741800201, PI P.~Kavanagh) during an observation of the nearby remnant MCSNR~J0512--6707 ($[$HP99$]$~483). Further details on these observations can be found in \citet{Kavanagh2015c}.
 
\begin{table*}
\caption{\xmm\ observations of LMC SNRs and candidates. The listed exposure times are flare-filtered. The PI for all observations was P.~Kavanagh.}
\begin{center}
\label{table:xray_observations}
\begin{tabular}{llllllll}
 \hline
\hline
Target & Obs. ID & Date & RA (J2000) & Dec (J2000) & \multicolumn{3}{c}{Exp. time (ks)} \\
       &         &      &            &             &  pn & MOS1 & MOS2 \\
\hline
$[$HP99$]$~483 & 0741800201 & 2014-05-17 & 05:12:28.0 & --67:07:27  & 42.1 & 48.2 & 45.9  \\
$[$HP99$]$~460 & 0764060101 & 2015-06-20 & 04:48:22.5 & --67:00:04  & 27.6 & 32.3 & 32.3  \\
$[$HP99$]$~529 & 0764060201 & 2015-08-19 & 05:04:49.5 & --67:24:15 & 26.7 & 31.2 & 31.3   \\
0527--7134 & 0780620101 & 2016-10-16 & 05:27:48.5 & --71:34:06 & 18.6 & 31.3 & 29.3  \\
0457--6739 & 0801990201 & 2017-10-27 & 04:57:33.0 & --67:39:06 & 16.6 & 43.2 & 42.8  \\
0449--6903 & 0801990401 & 2017-10-26 & 04:49:34.0 & --69:03:34 & 3.0 & 12.0 & 12.3   \\
0447--6918 & 0801990301 & 2017-09-28 & 04:47:09.7 & --69:18:58 & 17.8 & 37.1 & 35.9 \\
0456--6950 & 0841660501 & 2019-10-08 & 04:56:30.3 & --69:50:47 & 31.1 & 45.6 & 45.6   \\
0449--6903 & 0841660101 & 2019-10-15 & 04:49:34.0 & --69:03:33 & 34.7 & 52.5 & 52.5   \\
0510--6708 & 0841660401 & 2020-04-07 & 05:10:11.4 & --67:08:03 & 28.4 & 39.8 & 39.2   \\
0542--7104 & 0841660201 & 2020-04-19 & 05:42:45.0 & --71:04:00 & 28.6 & 42.4 & 41.9   \\

\hline
\end{tabular}
\end{center}
\end{table*}

\subsection{Optical}
To search for optical emission signatures of SNRs we made use of both imaging and, in one case, spectroscopic observations.

\subsubsection{Imaging}
We used the MCELS \citep{Smith2006} for our imaging analysis. The observations were taken with the 0.6 m University of Michigan/Cerro Tololo Inter-American Observatory (CTIO) Curtis Schmidt Telescope producing individual images of $1.35^{\circ} \times 1.35^{\circ}$ at a scale of 2.3\arcsec\ pixel$^{-1}$ in narrow bands covering \oiii$\lambda$5007\AA, \halpha, and \sii$\lambda$6716,6731\AA, along with matched green and red continuum bands. These individual images were flux calibrated and combined to create the final Magellanic Cloud mosaics. We extracted cutouts from the LMC mosaic centred on each of our objects. We then subtracted the appropriate continuum images from the emission line images to remove stellar continuum and reveal the full extent of the faintest diffuse emission. Finally, we produced \ratio\ maps for each of our objects by dividing the continuum-subtracted \sii\ image by the continuum-subtracted \ha\ image. The emission line images and contours derived from the \ratio\ maps for each remnant are shown in the top-right of Figs.~\ref{figures:0447-6918_4panel}--\ref{figures:0542-7104_4panel}.

\subsubsection{MCSNR~J0512--6717 spectroscopy}
\label{0512-eli}
The \ratio\ map determined from the MCELS images of MCSNR~J0512--6717 revealed sporadic regions \ratio$\gtrsim0.4$ that were not coincident with the X-ray SNR shell. Therefore, we identified two \ha\ and \sii\ bright filaments with \ratio$\gtrsim0.4$ which were located in the X-ray shell (see Fig.~\ref{figures:0512-6717_4panel} top-right) and performed follow-up observations using the Wide-Field Spectrograph \citep[WiFeS,][]{Dopita2007,Dopita2010} integral field unit (IFU) mounted on the Australian National University 2.3~m Telescope at the Siding Springs Observatory.
WiFeS provides optical spectra in a $25\arcsec\times38\arcsec$ field of view in the 3300-9200~\AA\ range. This field is divided into twenty-five 1\arcsec\ wide slitlets with 0.5\arcsec\ sampling along the 38\arcsec\ lengths. We used the R3000 and B3000 dispersers for the red and blue cameras, respectively, to achieve a spectral resolution of R=3000 across the wavelength range.
Both fields in MCSNR~J0512--6717 were observed on 2015-10-09 (Prop.~ID: 3150070). Field~1 (RA=05:12:48.8, Dec=--67:16:29) was observed for 180~s and field~2 (RA=05:12:27.8, Dec=--67:18:32) for 120~s. We reduced the raw data using the PyWiFeS\footnote{Available from \url{http://www.mso.anu.edu.au/pywifes/doku.php}} data reduction pipeline \citep{Childress2014}.

\subsection{Infrared}
To aid in the discussion of the multi-wavelength morphology of our objects, we made use of data from the SAGE survey of the LMC \citep{Meixner2006} with the \spitzer\ \textit{Space Telescope} \citep{Werner2004}. The SAGE data set comprises a $7\times7$ degree survey of the LMC with the Infrared Array Camera \citep[IRAC,][]{Fazio2004} in the 3.6~$\mu$m, 4.5~$\mu$m, 5.8~$\mu$m, and 8~$\mu$m bands, and with the Multiband Imaging Photometer \citep[MIPS,][]{Rieke2004} in the 24~$\mu$m, 70~$\mu$m, and 160~$\mu$m bands. For this work we extracted cutouts from the MIPS 24~$\mu$m mosaic for each of our objects using the NASA/IPAC Infrared Science Archive\footnote{See \url{http://irsa.ipac.caltech.edu/data/SPITZER/SAGE/}} tools, which are shown in the bottom right panels of Figs.~\ref{figures:0447-6918_4panel}--\ref{figures:0542-7104_4panel}. These images reveal the stochastically, thermally, and radiatively heated dust in the vicinity of our objects at a comparable resolution to \xmm.

\section{Analysis}
\label{sect:analysis}

\subsection{X-ray imaging}
\label{x-ray-imaging}
\begin{figure*}
\begin{center}
\resizebox{6.5in}{!}{\includegraphics[trim= 0cm 0cm 0cm 0cm, clip=true, angle=0]{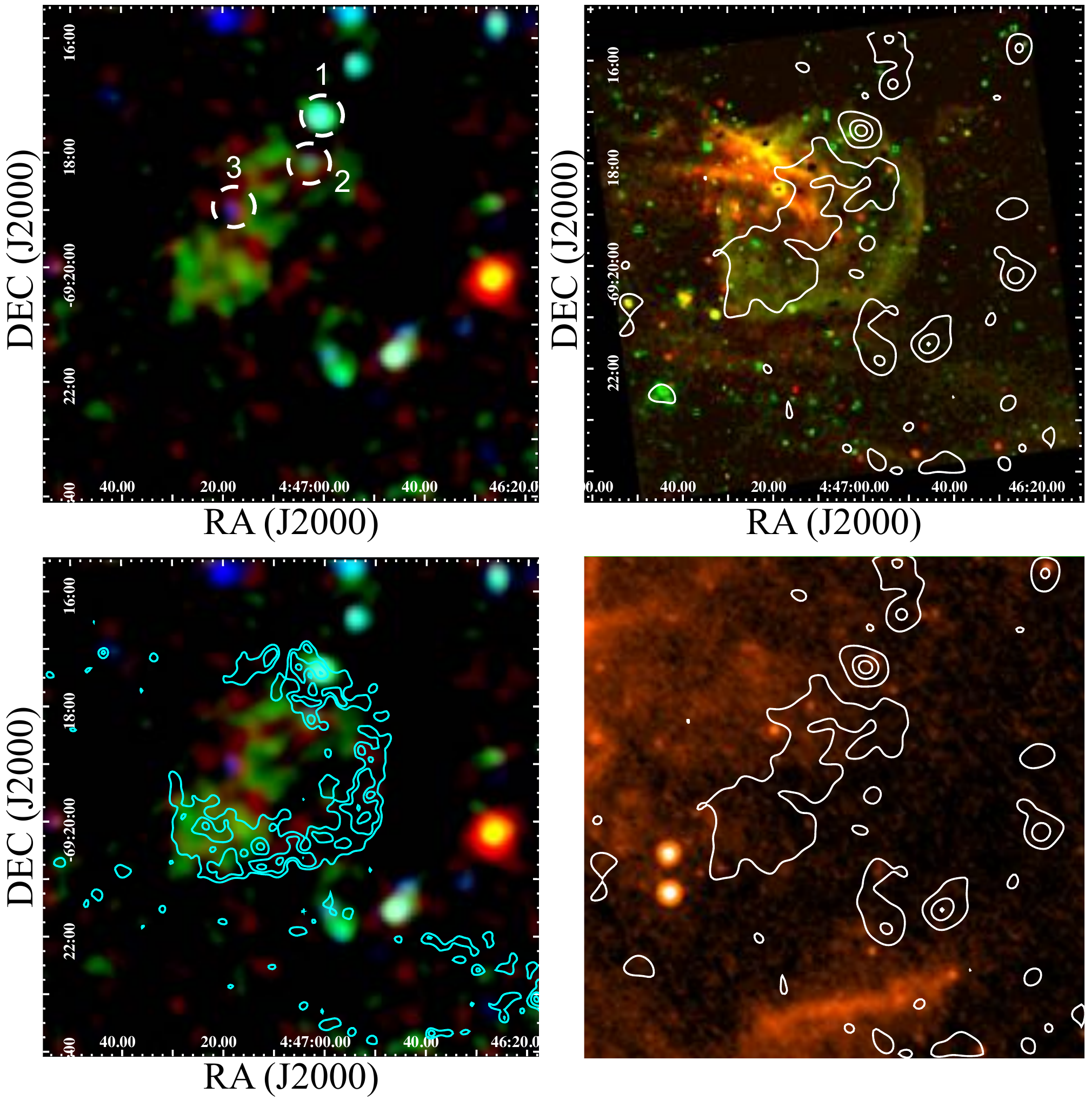}}
\caption{{\it Top left:} \xmm\ EPIC image of MCSNR~J0447--6919 in false colour with RGB corresponding to 0.3--0.7~keV, 0.7--1.1~keV, and 1.1--4.2~keV. Detected X-ray point sources are shown by the dashed white circles. {\it Top right:} Continuum-subtracted MCELS image of MCSNR~J0447--6919 with \ha\ in red and \sii\ shown in green overlaid with 0.3--0.7~keV contours. The lowest contour level represents 3$\sigma$ above the average background surface brightness, with the remaining levels marking 25\%, 50\%, and 75\% of the maximum above this level. {\it Bottom left:} Same as top left but with \ratio\ contours with the lowest level corresponding to \sii/\ha~=~0.4, and the remaining levels at 25\%, 50\%, and 75\% of the maximum above this level overlaid.  {\it Bottom right:} \spitzer\ MIPS 24~$\mu$m image of the MCSNR~J0447--6919 region with the X-ray contours from top right in white. The image scale is the same as in all other panels.}
\label{figures:0447-6918_4panel}
\end{center}
\end{figure*}

\begin{figure*}
\begin{center}
\resizebox{6.5in}{!}{\includegraphics[trim= 0cm 0cm 0cm 0cm, clip=true, angle=0]{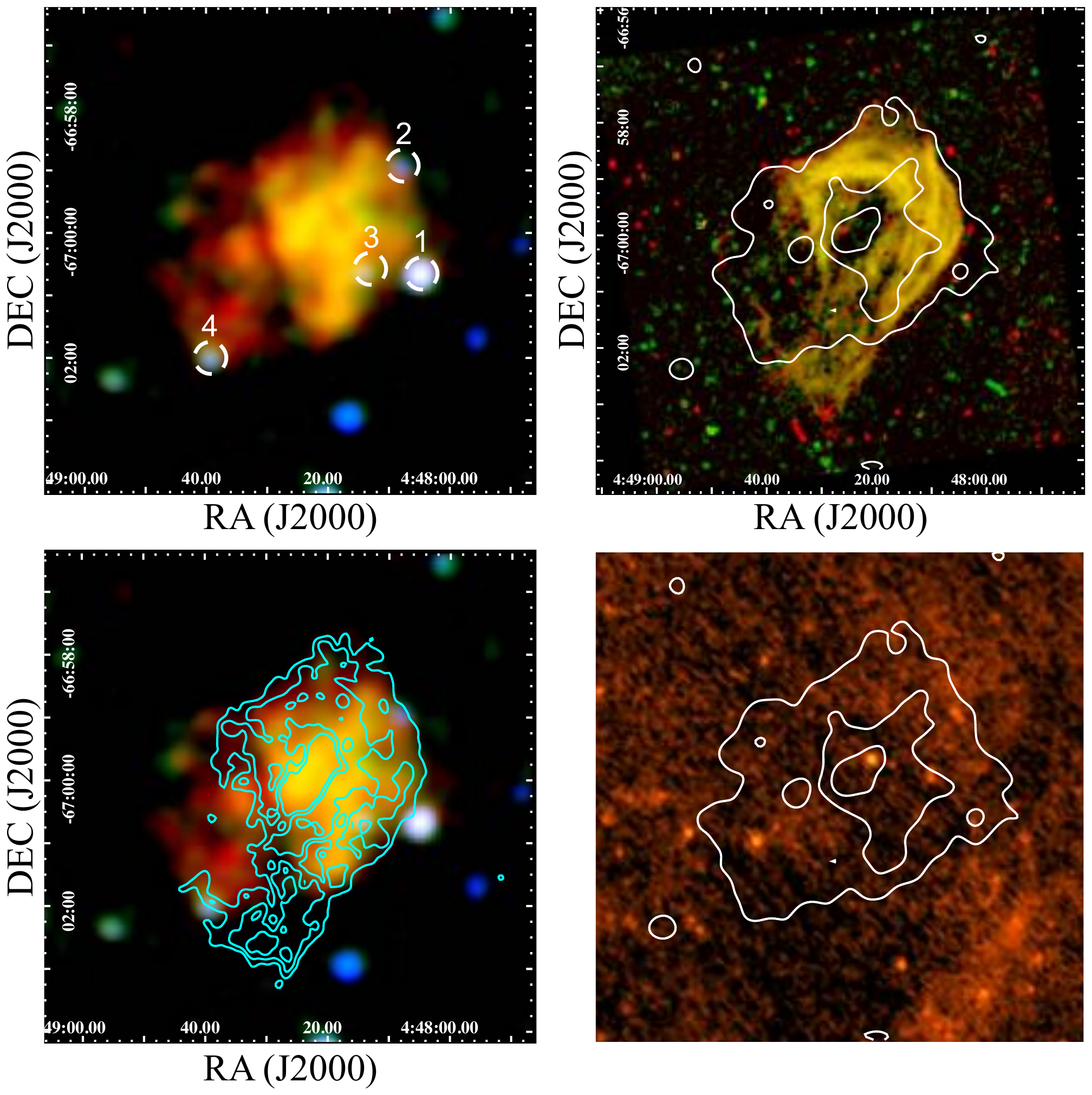}}
\caption{Same as Fig.~\ref{figures:0447-6918_4panel} but for MCSNR~J0448--6700.}
\label{figures:0448-6700_4panel}
\end{center}
\end{figure*}

\begin{figure*}
\begin{center}
\resizebox{6.5in}{!}{\includegraphics[trim= 0cm 0cm 0cm 0cm, clip=true, angle=0]{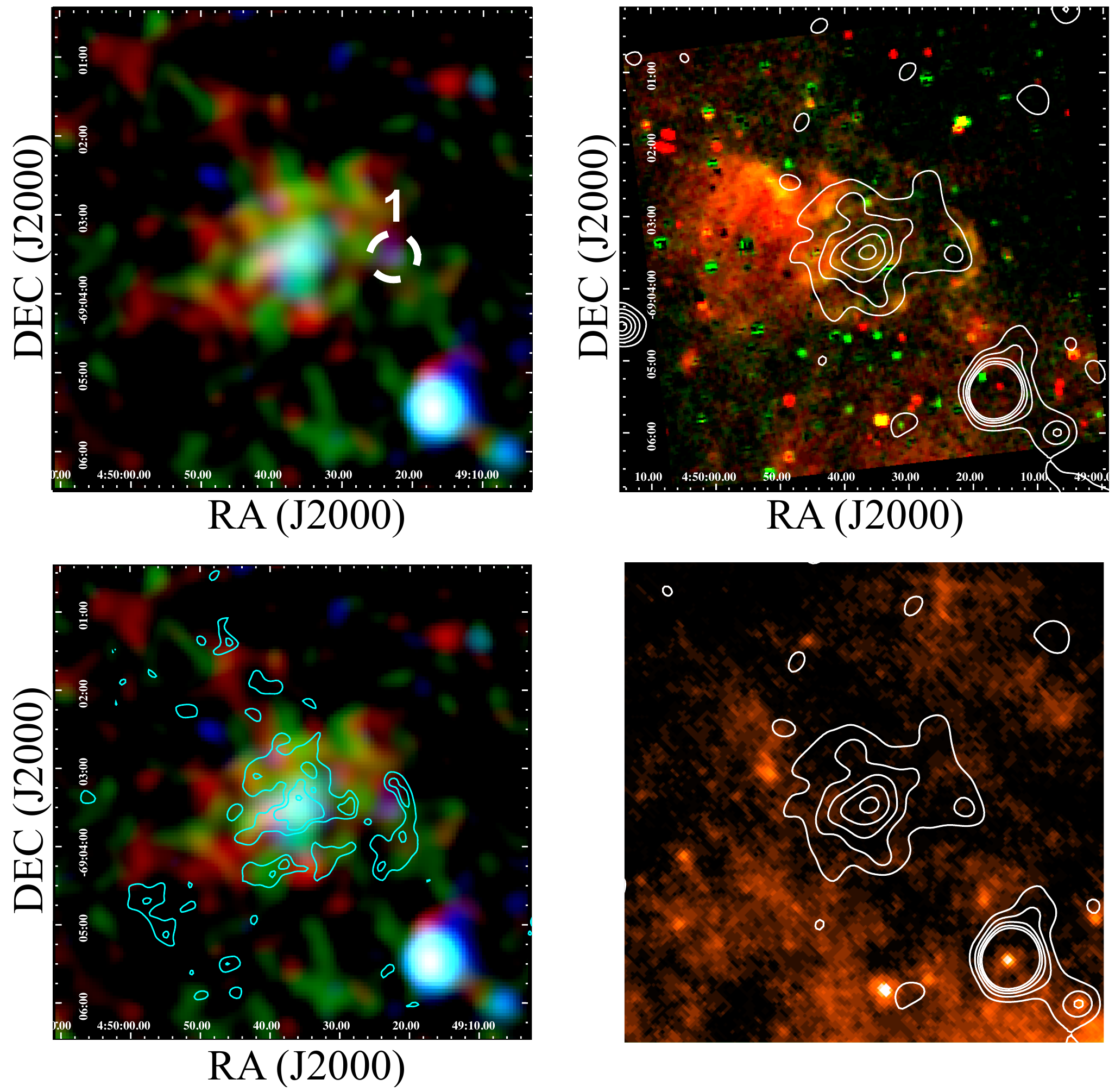}}
\caption{Same as Fig.~\ref{figures:0447-6918_4panel} but for MCSNR~J0449--6903.}
\label{figures:0449-6903_4panel}
\end{center}
\end{figure*}

\begin{figure*}
\begin{center}
\resizebox{6.5in}{!}{\includegraphics[trim= 0cm 0cm 0cm 0cm, clip=true, angle=0]{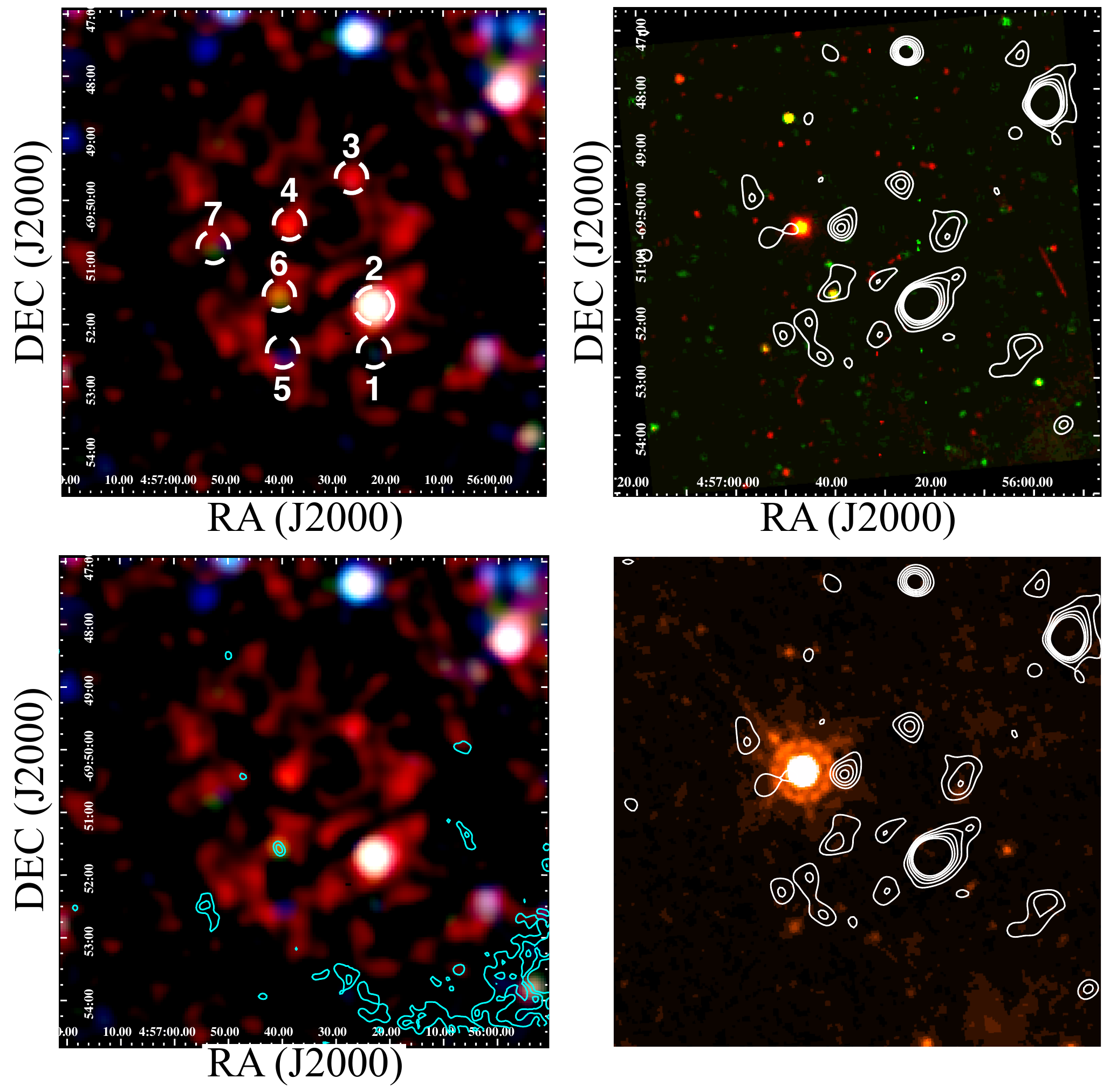}}
\caption{Same as Fig.~\ref{figures:0447-6918_4panel} but for MCSNR~J0456--6950.}
\label{figures:0456-6951_4panel}
\end{center}
\end{figure*}

\begin{figure*}
\begin{center}
\resizebox{6.5in}{!}{\includegraphics[trim= 0cm 0cm 0cm 0cm, clip=true, angle=0]{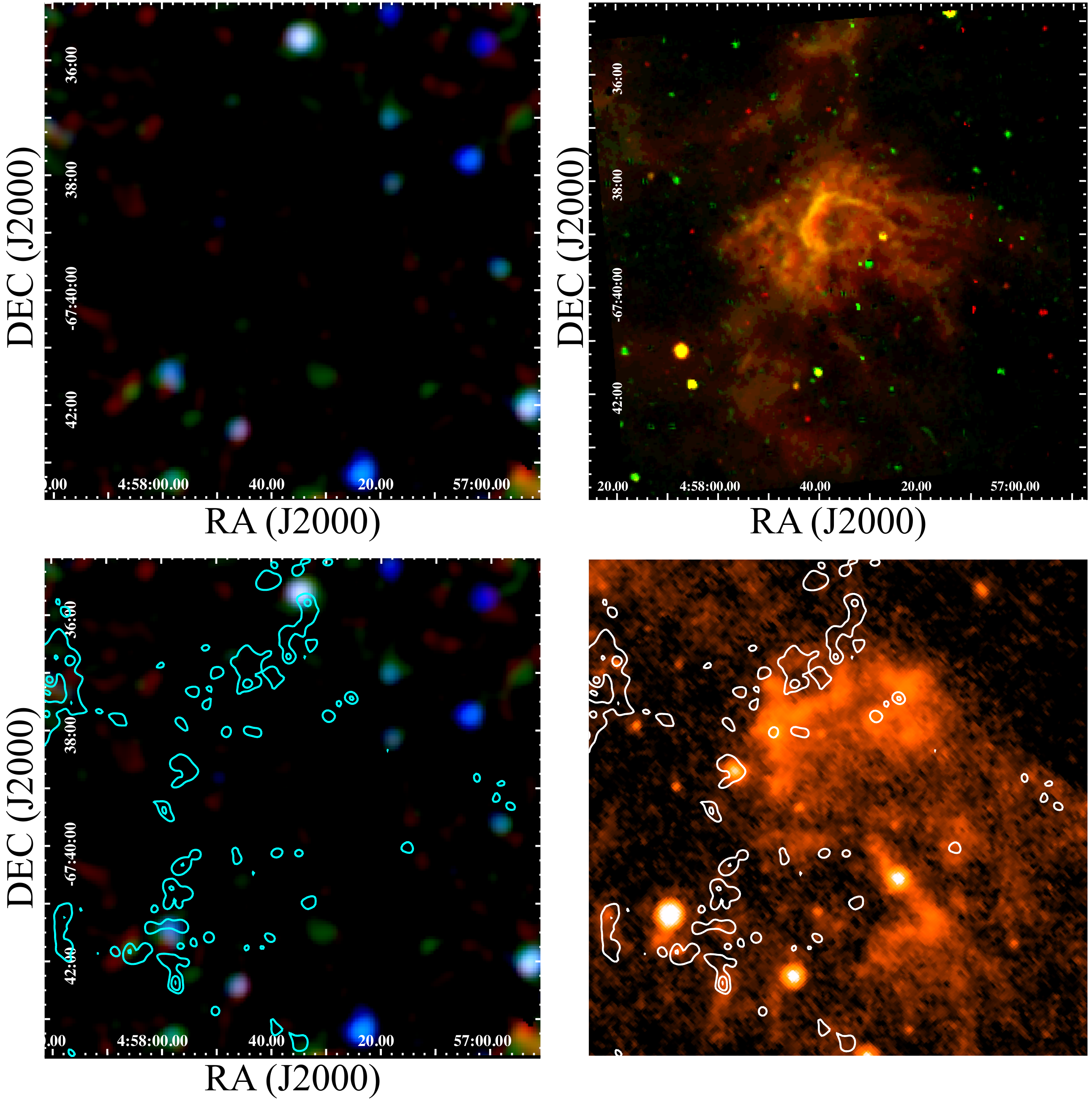}}
\caption{Same as Fig.~\ref{figures:0447-6918_4panel} but for 0457--6739.}
\label{figures:0457-6739_4panel}
\end{center}
\end{figure*}

\begin{figure*}
\begin{center}
\resizebox{6.5in}{!}{\includegraphics[trim= 0cm 0cm 0cm 0cm, clip=true, angle=0]{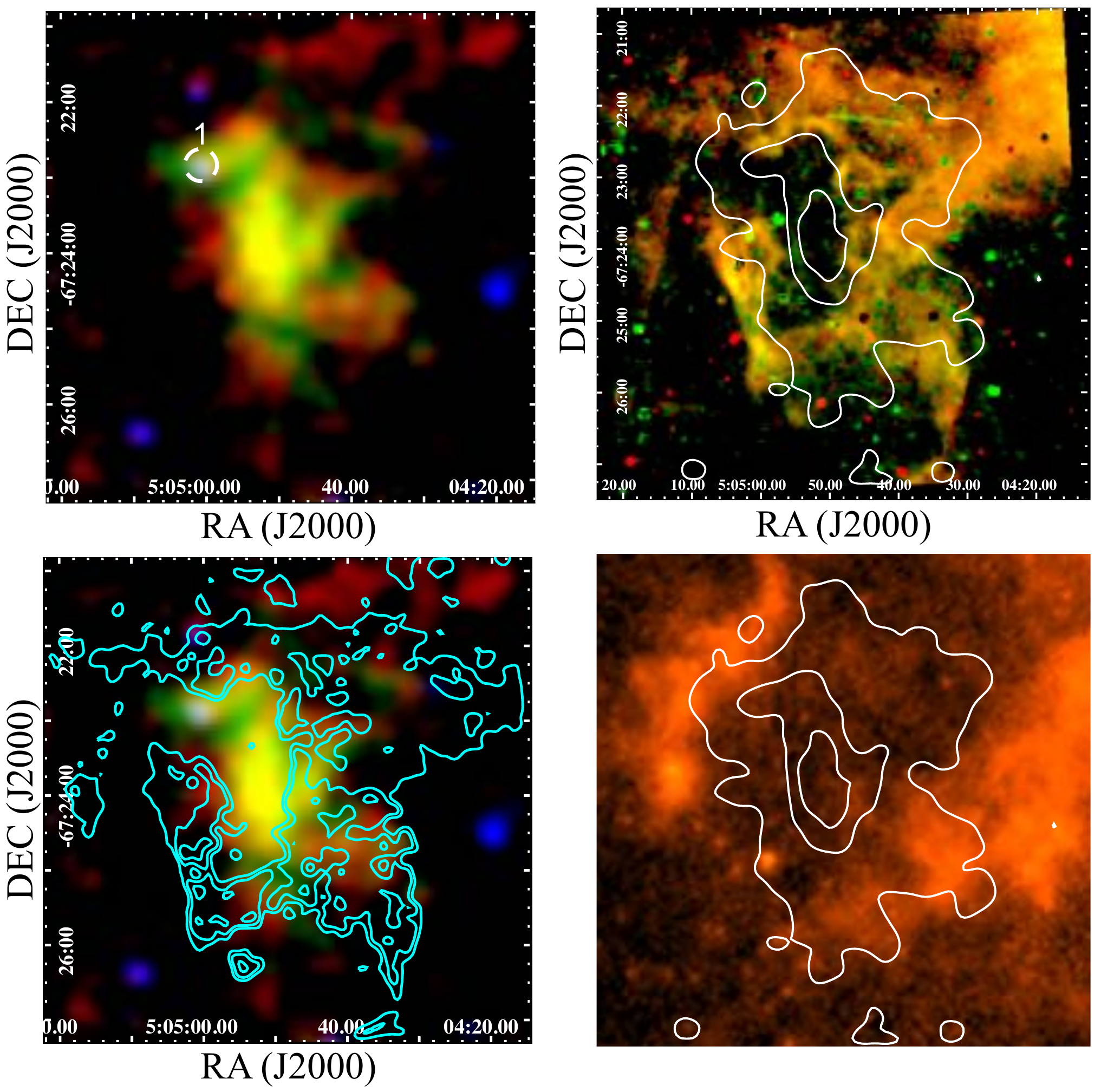}}
\caption{Same as Fig.~\ref{figures:0447-6918_4panel} but for MCSNR~J0504--6723.}
\label{figures:0504-6723_4panel}
\end{center}
\end{figure*}

\begin{figure*}
\begin{center}
\resizebox{6.5in}{!}{\includegraphics[trim= 0cm 0cm 0cm 0cm, clip=true, angle=0]{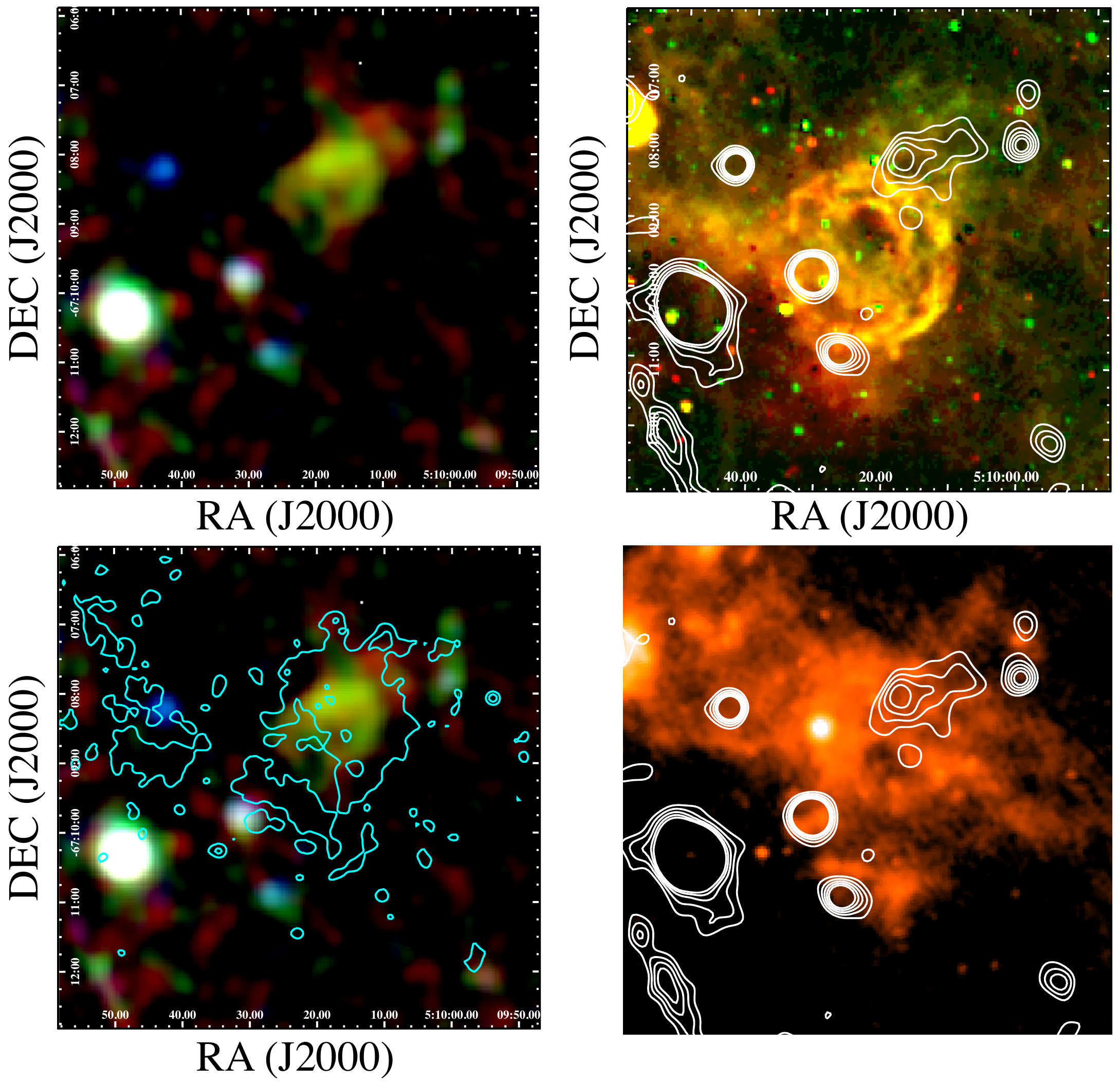}}
\caption{Same as Fig.~\ref{figures:0447-6918_4panel} but for MCSNR~J0510--6708.}
\label{figures:0510-6708_4panel}
\end{center}
\end{figure*}

\begin{figure*}
\begin{center}
\resizebox{6.5in}{!}{\includegraphics[trim= 0cm 0cm 0cm 0cm, clip=true, angle=0]{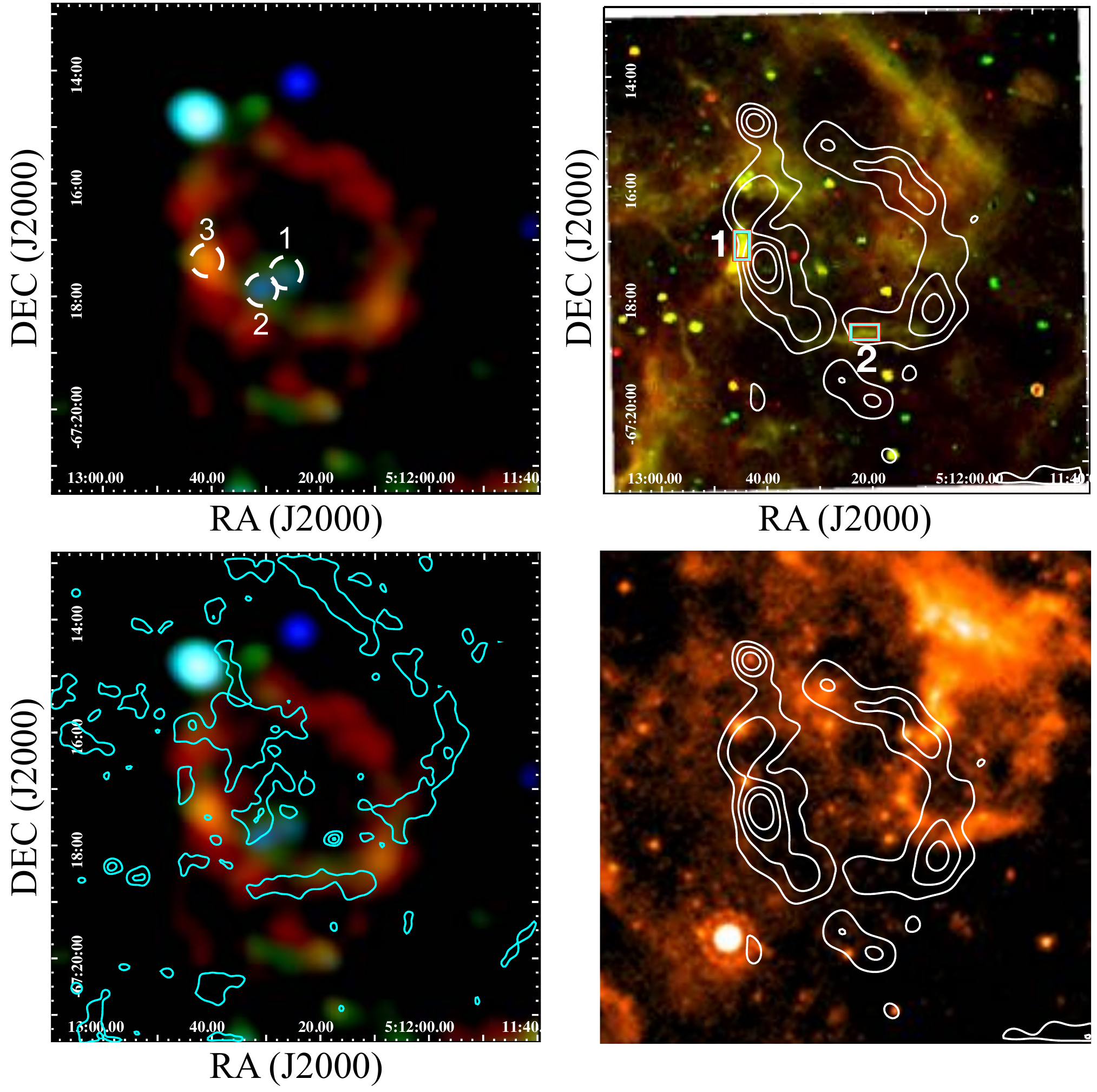}}
\caption{Same as Fig.~\ref{figures:0447-6918_4panel} but for MCSNR~J0512--6717. The numbered cyan boxes in top-right indicate the WiFeS fields from which spectroscopic data were obtained.}
\label{figures:0512-6717_4panel}
\end{center}
\end{figure*}

\begin{figure*}
\begin{center}
\resizebox{6.5in}{!}{\includegraphics[trim= 0cm 0cm 0cm 0cm, clip=true, angle=0]{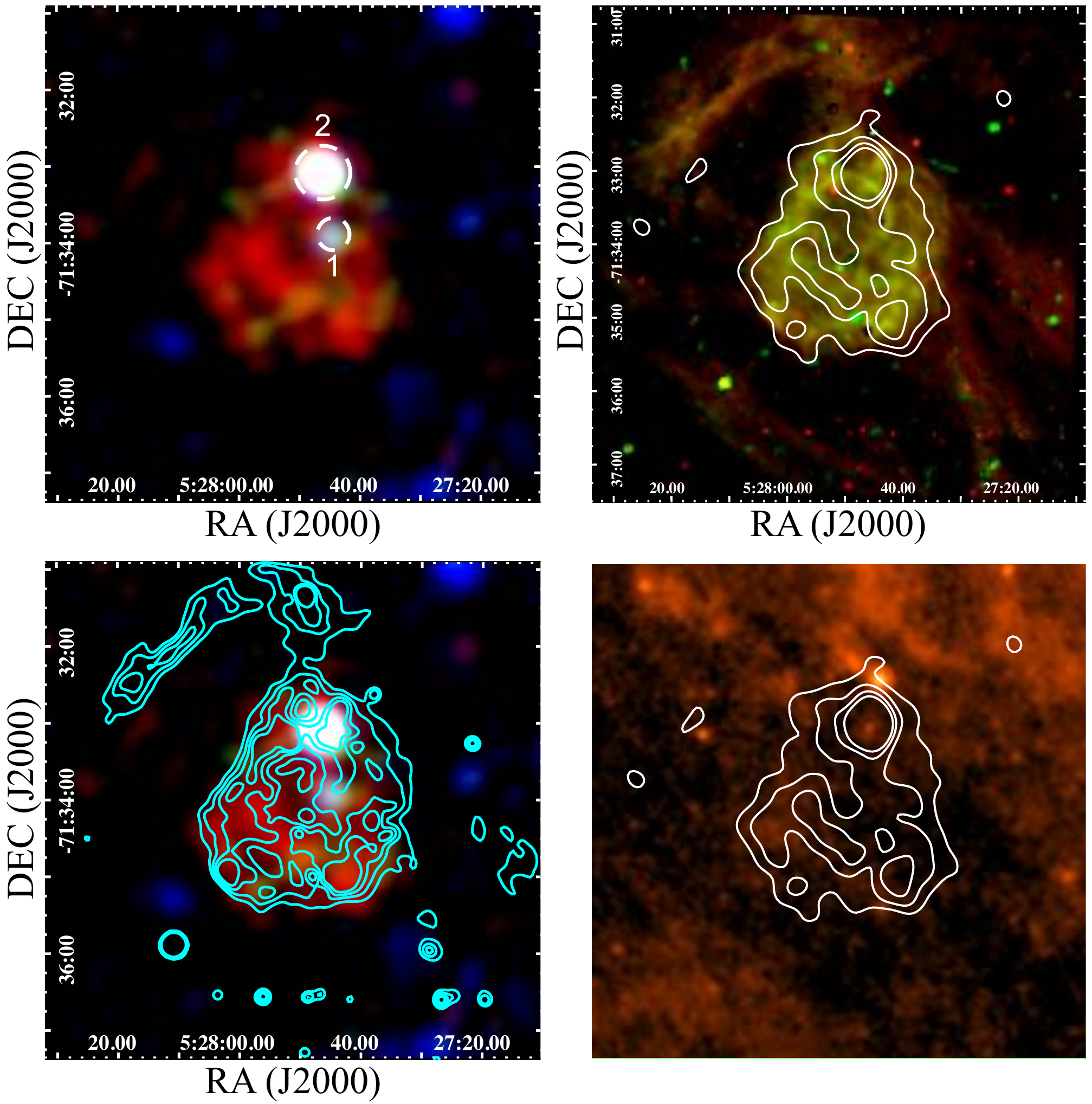}}
\caption{Same as Fig.~\ref{figures:0447-6918_4panel} but for MCSNR~J0527--7134.}
\label{figures:0527-7134_4panel}
\end{center}
\end{figure*}

\begin{figure*}
\begin{center}
\resizebox{6.5in}{!}{\includegraphics[trim= 0cm 0cm 0cm 0cm, clip=true, angle=0]{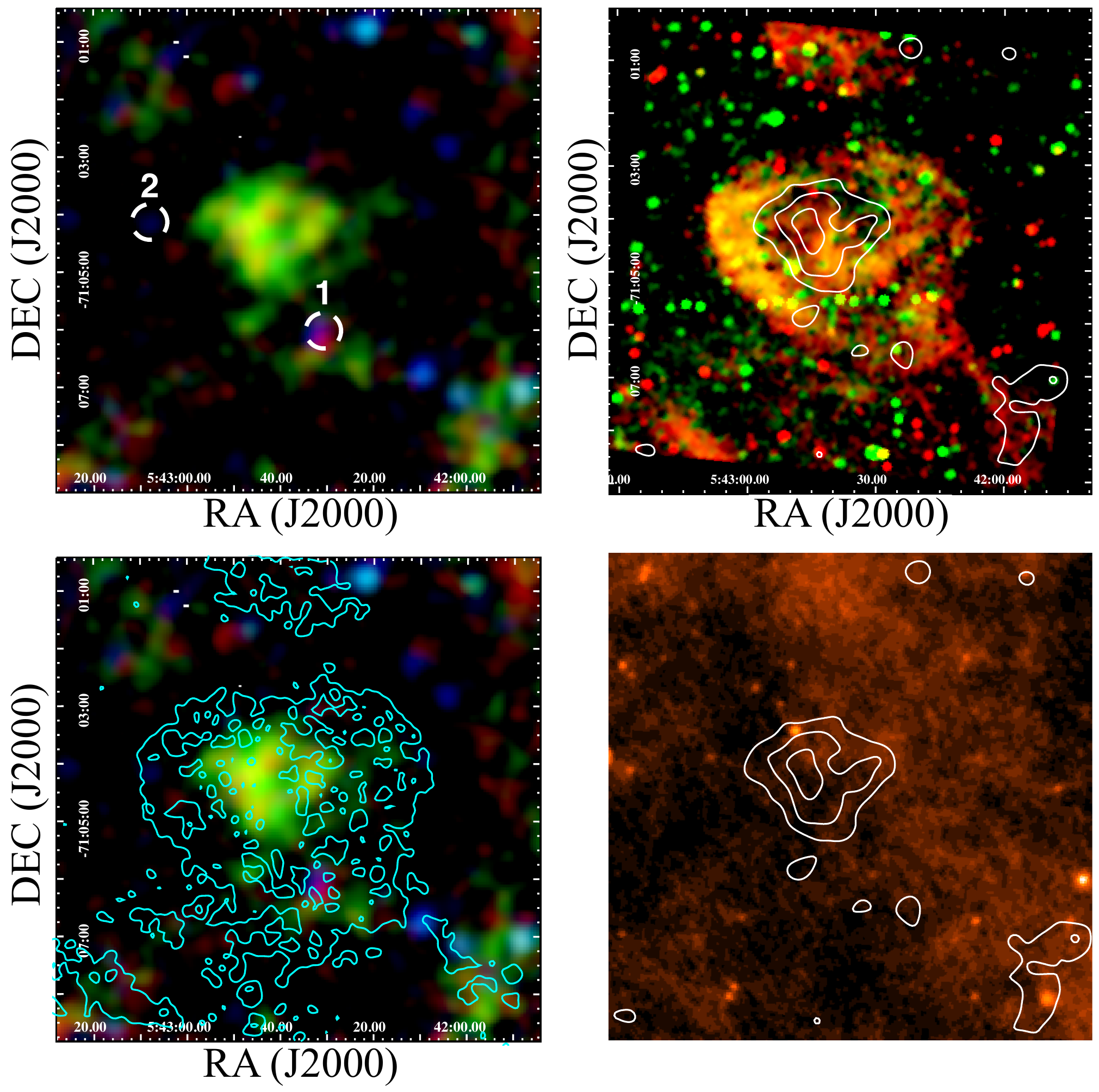}}
\caption{Same as Fig.~\ref{figures:0447-6918_4panel} but for MCSNR~J0542--7104.}
\label{figures:0542-7104_4panel}
\end{center}
\end{figure*}

We produced particle background subtracted, exposure corrected, and adaptively smoothed EPIC images in three energy bands particularly suited to the analysis of SNRs: a soft band from 0.3--0.7~keV that includes strong lines from O; a medium band from 0.7--1.1~keV which contains Fe L-shell lines as well as Ne He$\alpha$ and Ly$\alpha$ lines; and a hard band (1.1--4.2~keV) that includes lines from Mg, Si, S, Ca, Ar, as well as non-thermal continuum if present. To produce these images, we followed the same procedure outlined in detail in many of our previous works, e.g. \citet{Kavanagh2015b,Kavanagh2015c,Kavanagh2016,Maggi2016}. A three-colour image of each of our objects was created, and are shown in the top left panels of Figs.~\ref{figures:0447-6918_4panel}--\ref{figures:0542-7104_4panel}.

\subsection{X-ray point sources}
\label{sect:source_detect}
Point sources located within the extent of our objects may be associated with and provide evidence of the nature of the SN progenitor of our objects. For example, the identification of a remnant compact object suggests an SNR resulted from a core-collapse event. To search for point sources, we extracted images from each of the EPIC instruments in the standard energy bands 0.2--0.5~keV, 0.5--1~keV, 1--2~keV, 2--4.5~keV, and 4.5--12~keV to use as input to the SAS task \texttt{edetect\_chain}. For an adopted minimum-detection likelihood\footnote{The detection likelihood $L$ is defined by $L$~=~$-\mathrm{ln}~p$, where $p$ is the probability that a Poissonian fluctuation of the background is detected as a spurious source.} of 10, we detected several sources potentially associated with our objects. These sources were visually inspected and many that were identified as false sources due to structure in the extended emission of our objects were excluded. Detailed analysis of the remaining sources, including the identification of counterparts, is deferred to Appendix~A. None of our detected point sources could be confirmed as the remnant compact objects of the SN explosions.

\subsection{X-ray spectral analysis}
\label{spec_analysis}
Before extracting spectra we generated vignetting-weighted event lists for each EPIC instrument to correct for the effective area variation across FOV using the SAS task \texttt{evigweight}. We extracted source and background spectra from these corrected event lists with the source spectra extracted from elliptical regions encompassing the extended emission from our objects and the background spectra from a larger annular regions. Detected point sources were excluded in each case. To allow the use of the $\chi^{2}$ statistic in the spectral fitting process, all spectra were rebinned so that each bin contained a minimum of 30 counts. We fitted the EPIC-pn and EPIC-MOS source and background spectra simultaneously in XSPEC \citep{Arnaud1996} version 12.11.1 with abundance tables set to those of \citet[][hereafter W00]{Wilms2000}, the photoelectric absorption cross-sections of \citet{Bal1992}, and atomic data from ATOMDB~3.0.9\footnote{\url{http://www.atomdb.org/index.php}}.

\subsubsection{X-ray background}
\label{x-ray_background}
\begin{figure*}
\begin{center}
\resizebox{7in}{!}{\includegraphics[trim= 0cm 0cm 0cm 0cm, clip=true, angle=0]{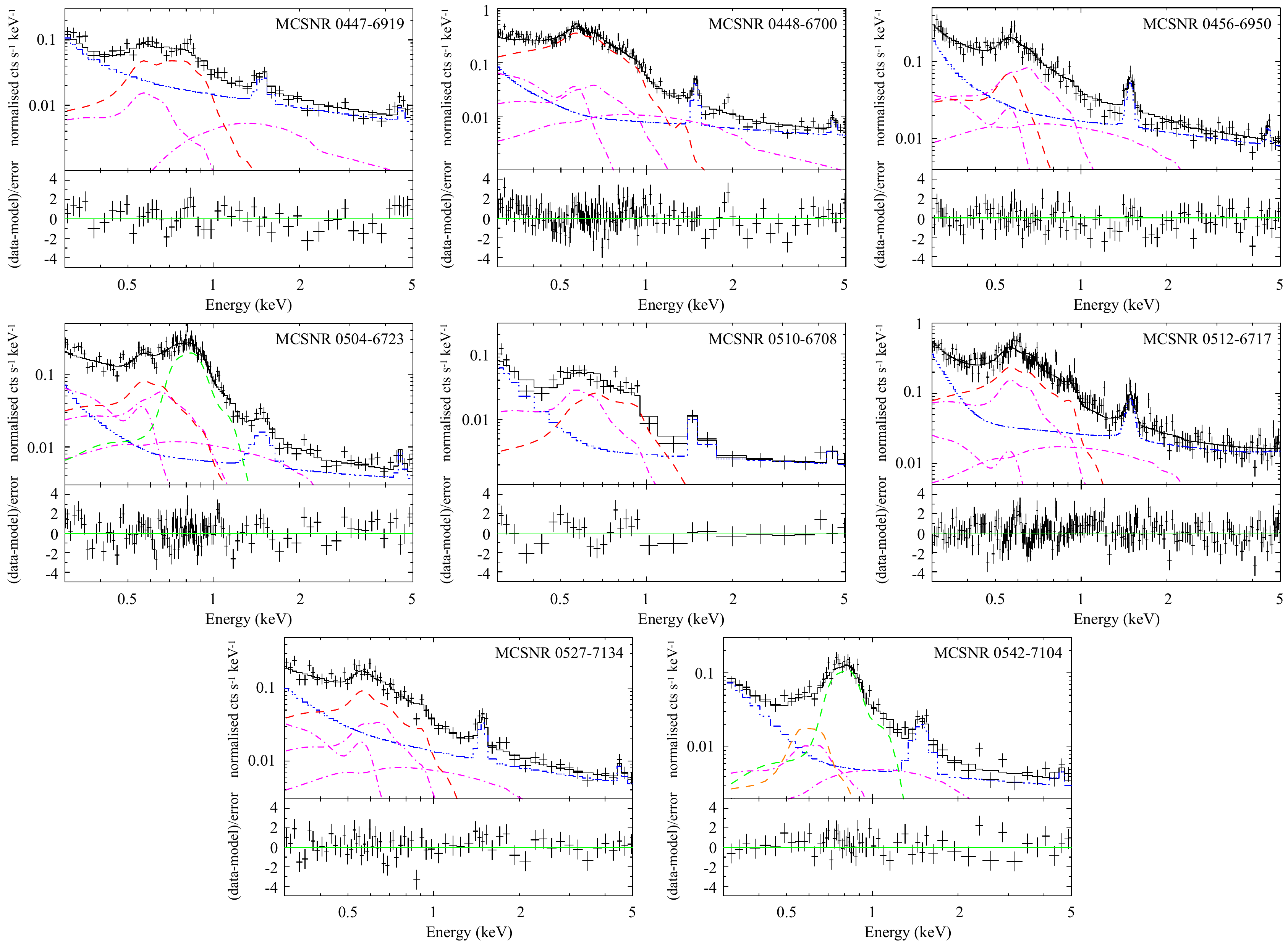}}
\caption{X-ray spectral fits to our SNR sample. The dashed red line shows the fitted thermal plasma component, the magenta dash-dot lines indicate the AXB components, and the blue dash-dot-dot-dot line shows the combined contributions of the QPB, instrumental fluorescence lines, and electronic noise. In the cases of MCSNR~J0504--6723 and MCSNR~J0542--7104, the green and orange dashed lines represent the pure Fe and pure O components, respectively.}\label{fig:spectra}
\end{center}
\end{figure*}

\begin{figure*}
\begin{center}
\resizebox{6.5in}{!}{\includegraphics[trim= 0cm 0cm 0cm 0cm, clip=true, angle=0]{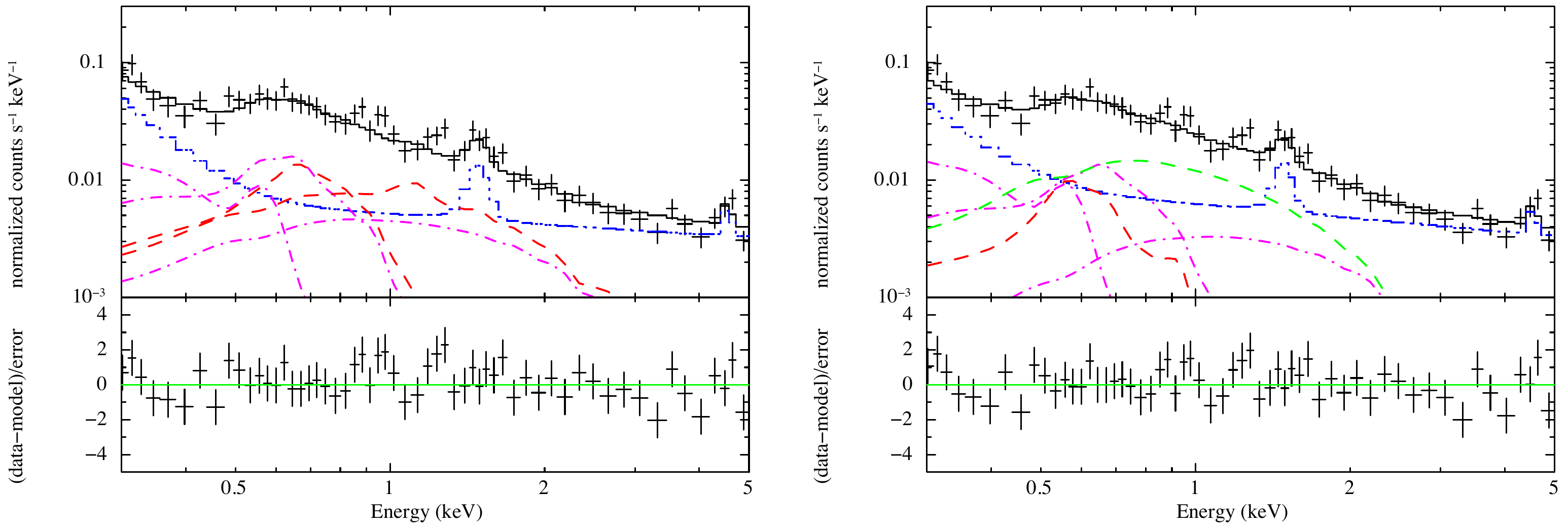}}
\caption{Two-component X-ray spectral fits to the spectrum of MCSNR~J0449--6903. {\it Left}: The absorbed two-temperature thermal plasma model (\texttt{vphabs*(vapec+vapec)}). The dashed red lines show the fitted thermal plasma components, the magenta dash-dot lines indicate the AXB components, and the blue dash-dot-dot-dot line shows the combined contributions of the QPB, instrumental fluorescence lines, and electronic noise. {\it Right}: The absorbed thermal plus non-thermal component fit (\texttt{vphabs*(vapec+pow)}). The lines have the same meaning as in {\it Left} except the green dashed line represents the non-thermal component.}\label{fig:0449_spectra}
\end{center}
\end{figure*}

Detailed descriptions of the X-ray background constituents and spectral modelling can be found in, e.g. \citet{Bozzetto2014,Maggi2014, Kavanagh2015b,Kavanagh2015c,Kavanagh2016,Maggi2016}. As it is important for the interpretation of the plots shown in Figs.~\ref{fig:spectra} and \ref{fig:0449_spectra}, we briefly describe each of the constituents here. The X-ray background consists of the astrophysical X-ray background (AXB) and particle induced background. The AXB contains contributions from the Local Hot Bubble, the Galactic halo, and unresolved active galactic nuclei (AGN). We modelled these using an unabsorbed thermal component for the Local Hot Bubble, absorbed cool and hot thermal components for the Galactic halo emission, and an absorbed power law for the unresolved background AGN \citep{Snowden2008,Kuntz2010}. The spectral parameters of the AGN component were fixed to the well known values of $\Gamma \sim 1.46$ and a normalisation equivalent to 10.5 photons~cm$^{-2}$~s$^{-1}$~sr$^{-1}$ at 1~keV \citep{Chen1997}. The absorbing components comprise both Galactic and LMC contributions. The foreground Galactic absorption was fixed based on the \citet{Dickey1990} HI maps, determined using the HEASARC $N_{\rm{H}}$ Tool\footnote{\url{http://heasarc.gsfc.nasa.gov/cgi-bin/Tools/w3nh/w3nh.pl}}. We allowed the the foreground LMC absorption component, with abundances set to those of the LMC, to vary in our fits.

The particle-induced background is made up of the quiescent particle background (QPB), instrumental fluorescence lines, electronic read-out noise, and residual soft proton (SP) contamination. We constrained the contributions of the first three using FWC spectra extracted from the same detector regions as the our source and background spectra. The EPIC-pn and EPIC-MOS FWC spectra were fitted with the empirical models of \citet{SturmPhD} and \citet{Maggi2016}, respectively, using a diagonal response in XSPEC as these contributions are not subject to the instrumental response. The resulting best-fit model was included and frozen in the fits to our objects' spectra, with only the widths and normalisations of the fluorescence lines allowed to vary and a multiplicative constant to normalise the continuum of the FWC spectra. The residual SP contamination was fitted by a power law not convolved with the instrumental response as described in \citet{Kuntz2008}.

\subsubsection{Extended emission}
For evolved SNRs expanding into an ISM with typical density, the spectra should contain a thermal plasma component of LMC abundance representing shock-heated ISM that has been swept-up by the blast wave. There may also be an additional component due to reverse shock-heated ejecta, which will be characterised by strong emission lines indicative of its metal content. Only the SNRs MCSNR~J0504--6723 and MCSNR~J0542--7104 show obviously enhanced Fe~L-shell emission, consistent with the class of evolved SNRs with centrally-peaked, Fe dominated ejecta \citep[e.g.,][]{Borkowski2006, Maggi2014, Bozzetto2014, Kavanagh2016}.

We fitted the swept-up ISM contribution using both collisional ionisation equilibrium (CIE) and non-equilibrium ionisation (NEI) thermal plasma models, which were absorbed by foreground Galactic and LMC material, namely the \vapec\ and plane-parallel \vpshock\ models \citep{Borkowski2001}, respectively. To model pure-Fe plasma, and in some cases pure-O plasma, we used pure-metal thermal plasma models, achieved by setting all abundances apart from Fe or O to 0. A detailed description of this technique is given in \citet{Kavanagh2016}. The results of our spectral fits are given in Sect.~\ref{results}.

\begin{table*}
\caption{Single component X-ray spectral fit results to our SNR sample. Fits with CIE under the $\tau$ column are \texttt{vphabs}*\vapec\ models while fits with a numeric $\tau$ value are \texttt{vphabs}*\vpshock\ models.}
\label{table:xray_fits}
\begin{footnotesize}
\begin{center}
\begin{tabular}{llllllll}
 \hline
\hline
$N_{\rm{H}}$~$^{\left(a\right)}$ & k$T$ & $\tau$ & Abundances & $\chi^{2}_{\nu}$ (d.o.f.) & $EM$ & log~$F_{\rm{X}}$~$^{\left(b\right)}$ & log~$L_{\rm{X}}$~$^{\left(c\right)}$ \\
 ($10^{22}$~cm$^{-2}$) & (keV) & ($10^{11}$~s~cm$^{-3}$) & ($Z$/Z$_{\sun}$) & & ($10^{57}$~cm$^{-3}$) & (erg~cm$^{-2}$~s$^{-1}$) & (erg~s$^{-1}$) \\
\hline
 & & & & & & & \\
 \multicolumn{8}{c}{\underline{\hspace{2cm}{\bf MCSNR~J0447--6919}\hspace{2cm}}} \\
\rule{0pt}{3ex}0.59 (0.46--0.73) & 0.23 (0.20--0.30) & 2.63 (0.90--8.62) & W00 & 1.10 (290) & 10.11 (4.17--21.43) & -13.57 & 35.16 \\
\rule{0pt}{3ex}0.70 (0.58--0.82) & 0.15 (0.14--0.19) & 5.26 ($>0.86$) & Fe: 14.15 ($>1.40$) & 1.03 (289) & 43.92 (18.14--122.26) & -13.38 & 35.49 \\
 & & & & & & & \\
 
\multicolumn{8}{c}{\underline{\hspace{2cm}{\bf MCSNR~J0448--6700}\hspace{2cm}}} \\
 \rule{0pt}{3ex}  & & & O: 0.25 (0.20--0.33) & & & & \\
 0.14 (0.09--0.19) & 0.19 (0.18--0.21) & $>17.85$ & Ne: 0.32 (0.21--0.47) & 1.28 (357) & 32.71 (21.22--48.71) & -12.81 & 35.18 \\
  & & & Mg: 0.83 (0.21--1.74) & & & & \\
 & & & Fe: 0.61 (0.35--0.84) & & & & \\
 & & & & & & & \\

 \multicolumn{8}{c}{\underline{\hspace{2cm}{\bf MCSNR~J0456--6950}\hspace{2cm}}} \\
 \rule{0pt}{3ex}$<0.04$ & 0.14 (0.12--0.16) & CIE & W00 & 1.15 (479) & 3.05 (2.24--5.17) & -13.65 & 34.11 \\
 & & & & & & & \\

\multicolumn{8}{c}{\underline{\hspace{2cm}{\bf MCSNR~J0510--6708}\hspace{2cm}}} \\

 \rule{0pt}{3ex}0.73 (0.28--0.99) & 0.20 (0.16--0.32) & CIE & W00 & 1.01 (70) & 9.83 (1.11--47.62) & -13.87 & 34.92 \\
 & & & & & & & \\
\multicolumn{8}{c}{\underline{\hspace{2cm}{\bf MCSNR~J0512--6717}\hspace{2cm}}} \\
\rule{0pt}{3ex}0.14 (0.09--0.21) & 0.19 (0.18--0.21) & CIE & O: 0.25 (0.19--0.31) & 1.14 (914) & 20.04 (13.80--32.93) & -12.97 & 35.02 \\
 & & & Fe: 0.60 (0.33--1.10) & & & & \\
 
 & & & & & & & \\
\multicolumn{8}{c}{\underline{\hspace{2cm}{\bf MCSNR~J0527--7134}\hspace{2cm}}} \\
 \rule{0pt}{3ex}0.02 ($<0.14$) & 0.53 (0.19--1.32) & 0.26 (0.09--12.66) & O: 0.36 (0.19--0.59) & 0.97 (271) & 0.71 (0.36--3.53) & -13.37 & 34.43 \\
 & & & Ne: 0.53 (0.21--1.00) & & & & \\

 \rule{0pt}{3ex}0.06 ($<0.17$) & 0.19 (0.18--0.21) & CIE & O: 0.22 (0.12--0.41) & 0.98 (272) & 7.19 (3.28--17.92) & -13.38 & 34.54 \\
 & & & Ne: 0.59 (0.28--1.19) & & & & \\
 
 & & & & & & & \\
\hline
\end{tabular}
\tablefoot{The numbers in parentheses are the 90\% confidence intervals.
$^{\left(a\right)}$ Absorption abundances fixed to those of the LMC.
$^{\left(b\right)}$ 0.3--8~keV absorbed X-ray flux. 
$^{\left(c\right)}$ 0.3--8~keV de-absorbed X-ray luminosity, adopting a distance of 50~kpc to the LMC. W00 refers to \citet[][see text]{Wilms2000}.}
\end{center}
\end{footnotesize}
\end{table*}%

\begin{table*}
\begin{scriptsize}
\caption{X-ray spectral fit results for the Fe-rich SNRs.}
\begin{center}
\label{Fe_fits}
\begin{tabular}{llllllllll}
 \hline
\hline
$N_{\rm{H}}$~$^{\left(a\right)}$ & k$T_{\mathrm{sh}}$ & $EM_{\mathrm{sh}}$ &
 k$T_{\mathrm{Fe}}$ & $EM_{\mathrm{Fe}}$ & k$T_{\mathrm{O}}$ & $EM_{\mathrm{O}}$ & $\chi^{2}_{\nu}$ (d.o.f.) & log~$F_{\rm{X}}$~$^{\left(b\right)}$ & log~$L_{\rm{X}}$~$^{\left(c\right)}$ \\
 $(10^{22}$~cm$^{-2}$) & (keV) & ($10^{57}$~cm$^{-3}$) & (keV) & ($10^{52}$~cm$^{-3}$) & (keV) & ($10^{53}$~cm$^{-3}$) & & (erg~cm$^{-2}$~s$^{-1}$) & (erg~s$^{-1}$) \\
\hline
 & & & & & & & & & \\
  \multicolumn{10}{c}{\underline{\hspace{2cm}{\bf MCSNR~J0504--6723}\hspace{2cm}}} \\
 \multicolumn{10}{c}{\rule{0pt}{3ex}Model: \vphabs*(\vpshock+\vnei$_{\mathrm{Fe}}$)} \\
\rule{0pt}{3ex}0.04 ($<0.11$) & 0.24 (0.20--0.34) & 1.80 (1.30--2.78) & 0.68 (0.66--0.71) & 9.23 (7.35--10.71) & -- & -- & 1.20 (324) & -13.00 & 34.63 \\

\multicolumn{10}{c}{\rule{0pt}{3ex}Model: \vphabs*(\vpshock+\vnei$_{\mathrm{Fe}}$+\vnei$_{\mathrm{O}}$)} \\

\rule{0pt}{3ex}0.04 ($<0.10$) & 0.24 (0.18--0.34) & 1.78 (1.25--2.78) & 0.68 (0.66--0.71) & 9.22 (7.56--10.73) & =k$T_{\mathrm{Fe}}$ & 0.03 ($<$19.53) & 1.20 (323) & -13.00 & 34.63 \\
 & & & & & & & & & \\
 \multicolumn{10}{c}{\underline{\hspace{2cm}{\bf MCSNR~J0542--7104}\hspace{2cm}}} \\
 \multicolumn{10}{c}{\rule{0pt}{3ex}Model: \vphabs*(\vnei$_{\mathrm{Fe}}$)} \\
\rule{0pt}{3ex}0.16 (0.08--0.26) & -- & -- & -- & 0.57 (0.55--0.60) & 7.55 (6.27--9.39) & -- & 1.25 (402) & -13.42 & 34.28 \\

\multicolumn{10}{c}{\rule{0pt}{3ex}Model: \vphabs*(\vnei$_{\mathrm{Fe}}$+\vnei$_{\mathrm{O}}$)} \\

\rule{0pt}{3ex}0.26 (0.16--0.37) & -- & -- & 0.58 (0.54--0.62) & 8.58~(6.89--10.91) & 0.19 (0.14--0.24) & 9.06~(4.33--21.03) & 1.02 (402) & -13.38 & 34.46 \\
\hline
\end{tabular}
\tablefoot{The numbers in parentheses are the 90\% confidence intervals.
$^{\left(a\right)}$ Absorption abundances fixed to those of the LMC.
$^{\left(b\right)}$ 0.3--8~keV absorbed X-ray flux.
$^{\left(c\right)}$ 0.3--8~keV de-absorbed X-ray luminosity, adopting a distance of 50~kpc to the LMC. The ionisation parameter for the \vpshock\ component in the fits ($\tau_{\rm{sh}}$) was fixed to $10^{12}$~s~cm$^{-3}$ representing a plasma in CIE (see text).}
\end{center}
\end{scriptsize}
\end{table*}%

\begin{table*}
\begin{scriptsize}
\caption{MCSNR~J0449--6903 two-component X-ray spectral fit results.}
\begin{center}
\label{0449_2T}
\begin{tabular}{llllllllll}
 \hline
\hline
$N_{\rm{H}}$~$^{\left(a\right)}$ & k$T_{\mathrm{1}}$ & $EM_{\mathrm{1}}$ &
 k$T_{\mathrm{2}}$ & $\Gamma$ &  $EM_{\mathrm{2}}$ & {\it norm.} & $\chi^{2}_{\nu}$ (d.o.f.) & log~$F_{\rm{X}}$~$^{\left(b\right)}$ & log~$L_{\rm{X}}$~$^{\left(c\right)}$ \\
 $(10^{22}$~cm$^{-2}$) & (keV) & ($10^{57}$~cm$^{-3}$) & (keV) &  &  ($10^{57}$~cm$^{-3}$) & ($10^{-5}$ cm$^{-5}$) & & (erg~cm$^{-2}$~s$^{-1}$) & (erg~s$^{-1}$) \\
\hline
 & & & & & & & & &  \\
  \multicolumn{9}{c}{\rule{0pt}{3ex}Model: \vphabs*(\vapec+\vapec)} \\
\rule{0pt}{3ex}0.11 (0.04--0.22) & 0.26 (0.21--0.37) & 0.55 (0.25--1.21) & 2.81 (2.06--4.60) & -- & 0.91 (0.74--1.07) & -- & 1.10 (224) & -13.44 & 34.20 \\
 \multicolumn{9}{c}{\rule{0pt}{3ex}Model: \vphabs*(\vapec+\pow)} \\
\rule{0pt}{3ex}0.27 (0.15--0.44) & 0.17 (0.05--0.26) & 1.11 (0.06--7.99) & -- & 2.82 (2.34--3.35) & -- & 1.54 (1.16--2.03) & 1.09 (224) & -13.46 & 34.46 \\
 & & & & & & & &  \\
\hline\noalign{\smallskip}
\end{tabular}\\
\tablefoot{The numbers in parentheses are the 90\% confidence intervals.
$^{\left(a\right)}$ Absorption abundances fixed to those of the LMC.
$^{\left(b\right)}$ 0.3--8~keV absorbed X-ray flux.
$^{\left(c\right)}$ 0.3--8~keV de-absorbed X-ray luminosity, adopting a distance of 50~kpc to the LMC.}
\end{center}
\end{scriptsize}
\end{table*}

\subsection{Extended source position and size}
\label{size_fits_sect}
\begin{figure*}
\begin{center}
\resizebox{6.5in}{!}{\includegraphics[trim= 0cm 0cm 0cm 0cm, clip=true, angle=0]{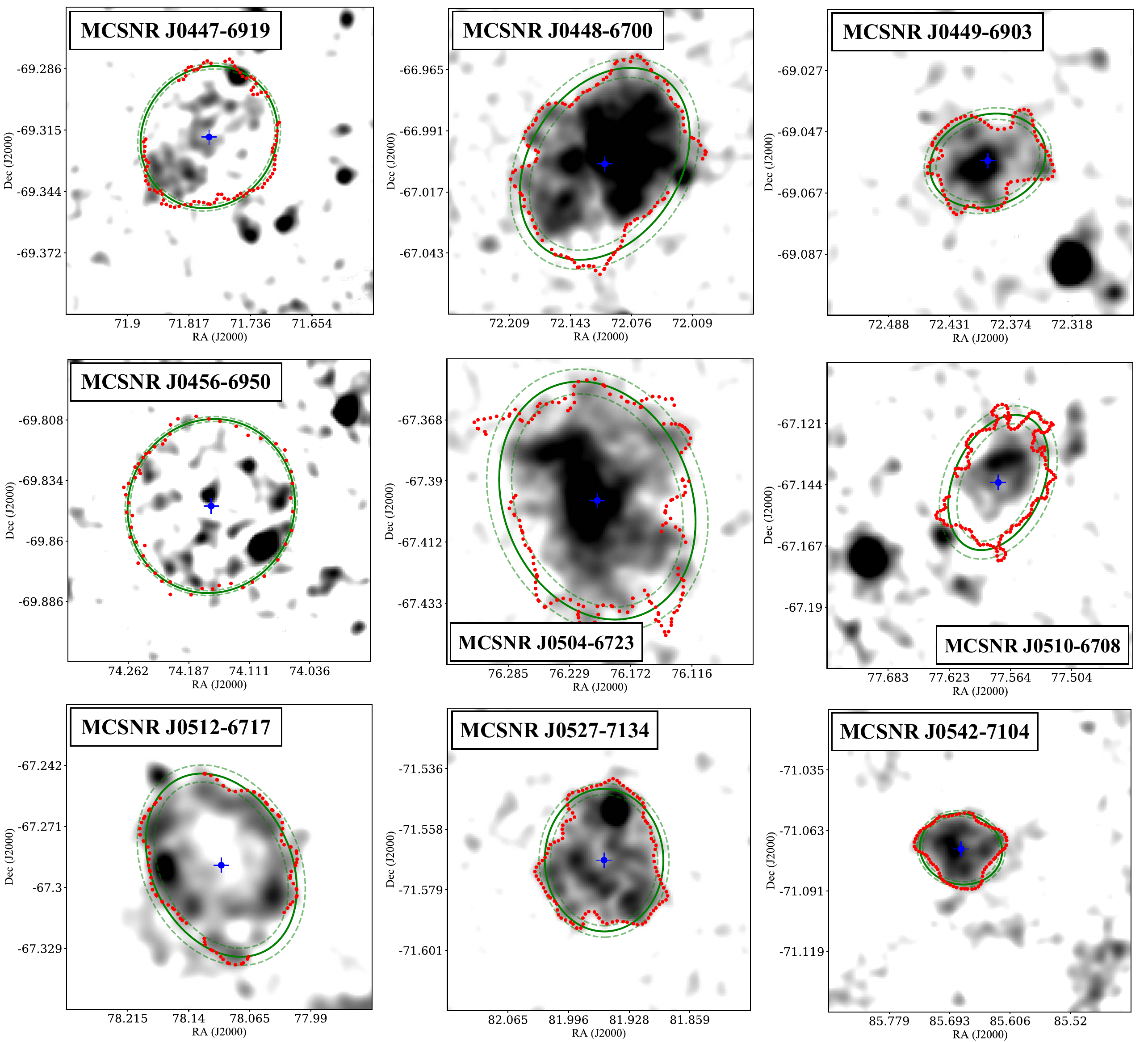}}
\caption{\xmm\ EPIC images of each of our new SNRs with the best fit dimensions. The red points delineate the contour level used in the fits (see Sect.~\ref{size_fits_sect}). The green solid line shows the best-fit ellipse to the contour, with the dashed lines indicating the 1$\sigma$ error on the fit. The blue plus-sign marks the best-fit centre of the SNRs. See Sect.~\ref{results} for details on individual fits.
}
\label{size}
\end{center}
\end{figure*}

To estimate the size of the remnants we used a similar method to that described in \citet{Kavanagh2015b}, which fits an ellipse to the outer X-ray contours of the SNR. While fitting the outer X-ray contours is applicable to some of our sample, it did not always represent the true extent of the object, particularly in those with no shell emission, which instead may be better traced by either the optical \ratio\ contour (e.g. MCSNR~J0447--6919), a hybrid outer contour comprising X-ray and \ratio\ contours in different shell regions (e.g. MCSNR~J0448--6700), or radio contours (e.g. MCSNR~J0456-6950). More details are provided on individual cases in Sect.~\ref{results}. The best-fit dimensions and errors are given in Table~\ref{table:size_fits} and shown on the X-ray images in Fig.~\ref{size}. 

\begin{table*}
\caption{Best fit position and dimensions of the SNR sample.
}
\begin{center}
\label{table:size_fits}
\begin{tabular}{llllll}
 \hline
\hline
ID & RA & Dec & $D_{\rm{maj}} \times D_{\rm{min}}$ & $D_{\rm{maj}} \times D_{\rm{min}}$ & PA \\
 & (J2000) & (J2000) & (arcsec) & (pc) & (degrees EoN) \\
\hline
MCSNR~J0447--6919 & 04:47:10.26 & --69:19:07.4 & $246.6\times225.0~\left(\pm9.6\right)$ & $59.76\times54.58~\left(\pm2.37\right)$ & 147.6  \\
MCSNR~J0448--6700 & 04:48:25.65 & --67:00:22.1 & $  319.2\times229.2~\left(\pm29.4\right)$ & $ 77.32\times55.54~\left(\pm7.15\right)$ & 146.0 \\
MCSNR~J0449--6903 & 04:49:35.47 & --69:03:25.6 & $ 140.4\times 108.6~\left(\pm13.8\right)$ & $33.97\times 26.34~\left(\pm3.37\right)$ & 108.1 \\
MCSNR~J0456--6950 & 04:56:38.55 & --69:50:44.1 & $ 271.5\times 256.8~\left(\pm7.6\right)$ & $65.80\times 62.26~\left(\pm1.80\right)$ & 151.5 \\
MCSNR~J0504--6723 & 05:04:49.18 & --67:23:51.0 & $ 312.0\times 239.4 ~\left(\pm31.8\right)$ & $75.62\times 58.00 ~\left(\pm7.71\right)$ & 21.2  \\
MCSNR~J0510--6708 & 05:10:18.59 & --67:08:38.0 & $ 193.8\times121.2~\left(\pm24.6\right)$ & $47.03\times29.46~\left(\pm5.92\right)$ & 155.9  \\
MCSNR~J0512--6717 & 05:12:24.35 & --67:17:24.7 & $ 328.2\times236.4~\left(\pm29.4\right)$ & $79.57\times 57.33~\left(\pm7.12\right)$ & 26.3  \\
MCSNR~J0527--7134 & 05:27:50.02 & --71:34:10.4 & $ 183.6\times152.4~\left(\pm14.4\right)$ & $44.52\times36.98~\left(\pm3.55\right)$ & 0.7  \\
\hline
\end{tabular}
\end{center}
\end{table*}%

\subsection{MCSNR~J0512--6717 optical spectroscopy}
\label{opt-spec}
\begin{figure*}
\begin{center}
\resizebox{6.5in}{!}{\includegraphics[trim= 0cm 0cm 0cm 0cm, clip=true, angle=0]{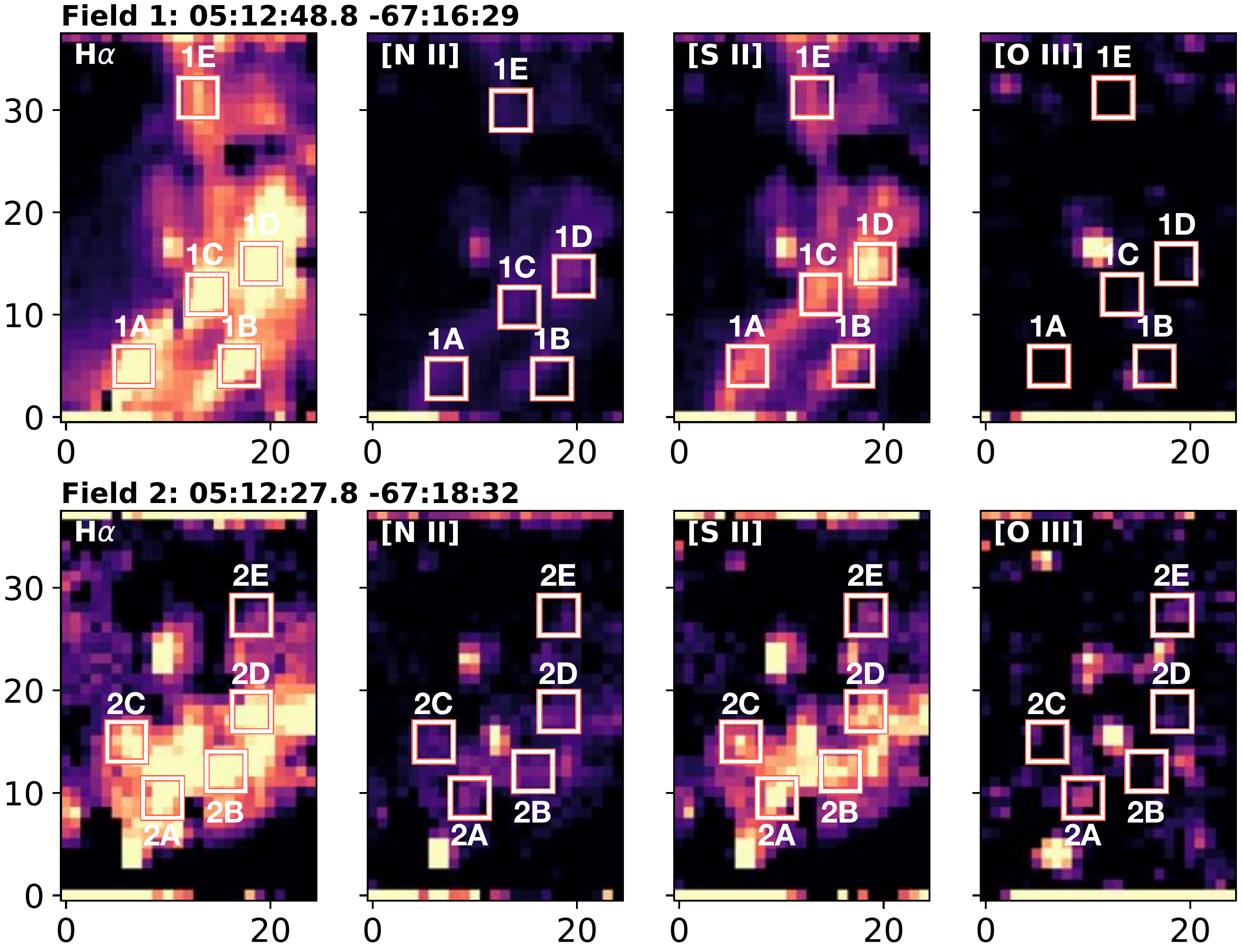}}
\caption{Emission line images produced by collapsing sub-cubes extracted from around \ha, the \sii$\lambda$6716\AA,$\lambda$6731\AA\ doublet, \nii$\lambda$6583\AA, and \oiii$\lambda$5007\AA. The fields and species are indicated in the panels. Also shown are the $4\arcsec\times4\arcsec$ apertures used for spectral extraction. The color scale is linear and the scale limits for each field are the same.}
\label{figures:wifes_apertures}
\end{center}
\end{figure*}

\begin{figure*}
\begin{center}
\resizebox{6.5in}{!}{\includegraphics[trim= 0cm 0cm 0cm 0cm, clip=true, angle=0]{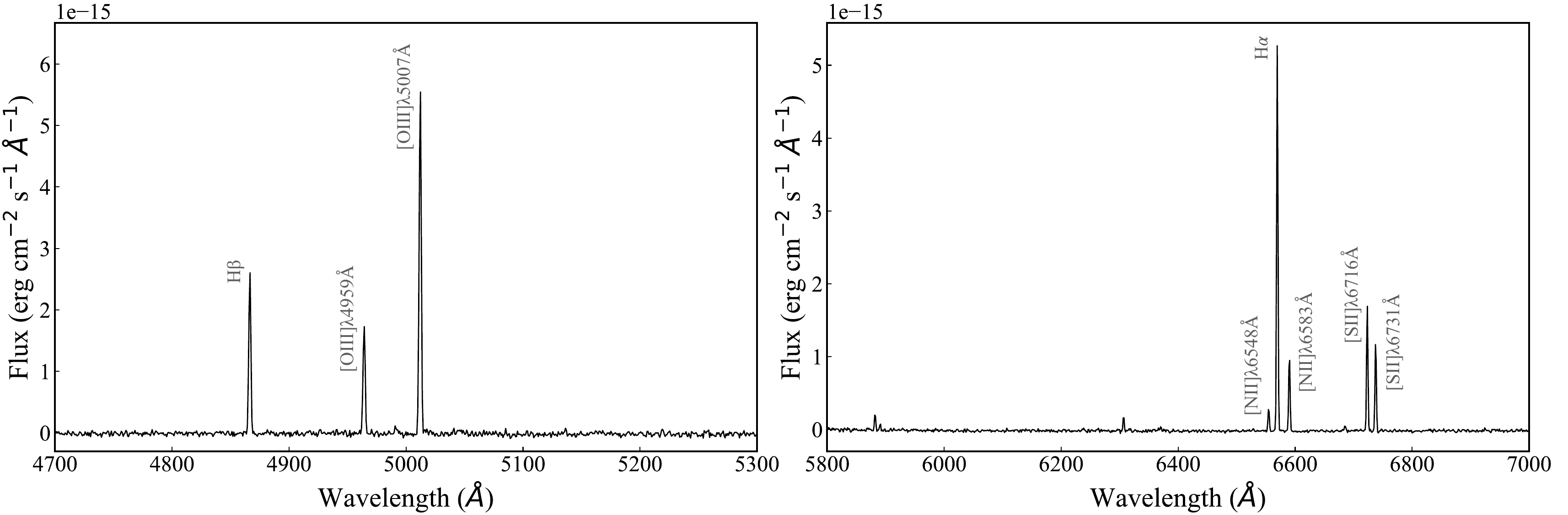}}
\caption{Continuum subtracted blue (left) and red (right) spectra from extraction aperture 1A (see Fig.~\ref{figures:wifes_apertures}). The observed emission lines are indicated on the plot.}
\label{figures:wifes_example}
\end{center}
\end{figure*}

To visualise the IFU data, we extracted and collapsed sub-cubes from around the \ha\ line, the \sii$\lambda$6716\AA,$\lambda$6731\AA\ doublet, \nii$\lambda$6583\AA, and \oiii$\lambda$5007\AA\ in the full-band red or blue camera data-cubes for each field. The resulting images are shown in Fig.~\ref{figures:wifes_apertures} with Field~1 shown in the top row and Field~2 in the bottom. There is clear filamentary structure evident in each field, with \ha\ and \sii\ images being particularly bright. We identified five regions in each field which are free from point source contamination (those sources which are evident in each emission line image of a given field) and extracted spectra from a $4\arcsec\times4\arcsec$ aperture centred on each position. These are shown as the boxes in Fig.~\ref{figures:wifes_apertures} and sample extracted spectra are shown in Fig.~\ref{figures:wifes_example}. We then identified the \hb, \ha, \sii, \nii, and \oiii\ emission lines in the spectra and fitted Gaussian profiles to each.

We estimated the line fluxes for each aperture and calculated the \nii/\ha, \sii/\ha, and \sii\ ratios, the values of which are indicative of the SNR nature of the filaments. The fluxes and ratios are shown for each aperture of each field in Table~\ref{table:optical_spectra_fits}.

\begin{table*}
\caption{Optical line intensities and ratios for extracted from the apertures of Fields 1 and 2 of MCSNR~J0512--6717. The aperture positions are shown in Fig.~\ref{figures:wifes_apertures}. All fluxes are in units of erg~cm$^{-2}$~s$^{-1}$~\AA$^{-1}$. Dispersion errors are as in \citet{Dopita2007,Dopita2010,Childress2014}. The \nii\ and \sii\ ratios relative to \ha\ correspond to the summation of their doublet lines. The last column shows the electron density ($n_{e}$) derived from the \sii\ ratio and the updated calibration curve of \citet{Proxauf2014} with LDL meaning $n_{e}$ is in the low-density limit. The fainter \oiii$\lambda$4959\AA\ and \nii$\lambda$6548\AA\ lines could not be measured in 2A, 2B, and 2C. To determine the \nii/\ha\ ratios in these cases we used the fact that this is set by atomic physics to be $\sim2.9$ \citep{Acker1989} and the measured \nii$\lambda$6583\AA\ to determine the ratio.}
\begin{center}
\begin{footnotesize}
\label{table:optical_spectra_fits}
\begin{tabular}{lcccccccccccc}
\hline\hline\noalign{\smallskip}
Aper. & \hb & \oiii & \oiii & \ha & \nii & \nii & \sii & \sii & \nii/\ha & \sii/\ha & \sii\ ratio & $n_{e}$ \\

 & & $\lambda$4959\AA & $\lambda$5007\AA &  & $\lambda$6548\AA & $\lambda$6583\AA & $\lambda$6716\AA & $\lambda$6731\AA & & &  & (cm$^{-3}$) \\

\noalign{\smallskip}\hline\noalign{\smallskip}
	1A &	5.69e-15 & 4.03e-15 & 1.23e-14 & 1.50e-14 & 8.75e-16 & 3.02e-15 & 4.85e-15 & 3.47e-15 & 0.26 & 0.55 & 1.40 & 55.2 \\

	1B &	4.05e-15 & 4.54e-15 & 1.40e-14 & 1.24e-14 & 6.07e-16 & 1.68e-15 & 2.60e-15 & 1.82e-15 & 0.18 & 0.36 & 1.43 & LDL \\
	
	1C &    3.18e-15 & 2.94e-15 & 8.82e-15 & 9.95e-15 & 4.47e-16 & 1.47e-15 & 2.54e-15 & 1.78e-15 & 0.19 & 0.43 & 1.43 & LDL \\
	
	1D &	4.83e-15 & 4.28e-15 & 1.28e-14 & 1.62e-14 & 8.76e-16 & 2.68e-15 & 4.06e-15 & 2.83e-15 & 0.22 & 0.43 & 1.44 & LDL \\
	
	1E &	2.82e-15 & 2.13e-15 & 6.36e-15 & 1.09e-14 & 6.65e-16 & 1.90e-15 & 3.84e-15 & 2.82e-15 & 0.24 & 0.61 & 1.36 & 74.6 \\
	
	2A &	3.84e-16 & -- & -- & 1.27e-15 & 2.49e-16 & 1.82e-16 & 3.57e-16 & 2.59e-16 & 0.45 & 0.49 & 1.38 & 65.0 \\
	
	2B &	2.95e-16 & -- & 5.76e-16 & 1.18e-15 & 3.68e-17 & 1.81e-16 & 2.82e-16 & 2.36e-16 & 0.24 & 0.44 & 1.19 & 210.8 \\	
	
	2C &	1.07e-15 & -- & -- & 2.96e-15 & 4.09e-16 & 5.51e-16 & 1.08e-15 & 8.21e-16 & 0.43 & 0.64 & 1.32 & 102.6 \\	
	
	2D &	8.50e-16 & 3.81e-16 & 1.28e-15 & 2.45e-15 & 1.93e-16 & 4.28e-16 & 9.16e-16 & 7.19e-16 & 0.25 & 0.67 & 1.27 & 134.5 \\
	
	2E &	1.22e-15 & 8.44e-16 & 2.29e-15 & 3.95e-15 & 1.74e-16 & 8.83e-16 & 1.54e-15 & 1.24e-15 & 0.27 & 0.70 & 1.25 & 159.0 \\

\hline
\end{tabular}
\end{footnotesize}
\end{center}
\end{table*}%

\section{Results}
\label{results}

\subsection{Non-confirmed SNRs}
\label{ncc-res}
\label{0457-res}
Only one of our SNR candidates, 0457--6739, failed to meet the criteria for SNR classification with the current data. This object was identified as a candidate SNR because faint radio continuum emission was detected at 20~cm \citep{Bozzetto2017}. Shell-like optical emission was also detected. The multi-wavelength images of 0457--6739 are shown in Fig.~\ref{figures:0457-6739_4panel}. No X-ray emission was detected from this candidate. Bright optical emission is seen but this is inconsistent with an SNR origin with \ratio$<0.4$. Sporadic regions of \ratio$>0.4$ are seen which seem to trace the edge of a higher density region evident in the $24~\mu$m image, though appear inconsistent with an SNR shell. The faint radio emission detected from this object in \citet{Bozzetto2017} could not be determined to be non-thermal in origin. Because of the lack of a clear optical shell with enhanced \sii\ and X-ray emission, we cannot confirm this candidate as an SNR. We searched for this SNR candidate in our new ASKAP \citep{2021MNRAS.506.3540P} and ATCA \citep{2021MNRAS.507.2885F} high resolution and sensitivity images and we did not find any signature that may reflect the presence of SNR.

\subsection{Confirmed SNRs}
\label{cc-res}
\subsubsection{MCSNR~J0447--6919}
\label{0447-res}
The candidate 0447--6918 was selected for our observations due to its shell-type morphology in the optical with enhanced \sii\ and \ratio$>0.4$, indicative of an SNR. Very faint 20~cm radio continuum emission was detected in the western region, however, no reliable flux density estimate or spectral index determination was possible \citep[][their Fig.~1, top-left]{Bozzetto2017}. The multi-wavelength images of MCSNR~J0447--6919 are shown in Fig.~\ref{figures:0447-6918_4panel}. Faint extended X-ray emission, primarily in the 0.7--1.1~keV range, is detected with a northwest-southeast elongated morphology. Bright \halpha\ emission is seen in the east of the remnant, along with faint, shell-like \sii\ emission in other regions. The \ratio=0.4 contour shows that the eastern region bright in \halpha\ is not consistent with an SNR origin. However, the distinct shell seen in all other regions shows \ratio$>0.4$, indicating the presence of an SNR shock. This shell also surrounds the extended X-ray emission, appearing to confine the hot gas in the northwest and southeast regions. This observed multi-wavelength morphology is somewhat similar to MCSNR~J0527--7104 \citep{Kavanagh2016}, though the X-ray emission is much fainter in MCSNR~J0447--6919. There is a slight southwest-northeast ambient density gradient evident from the $24~\mu$m image. This may explain the slight ovality and brightness variation in the optical morphology, with the shell appearing fainter and more extended into the lower density region. We estimated the SNR position and size using the outer \ratio$>0.4$ contour, omitting the region bright in \halpha\ to the east, the results of which are given in Table~\ref{table:size_fits}. The size was found to be $54.58\times59.76~(\pm2.37)$~pc, which is consistent with evolved SNRs in the LMC.

Given the emission from MCSNR~J0447--6919 is quite faint, we initially fitted the X-ray spectrum with an NEI thermal plasma model (\vpshock\ in XSPEC) of LMC abundance. While this provided a good fit to the data ($\chi^{2}_{\nu}$~=~1.10, see Table~\ref{table:xray_fits}), there were obvious residuals $\gtrsim0.8$~keV where the Fe L-shell complex might dominate. We therefore performed trial fits with the Fe abundances allowed to vary to determine if it is enhanced compared to the LMC value. This provided a slightly better fit to the data ($\chi^{2}_{\nu}$~=~1.03) and suggests an overabundance of Fe (see Table~\ref{table:xray_fits} and Fig.~\ref{fig:spectra}). However, the X-ray emission in MCSNR~J0447--6919 is too faint to perform a more detailed multi-component spectral analysis. The best fit plasma temperatures in both fits are low k$T<$0.3~keV which is consistent with an evolved remnant. For the model with LMC abundance, $\tau \sim10^{11}$~s~cm$^{-3}$ which suggests the plasma is close to, but yet to reach CIE. However, this value could misrepresent the true state of the plasma given the better fit with variable Fe abundance. The value of $\tau$ in this variable Fe fit ($\tau \gtrsim10^{11}$~s~cm$^{-3}$) indicates a plasma close to or in CIE. Unfortunately, the poor statistics do not allow for better constraints on $\tau$ and a more definitive conclusion. The detection of extended, thermal X-ray emission and the shell with \ratio$>0.4$ confirms this object as an SNR. 

\subsubsection{MCSNR~J0448--6700}
\label{0448-res}
The object [HP99]~460 was detected by \rosat\ and classified as an SNR candidate based on its extent of 46.3\arcsec\ \citep{Haberl1999}. It was firmly detected in radio as an extended and circular object of diameter $\sim4$\arcmin, though with a somewhat flat spectral index $\alpha$~=~--0.11$\pm$0.05. Because of the radio and X-ray detections, this object was classified as an SNR in both \citet{Maggi2016} and \citet{Bozzetto2017}, now carrying the identifier MCSNR~J0448--6700. The multi-wavelength images of this object are shown in Fig.~\ref{figures:0448-6700_4panel}. Extended X-ray emission is clearly observed with the emission being brighter towards the west. A similar brightening is seen in the optical emission line images, though the morphology is more shell-like than the X-ray. The \ratio=0.4 contour shows the characteristic SNR value towards the west and south. There is no appreciably variation in the ambient density from the $24~\mu$m image though the optical and X-ray morphologies do suggest some ambient density variation. Such a variation could have been induced if the SN progenitor was a runaway massive star with a proper motion in an approximate east-to-west direction, with the bow-shock of the SN progenitor wind providing the higher density in the west \citep{Meyer2015}. We used a hybrid contour combining the outer 0.3--0.7~keV and \ratio=0.4 contours  to determine the position and size of the remnant, with the results shown in Table~\ref{table:size_fits}. The size was found to be $55.54\times77.32~(\pm7.15)$~pc, which is consistent with evolved SNRs in the LMC. 

We fitted the X-ray emission from MCSNR~J0448--6700 using the \vpshock\ model allowing the abundances of O, Ne, Mg, and Fe to vary, and all others fixed to the LMC abundance. Trial fits with other elemental abundances free did not allow for constraints on the parameters. This yielded an acceptable fit with $\chi^{2}_{\nu}$~=~1.28 (see Table~\ref{table:xray_fits} and Fig.~\ref{fig:spectra}). The best fit k$T\sim$0.2~keV and $\tau \gtrsim10^{12}$~s~cm$^{-3}$ are consistent with an evolved SNR with CIE well established. The fitted abundances are consistent with those in the swept-up ISM in other LMC SNRs \citep[e.g.][their Table~5]{Maggi2016}. 
\subsubsection{MCSNR~J0449--6903}
\label{0449-res}
The candidate 0449--6903 was detected in 36~cm, 20~cm, and 6~cm radio continuum, with a spectral index $\alpha$~=~--0.50$\pm$0.01, typical of evolved SNRs \citep[][their Fig.~1, top-left]{Bozzetto2017}. Optical emission was also detected at the northeastern edge of the radio emission. The multi-wavelength images of this object are shown in Fig.~\ref{figures:0449-6903_4panel}. Extended X-ray emission is detected with a centrally filled morphology. Interestingly, hard X-ray emission ($>1.1$~keV) is detected from the core region. Sporadic optical emission with \ratio$>0.4$ was detected though did not trace a typical SNR shell morphology, though there is some evidence of a shell to the south and southeast which encloses the X-ray emission. There is no indication of ambient density variation from the $24~\mu$m image. We used the outer 0.3--1.1~keV contour to determine the position and size of the remnant, given in Table~\ref{table:size_fits}. The size was found to be $33.97\times26.34~(\pm3.37)$~pc, the smallest SNR in our sample but also one of the faintest.

We performed trial fits of the spectra using single component thermal plasma models of LMC abundance in both CIE and NEI. The CIE fit was poor with significant residuals in the 0.5--0.7~keV range. While the NEI fit was statistically better, the best fit plasma temperature was much higher than expected from a faint, evolved SNR (k$T>4$~keV). Further fits with selected $\alpha$-group and Fe free showed some minor improvements but the abundance parameters could not be constrained to any reasonable degree. Given that the morphology suggests two distinct emission structures, i.e. the shell and the core, we tried fitting two thermal plasma components. Trial fits with two NEI plasmas with the ionisation parameters of one or both components free did not provide constraints on many of the fit parameters. For this reason, we fitted a two component CIE model to the spectra. This provided a good fit to the data ($\chi^{2}_{\nu}$~=~1.10), the results of which are presented in Table~\ref{0449_2T} and shown in Fig.~\ref{fig:0449_spectra} left. The low temperature component with k$T$=0.26~(0.21--0.37)~keV is consistent with an evolved SNR shell. The higher temperature component with k$T$=2.81~(2.06--4.60)~keV suggests hotter ejecta emission but there is no evidence for enhanced abundances which should be present if the emission is due to hot ejecta. We also tried a thermal plasma (CIE) plus non-thermal (power law) model to fit the spectrum of the SNR. This also provided a good fit to the data ($\chi^{2}_{\nu}$~=~1.09), the results of which are given in Table~\ref{0449_2T} and shown in Fig.~\ref{fig:0449_spectra} right. Again, the temperature of the soft thermal model suggests an evolved SNR shell (k$T$=0.17~(0.05--0.26)~keV). It is unclear as to what could be the origin of non-thermal emission in the SNR. 

\begin{figure}
\begin{center}
\resizebox{\hsize}{!}{\includegraphics[trim= 0cm 0cm 0cm 0cm, clip=true, angle=0]{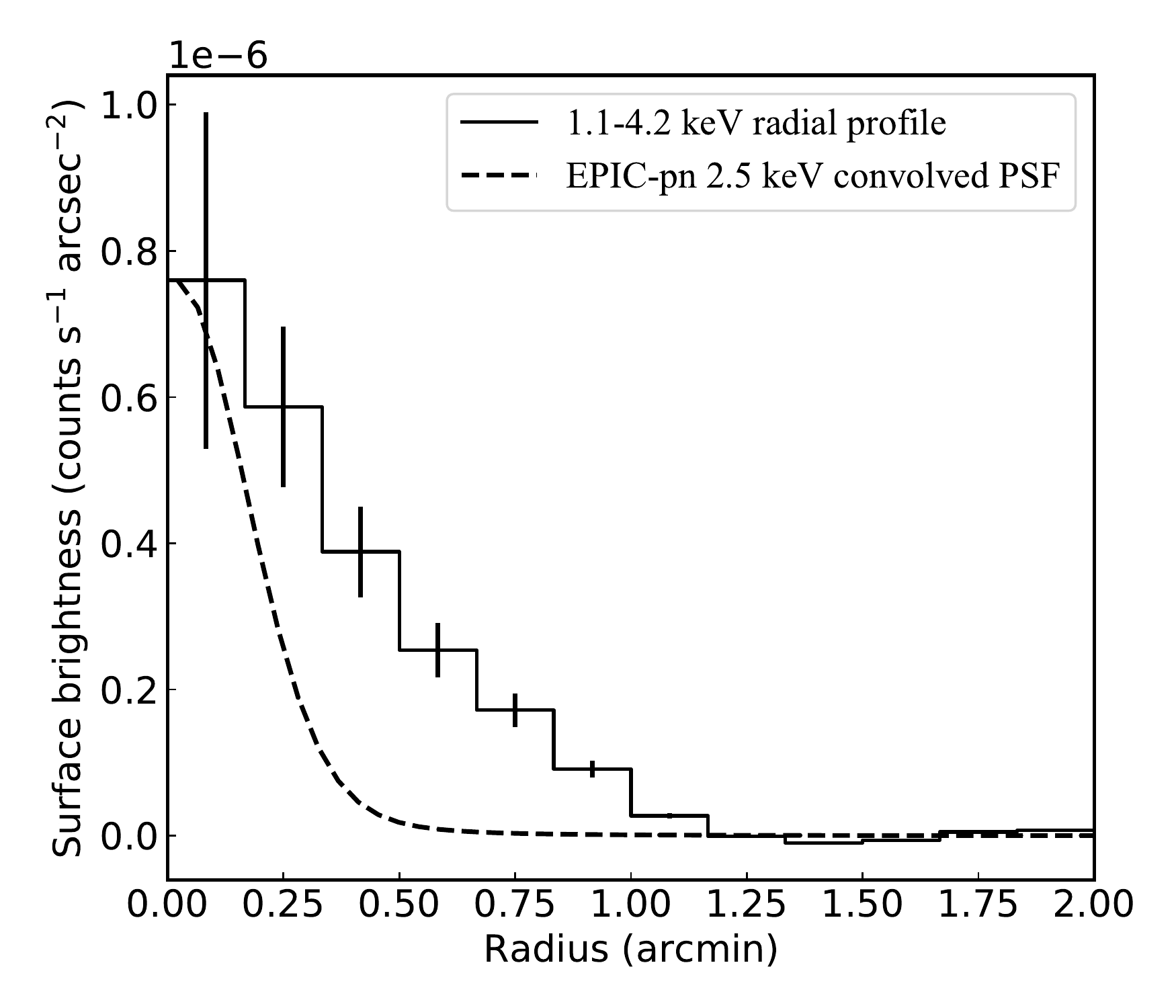}}
\caption{Radial profile plot of the hard emission in MCSNR~J0449-6903, extracted from the 1.1--4.2~keV image, centred on the brightest emission region, and in $10\arcsec$ bins. The dashed line shows the EPIC-pn PSF at 2.5~keV for comparison (see text) which has been scaled to the peak of the radial profile.}
\label{0449_ext}
\end{center}
\end{figure}

We investigated the extent of the hard X-ray emission by extracting a radial profile from the 1.1--4.2~keV image, centred on the brightest region of the hard emission, in $10\arcsec$ bins. To show that the source is indeed extended, we generated an EPIC-pn PSF using the SAS task {\texttt psfgen} at 2.5~keV and applied the same convolution as used for the 1.1--4.2~keV image. Both the radial profile and the PSF are shown in Fig.~\ref{0449_ext}. The hard source is clearly larger than the PSF, with a profile radius of $1.16~(\pm0.08)\arcmin$, which corresponds to $\sim~16.9~(\pm1.1)$~pc at the LMC distance. This size is comparable to the dimensions of the SNR (Table~\ref{table:size_fits}) so the hard emission appears to fill the SNR shell. The nature of this source is discussed in more detail in Sect.~\ref{atypsnrs-0449}.
 
Given the X-ray and optical detection in this work, and the radio detection reported in \citet{Bozzetto2017}, we confirmed the SNR nature of this object.

\subsubsection{MCSNR~J0456--6950}
\label{0456-res}
\begin{figure}
\begin{center}
\resizebox{\hsize}{!}{\includegraphics[trim= 0cm 0cm 0cm 0cm, clip=true, angle=0]{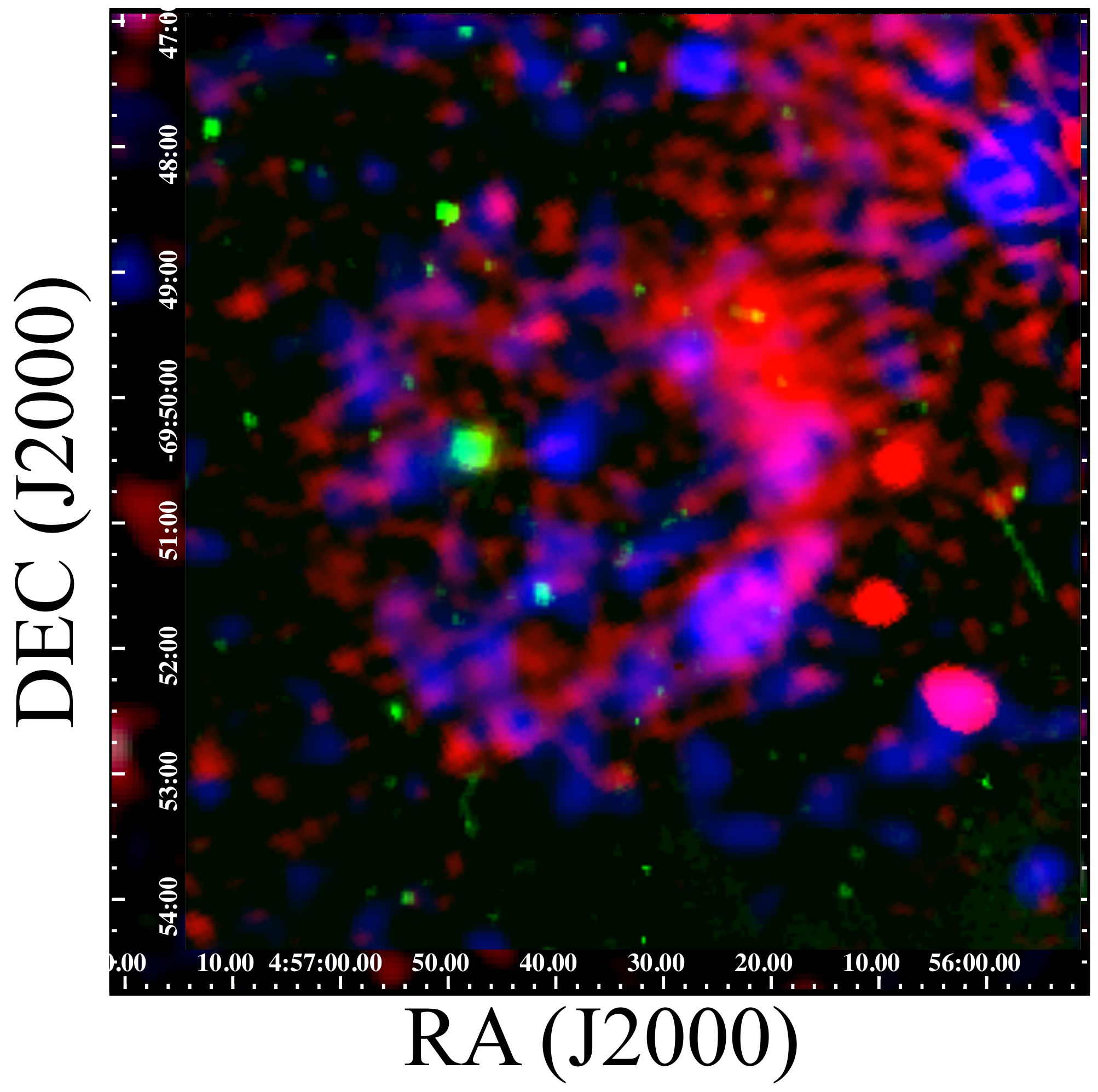}}
\caption{False color image of MCSNR~J0456-6950 with RGB=ASKAP 888~MHz, \ha, 0.3--0.7~keV. The ASKAP image reveals a large circular shell coincident with the faint X-ray emission. }
\label{0456_askap}
\end{center}
\end{figure}

The candidate 0456--6951 was a identified as a potential radio SNR based on a shell-like radio structure \citep{Bozzetto2017}. No clear optical counterpart to the shell could be found though there is extended optical emission around the candidate. The multi-wavelength images of MCSNR~J0456--6950 are shown in Fig.~\ref{figures:0456-6951_4panel}. A very faint extended source is evident in the 0.3-0.7~keV X-ray band. However, the emission is so faint it largely falls below the $3\sigma$ threshold used in Sect.~\ref{size_fits_sect} to estimate the SNR dimensions. Similarly, no optical shell or filaments are evident from the MCELS images. The \ratio\ contour does reveal some patchy regions with \ratio$>0.4$ to the south and southwest which appear to delineate the edge of the faint extended X-ray emission and suggests an association. We searched for radio emission from this object in our new ASKAP images \citep{2021MNRAS.506.3540P} which revealed a circular radio shell coincident with the X-ray emission, shown in Fig.~\ref{0456_askap}. This radio emission offers a much improved view of the size and shape of this SNR. We therefore determine its position and size using the radio contour, given in Table~\ref{table:size_fits} with the fit shown in Fig.~\ref{size}. The size was found to be $65.80\times62.26~(\pm1.80)$~pc, which is consistent with an evolved LMC SNR.

We initially fitted the X-ray emission from MCSNR~J0456--6950 using the \vpshock\ model, however we could not constrain the ionisation timescale . Given the large size of the remnant, it is likely a very evolved SNR and we assumed that the plasma has reached CIE and fixed the ionisation timescale in our fits to $10^{12}$~cm~s$^{-3}$. This model provided an acceptable fit ($\chi^{2}_{\nu}$~=~1.15, see Table~\ref{table:xray_fits}). The best fit temperature of $0.14~(0.13-0.16)$~keV is consistent with an evolved SNR. With the detection of extended thermal X-ray emission, sparse regions with \ratio$>0.4$, and the clear ASKAP detection, we confirmed this object as an SNR.

\subsubsection{MCSNR~J0504--6723}
\label{0504-res}
The object [HP99]~529 was detected by \rosat\ and classified as an SNR candidate based on its extent and hardness ratios \citep{Haberl1999}. No radio emission from this object was detected in the study of \citet{Bozzetto2017}. The multi-wavelength images of this object are shown in Fig.~\ref{figures:0504-6723_4panel}. Significant extended X-ray emission is detected with a centrally-filled morphology. The emission is elongated in a southwest-northeast direction. Interestingly, the 0.7--1.1~keV emission is mostly detected at the centre while the softer 0.3--0.7~keV emission is detected in the outer regions. This suggests that MCSNR~J0504--6723 might belong to the class of evolved SNRs with enhanced Fe emission \citep[e.g.,][]{Bozzetto2014,Maggi2014,Kavanagh2016}. However, the morphology of MCSNR~J0504--6723 appears far more dispersed that other remnants in the class with the Fe-rich material less confined to the core region. Significant optical emission is detected around the X-ray emission with \ratio$>0.4$, indicative of an SNR origin. The optical emission has an unusual morphology, appearing more rectangular than circular. The $24~\mu$m image reveals the likely reason for this with higher density regions appearing to confine the SNR in the northeast and southwest quadrants. The optical and X-ray morphology suggests the southeast emission is also confined while the remnant is free to expand in the northwest through a low density channel, similar to the asymmetric expansion proposed for MCSNR~J0517--6759 \citep{Maggi2014}. We used a hybrid contour combining the outer 0.3--1.1~keV and \ratio=0.4 contours to determine the position and size of the remnant, with the results shown in Table~\ref{table:size_fits} and shown in Fig.~\ref{size}. The size was found to be $58.00\times75.62~(\pm7.71)$~pc, which is consistent with evolved SNRs in the LMC.

To demonstrate the overabundance of Fe we initially fitted the X-ray spectrum with the \vpshock\ model, which resulted in a poor fit ($\chi^{2}_{\nu}$~=~1.56) with significant residuals between $\sim$0.8--0.9~keV. 
We then allowed the Fe abundance to vary, resulting in an acceptable fit ($\chi^{2}_{\nu}$~=~1.29) with an Fe abundance of ($Z/\mathrm{Z}_{\sun})_{\mathrm{Fe}}$~=~2.44~(1.90-4.32) and a high k$T$~=~0.72~(0.68-0.99)~keV. We also tried freeing Ne on its own and in combination with Fe as it can be hard to distinguish between enhanced Ne and Fe in the 0.8--0.9~keV range. The fit with Ne free resulted in a very poor fit. Freeing both Ne and Fe resulted in as statistically good a fit as with only Fe free ($\chi^{2}_{\nu}=1.20$) though the abundance of Fe was clearly enhanced ($Z/\mathrm{Z}_{\sun})_{\mathrm{Fe}}\gtrsim1.9$ while the Ne abundance was consistent with the LMC value. For this reason, we proceed under the assumption that the emission is Fe dominated.

We then took the same approach as for other similar objects of fitting the ejecta emission with spectral components comprising pure Fe, and possibly O \citep[e.g.][]{Maggi2014,Maggi2016,Kavanagh2016}, represented by the \vnei\ model with Fe abundance set to 1000 and all others to 0 \citep[see][for a description of the model employed]{Kavanagh2016}. The shell emission was modelled using a \vpshock\ model of LMC abundance. We initially tried allowing the ionisation timescales of both the shell and ejecta components and the shell abundances to vary, however we could not constrain any of these parameters. We therefore assumed that both the shell and ejecta components were in CIE, fixing $\tau$~=~$10^{12}$~cm~s$^{-3}$, and set the shell \vpshock\ model to the LMC abundance. The shell plus pure-Fe model improved the fit with $\chi^{2}_{\nu}$~=~1.20, the results of which are given in Table~\ref{Fe_fits}. We also tried fitting an additional pure-O component, which gave the same $\chi^{2}_{\nu}$~=~1.20 (Table~\ref{Fe_fits} and Fig.~\ref{fig:spectra}).

The detection of extended X-ray emission and significant optical emission with \ratio$>0.4$ confirmed this object as an SNR. 

\subsubsection{MCSNR~J0510--6708}
\label{0510-res}
\begin{figure}
\begin{center}
\resizebox{\hsize}{!}{\includegraphics[trim= 0cm 0cm 0cm 0cm, clip=true, angle=0]{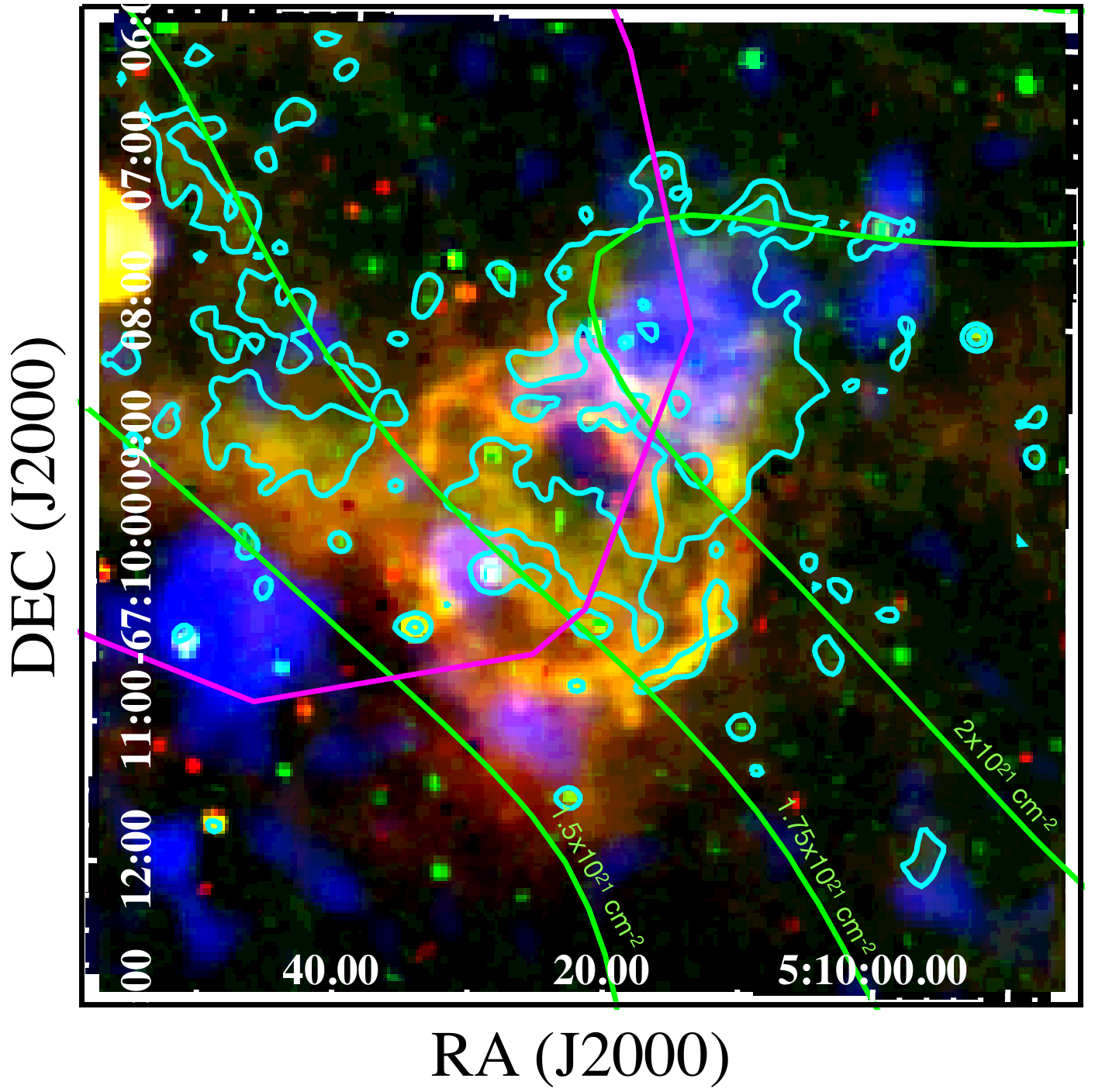}}
\caption{False color image of MCSNR~J0510--6708 with RGB=\ha, \sii, 0.7--1.1~keV. The cyan contour represents \ratio$>0.4$ (see Fig.~\ref{figures:0510-6708_4panel}). The green contours represent HI and are shown in equivalent hydrogen column (\nh) which were derived from the HI maps of \citet{Kim2003}. The magenta contour is $1\sigma$ level of the 12CO~(J=1-0) line observed with the NANTEN telescope \citep{Fukui2008}. }
\label{0510_hi}
\end{center}
\end{figure}

The candidate SNR 0510--6708 was classified as such as it exhibited optical emission with a \ratio\ $>0.4$, indicative of an SNR. Radio-continuum emission was found to be centrally peaked but very weak with a flux density of 2.5~mJy at 1377~MHz \citep{Bozzetto2017}. The multi-wavelength images of MCSNR~J0510--6708 are shown in Fig.~\ref{figures:0510-6708_4panel}. The multi-wavelength morphology of this object is somewhat puzzling. There is a large, circular optical shell in the MCELS image, bright in both \ha\ and \sii, but this does not show \ratio$>0.4$, and is therefore inconsistent with an SNR origin. In addition, the X-ray emission is not correlated with this large, circular shell, but projected outside the northwestern shell (Fig.~\ref{figures:0510-6708_4panel} top-right) and is quite faint. Interestingly, there are regions with \ratio$>0.4$ which surround the X-ray emission (Fig.~\ref{figures:0510-6708_4panel} bottom-left) and follow a different morphology than the large, circular shell. The $24~\mu$m image shows a quite complex density structure and suggests that foreground absorption may play a significant role in the observed morphology. To investigate this further we compared the observed optical and X-ray morphologies to the HI and CO maps of \citet{Kim2003} and \citet{Fukui2008}, respectively, shown in Fig.~\ref{0510_hi}. From the HI contours, we expected the foreground \nh\ value be $~2\times10^{22}$~cm$^{-2}$ at most. However, this was determined to be $\gtrsim~3\times10^{22}$~cm$^{-2}$ in our X-ray spectral fits, which suggests some additional absorption. The NANTEN map suggests that some molecular material could also be absorbing the X-ray emission in the east and south of the remnant. In any case, the fitted \nh\ is sufficient to mask out any faint soft emission that could come from the south-east region and/or inside the large, circular optical shell so the observed X-ray morphology could be due to absorption effects.

The location of the X-ray emission and the extent of the \ratio\ contour outside of the large, circular shell suggests that either we are seeing an SNR projected against a stellar wind bubble or superbubble, or that an SN has occurred near the bubble or superbubble shell and blown out into the surroundings. We determined the position and size of the remnant using the \ratio=0.4 contour, with the results shown in Table~\ref{table:size_fits} and shown in Fig.~\ref{size}. The size was found to be $29.46\times47.03~(\pm5.92)$~pc, which is consistent with evolved SNRs in the LMC.

The X-ray emission detected in MCSNR~J0510--6708 is rather faint. Due to the low number of counts in the MOS1 and MOS2 spectra, we consider the pn spectrum only. We initially fitted the X-ray emission using the \vpshock\ model, however, we could not constrain the ionisation parameter with only a lower limit of $\gtrsim10^{10}$~cm~s$^{-3}$ found. We therefore assumed CIE and fixed the parameter to $10^{12}$~cm~s$^{-3}$. This provided a good fit to the data ($\chi^{2}_{\nu}$~=~1.01) and constraints on the free parameters (see Table.~\ref{table:xray_fits}). We also tried fits with some abundances allowed to vary, however, these did not provide constraints on the parameters. This is not surprising given the low number of counts and degrees of freedom.

Given that the extended X-ray emission appears to be associated with enhanced \sii\ and not the large optical shell, we confirmed this object as an SNR.

\subsubsection{MCSNR~J0512--6717}
\label{0512_res}
This object was detected serendipitously in an \xmm\ observation of another candidate, and subsequently confirmed SNR, MCSNR~J0512--6707 \citep{Kavanagh2015b}. Located in the south of the FOV, 0512--6717 presented a large circular morphology in soft X-rays. Faint 20~cm radio continuum emission was detected along the western side of the object \citep[][their Fig.~2, middle-left]{Bozzetto2017}, though it was not possible to construct a radio continuum spectral energy distribution and estimate the spectral index. Therefore, this object was classed as a candidate in \citet{Bozzetto2017}. The multi-wavelength images of this object are shown in Fig.~\ref{figures:0512-6717_4panel}. Extended X-ray emission was detected in a distinct shell-like morphology. The optical emission line images of the object did not show any morphological consistency with the X-ray shell with some regions of \ratio$>0.4$ that were not coincident with the X-ray emission. Interestingly, the \ratio\ image reveals a shell-like region with \ratio$>0.4$ of similar size to the X-ray shell but shifted northwards. It is unclear if this is related to the X-ray shell with the observed displacement being due to a projection effect. To confirm the SNR nature of the object, we performed follow-up optical spectroscopic observations on regions of the X-ray shell that showed clearly enhanced and filamentary \sii\ emission. The spectral analysis, described in Sect.~\ref{opt-spec}, showed that the optical emission in these regions was consistent with an SNR origin with \ratio$>0.4$ (see Table~\ref{table:optical_spectra_fits}). The $24~\mu$m image shows an inhomogeneous ambient density surrounding the SNR though this does not appear to have significant influence on the X-ray or optical morphologies. We fitted the 0.3-0.7~keV X-ray contour of MCSNR~J0512--6717 to determine a size of $57.33\times79.57~(\pm7.12)$~pc (see Table~\ref{table:size_fits} and Fig.~\ref{size}) which suggests an evolved SNR.

We tried to fit the X-ray emission from MCSNR~J0512--6717 using the \vpshock\ model. However, the resulting fit could not constrain abundance parameters or the ionisation timescale. Given the large size of the object, we assumed that the plasma has most likely reached CIE and so fixed the ionisation timescale to $10^{12}$~cm~s$^{-3}$. Trial fits with elemental abundances allowed to vary showed that constraints could be placed on O and Fe abundances. This model provided an acceptable fit ($\chi^{2}_{\nu}$~=~1.14, see Table~\ref{table:xray_fits}). The best fit temperature of $0.19~(0.18-0.21)$~keV is consistent with an evolved SNR and the fitted abundances are consistent with those in the swept-up ISM in other LMC SNRs \citep[e.g.][their Table~5]{Maggi2016}.

Based on the detection of extended thermal X-ray emission and optical signatures of an SNR, including \ratio$>0.4$, we confirmed this object as an SNR. 

\subsubsection{MCSNR~J0527--7134}
\label{0527_res}
Radio continuum emission was detected from the candidate 0527--7134 at 20~cm and 36~cm with a shell-like morphology that is correlated with an \sii\ dominated optical shell \citep[][their Fig.~2, bottom-right]{Bozzetto2017}. The enhanced \sii\ optical shell and determined spectral index $\alpha$~=~--0.52 both indicated an SNR nature for the object. Our multi-wavelength images are shown in Fig.~\ref{figures:0527-7134_4panel}. Soft extended X-ray emission is clearly detected and associated with a significant \sii\ enhancement with \ratio$>0.4$. The X-ray and enhanced \sii\ emission are very well correlated and suggest an obvious association. A very slight ambient density gradient is apparent in the $24~\mu$m image, though this appears to have little influence on the optical or X-ray morphology. We used the outer 0.3--0.7~keV contour to determine the position and size of the remnant, given in Table~\ref{table:size_fits} and shown in Fig.~\ref{size}. The size was found to be $36.98\times44.52~(\pm3.55)$~pc, typical of an evolved LMC SNR. 

We fitted the X-ray emission from MCSNR~J0527--7134 using the \vpshock\ model allowing the abundances of O and Ne to vary, and all others fixed to the LMC abundance. Trial fits with other elemental abundances free did not allow for constraints on the parameters. This model provided a good fit to the data with $\chi^{2}_{\nu}$~=~0.97 (see Table~\ref{table:xray_fits}). However, some parameters were not well constrained, i.e. k$T$~=~0.53~(0.19--1.32)~keV) and $\tau$~=~0.26~(0.09--12.66)~10$^{11}$~s~cm$^{-3}$. Since the low and high 90\% confidence limit on k$T$ and $\tau$, respectively, are consistent with an evolved SNR, we also assumed that the plasma was in CIE and ran a trial fit with $\tau$ fixed to $10^{12}$~cm~s$^{-3}$, which is shown in Table~\ref{table:xray_fits}. The CIE model also provided a good fit to the data with $\chi^{2}_{\nu}$~=~0.98 (see Table~\ref{table:xray_fits}), with tighter constraints on the plasma temperature (k$T~0.19~(0.19-0.21)$~keV). In both models, the fitted abundances of O and Ne are consistent with those in the swept-up ISM in other LMC SNRs \citep[e.g.][their Table~5]{Maggi2016}. The interpretation of the results of both fits are discussed in more detail in Sect.~\ref{sect:0527_disc} where SNR properties are derived using the Sedov evolutionary model.

Based on the detection of extended thermal X-ray emission, correlated \ratio$>0.4$, and the previously reported non-thermal radio emission \citep{Bozzetto2017}, we confirmed this object as an SNR. 

\subsubsection{MCSNR~J0542--7104}
\label{0542-res}
The candidate 0542--7104 showed a shell-like optical emission with enhanced \sii. Radio emission was also detected at the brightest region of \sii\ emission with a possible correlation to the X-ray source [HP99]~1235 \citep{Bozzetto2017}. Recently, this object was confirmed as an SNR by \citet{Yew2021} based on its optical spectrum. They measured a size of $73\times51$~pc for the optical shell. Our multi-wavelength images of MCSNR~J0542--7104 are shown in Fig.~\ref{figures:0542-7104_4panel}, with the \ratio\ contours from \citet{Yew2021}. Extended X-ray emission, primarily in the 0.7--1.1~keV range, is detected towards the centre of the optical shell. Similar morphology has been observed in, e.g. MCSNR~J0508--6830, MCSNR~J0511--6759 \citep{Maggi2014}, and MCSNR~J0527--7104 \citep{Kavanagh2016} and suggests that this SNR belongs to the class of evolved SNRs with enhanced Fe emission, similar to MCSNR~J0504--6723 discussed above. The observed Fe-rich core morphology has been interpreted as the X-ray emission from the shocked ISM being too faint to detect whereas the hot Fe gas in the core remains detectable. We used the 0.7--1.1~keV contour to determine the size of the core emission region to be $33.46\times28.57~(\pm2.29)$~pc (see Fig.~\ref{size}). Knots of emission are also observed outside the core to the south, similar to MCSNR~J0511--6759. \citet{Maggi2014} interpret these features being clumps of X-ray emitting ejecta shrapnel. There appears to be a marginal ambient density gradient in the $24~\mu$m image in the northeast-southwest direction. However, the optical shell is brighter and fainter in the lower and higher density regions, respectively. If indeed the western shell is evolving into a higher density region, we would expected the optical emission to be brighter in the west. For this reason, it is likely the density gradient is a projection effect and does not represent the real ambient conditions of the SNR. 

To demonstrate the overabundance of Fe in the core we fitted the X-ray spectrum with the \vpshock\ model, allowing the Fe abundance to vary, resulting in a best fit abundance of ($Z/\mathrm{Z}_{\sun})_{\mathrm{Fe}}\gtrsim2$ and a high k$T\gtrsim0.5$~keV for a $\chi^{2}_{\nu}=1.21$. As with MCSNR~J0504--6723, we also tried freeing Ne on its own and in combination with Fe. The fit with Ne free resulted in a poorer fit with $\chi^{2}_{\nu}=1.35$. Freeing both Ne and Fe resulted in a similarly good fit as with only Fe free ($\chi^{2}_{\nu}=1.20$) though neither abundance was well constrained but the Fe was clearly enhanced ($Z/\mathrm{Z}_{\sun})_{\mathrm{Fe}}\gtrsim1.5$ while the lower limit of the Ne abundance was consistent with the LMC value. For this reason, we proceed under the assumption that the emission is Fe dominated.

We then took the same approach for other similar objects, including MCSNR~J0504--6723 in this work, of fitting the core emission with spectral components comprising pure Fe, and possibly O \citep[e.g.][]{Maggi2014,Maggi2016,Kavanagh2016}. We initially allowed the ionisation timescale to vary, however we could not constrain this parameter so instead fixed it to $10^{12}$~cm~s$^{-3}$. We first tried a pure Fe only model which provided a reasonable fit to the data ($\chi^{2}_{\nu}$~=~1.25, see Table~\ref{Fe_fits}). However, there were clearly residuals between 0.5--0.6~keV. For this reason we added a pure O component. Even though type Ia SNe can produce little O, its mass fraction could be as much as $\sim30$\% ($\sim0.3$~M$_{\sun}$) if most of the outer ejecta layers consist of O \citep[see, e.g.][and references therein]{Kosenko2010}. For low plasma temperatures (<1~keV), the emissivity of the \ion{O}{VII} and \ion{O}{VIII} lines is high, leading to significant line emission in the 0.5-0.6~keV range. We initially linked the Fe and O plasma temperatures, which is expected if the Fe and O are co-spatial and the plasmas have reached CIE. However, this did not properly account for the residuals in the 0.5--0.6~keV range ($\chi^{2}_{\nu}$~=~1.22, see Table~\ref{Fe_fits}). We therefore allowed the O plasma temperature to vary independently from Fe. This provided a much improved fit to the data ($\chi^{2}_{\nu}$~=~1.02, see Table~\ref{Fe_fits} and Fig.~\ref{fig:spectra}). The different values of k$T$ for the Fe ($0.58~(0.54-0.62)$~keV) and O ($0.19~(0.14-0.24)$~keV) components suggested these plasmas are not co-spatial.

\section{Discussion}
\label{discussion}
\subsection{SNR evolutionary properties}
The X-ray spectral models for several of our SNRs contained a soft thermal component interpreted as emission from an evolving Sedov-phase shell. We estimated the physical properties of these SNRs using the Sedov dynamical model and the SNR dimensions determined in Sect.~\ref{size_fits_sect} where we assumed that the measured semi-major and semi-minor axes of each of our objects were the first and second semi-principal axes of an ellipsoid, and that the third semi-principal axis was in the range between the two. The following procedure has been used previously in, e.g. \citet{Sasaki2004,Kavanagh2015b,Maggi2016,Kavanagh2016}. However, we reiterate the procedure here and introduce an additional term representing the hot gas filling factor to convey how our derived SNR properties may vary depending on this term.

From our estimates on the dimensions of the SNRs, we determined their volume ($V$) limits and corresponding effective radii ($R_{\rm{eff}}$). The best fit shell X-ray temperatures determined in our spectral fits (see Tables~\ref{table:xray_fits}, \ref{Fe_fits}, \ref{0449_2T}) correspond to a shock velocity

\begin{equation}
v_{s}=\left(\frac{16kT_{s}}{3\mu}\right)^{0.5},
\end{equation}

\noindent where k$T_{s}$ is the post-shock temperature and $\mu$ is the mean mass per free particle which, for a fully ionised plasma with LMC abundances, is $\mu=0.61\rm{m_{p}}$. The ages of our SNRs were then determined using the Sedov solution:

\begin{equation}
v_{s} = \frac{2R}{5t},
\end{equation}

\noindent where $R=R_{\rm{eff}}$ and $t$ is the age of the remnant. The pre-shock H densities ($n_{\rm{H},0}$) in front of the blast wave of each SNR were then determined from the emission measure ($EM$, see Tables~\ref{table:xray_fits}, \ref{Fe_fits}, \ref{0449_2T}). An expression for the emission measure can be determined by evaluating the emission integral over the Sedov solution using the approximation for the radial density distribution of \citet{Kahn1975}, which gives

\begin{equation}
\int n_{e}n_{\rm{H}}dV=EM=2.07\left(\frac{n_{e}}{n_{\rm{H}}}\right)n_{\rm{H},0}^{2}Vf,
\end{equation}

\noindent where $n_{\rm{e}}$ and $n_{\rm{H}}$ are electron and hydrogen densities, respectively, $V$ is the volume \citep[e.g.,][]{Hamilton1983}, $f$ is a filling factor, and $n_{e}/n_{\rm{H}}$~=~1.21. We then estimated the pre-shock density from $n_{0}\sim1.1 n_{\rm{H},0}$. The initial explosion kinetic energies ($E_{0}$) were determined from the equation:

\begin{equation}
\label{eq:exp_energy}
R=\left(\frac{2.02E_{0}t^{2}}{\mu_{n}n_{0}f^{-1/2}}\right)^{1/5},
\end{equation}

\noindent where $\mu_{n}$ is the mean mass per nucleus (=$1.4m_{p}$). The swept-up masses contained in the shell of each SNR is given by $M=V\mu_{n}n_{0}f^{1/2}$. The determined properties for each SNR are listed in Table~\ref{table:snr_properties} for an assumed $f=1$. 
\begin{table*}
\caption{Derived properties of our SNR sample. All quoted values assume a filling factor $f$ of unity.}
\begin{center}
\label{table:snr_properties}
\begin{tabular}{llllll}
 \hline
\hline
Name & $n_{0}$ & $v_{s}$ & Age & $M$ & $E_{0}$ \\
 & ($f^{-1/2} 10^{-2}$~cm$^{-3}$) & (km~s$^{-1}$) & (kyr) & ($f^{1/2} $~M$_{\sun}$) & ($f^{-1/2} 10^{51}$~erg) \\

\hline
MCSNR~J0447--6919 & 2.5--6.3 & 409--501 & 22--28 & 78--230 & 0.17--0.95 \\
MCSNR~J0448--6700 & 4.4--8.8 & 388--419 & 27--35 & 157--550 & 0.24--2.06 \\
MCSNR~J0449--6903 & 1.6--4.4 & 419--572 & 9--15 & 6--25 & 0.01--0.17 \\
MCSNR~J0456--6950 & 1.6--2.6 & 317--366 & 34--40 & 71--129 & 0.10--0.27 \\

MCSNR~J0504--6723 & 0.6--1.8 & 399--755 & 15--35 & 22--115 & 0.04--1.35 \\
MCSNR~J0510--6708 & 2.3--23.0 & 366--518 & 12--22 & 12--270 & 0.01--1.82 \\ 
MCSNR~J0512--6717 & 3.4--6.9 & 388--419 & 28--36 & 134--470 & 0.21--1.75 \\
MCSNR~J0527--7134 & 1.2--4.5 & 399--1051 & 7--21 & 12--64 & 0.02--1.32 \\

\hline
\end{tabular}
\end{center}
\end{table*}%

\subsection{Typical Sedov-phase SNRs}

\subsubsection{MCSNR~J0448--6700}
\label{sect:0448_disc}
The multi-wavelength properties of MCSNR~J0448--6700 suggested either an SNR evolving into an ambient density gradient or that the SN progenitor had a significant proper motion in an approximate east-to-west direction (see Sect.~\ref{0448-res}). The size was found to be $55.54\times77.32~(\pm7.15)$~pc, which is consistent with large evolved SNRs in the LMC, which is further supported by the X-ray spectrum being consistent with a cool thermal plasma (k$T\sim$0.2~keV) in CIE. From the spectral fit results and size estimate, we determined that this SNR is consistent with the Sedov dynamical model with an initial explosion energy in the range $(0.24-2.06)\times10^{51}$~erg and age of 27--35~kyr (see Table~\ref{table:snr_properties}).

\subsubsection{MCSNR~J0456--6950}
\label{sect:0456_disc}
MCSNR~J0456--6950 was found to have an X-ray faint shell without marginal evidence for an optical counterpart along the south of the shell. However, the ASKAP 888~MHz image (see Fig.~\ref{0456_askap}) revealed a large circular shell which led us to confirm this object as an SNR (Sect.~\ref{0456-res}). The size was found to be $65.80\times62.26~(\pm1.80)$~pc and the X-ray spectrum was consistent with a soft thermal plasma (k$T\sim$0.15~keV) in CIE. Comparison with the Sedov dynamical model shows an underestimate of the explosion energy $(0.10-0.27)\times10^{51}$~erg (see Table~\ref{table:snr_properties}). This could be due to the assumed filling factor of unity. If $f<1$, the determined explosion energy would be higher (see Eq.~\ref{eq:exp_energy}). The estimated age from the Sedov model was 34--40~kyr.

\subsubsection{MCSNR~J0512--6717}
\label{sect:0512_disc}
Our multi-wavelength study of MCSNR~J0512--6717 revealed a large evolved SNR with a distinct ring-like morphology in X-rays. Its size was found to be $57.33\times79.57~(\pm7.12)$~pc, which is consistent with large evolved SNRs in the LMC. Its X-ray spectrum was best described with a soft thermal plasma (k$T\sim$0.2~keV) in CIE (Sect.~\ref{0512_res}). From the spectral fit results and size determination, we found that MCSNR~J0512--6717 is consistent with the Sedov dynamical model with an initial explosion energy in the range $(0.21-1.75)\times10^{51}$~erg and age of 28--36~kyr (see Table~\ref{table:snr_properties}). 

\subsubsection{MCSNR~J0527--7134}
\label{sect:0527_disc}
The multi-wavelength morphological study of MCSNR~J0527--7134 revealed a bright and uniform X-ray shell, which was significantly correlated with the \ratio\ morphology. The size was found to be $36.98\times44.52~(\pm3.55)$~pc, typical of an evolved LMC SNR. Its X-ray spectrum could be described with either a CIE or NEI thermal plasma model. However, the NEI model was not as well constrained as the CIE model (Sect.~\ref{0527_res}) but both are consistent with a plasma CIE. From the spectral fit results and size determination, we found that MCSNR~J0527--7134 is consistent with the Sedov dynamical model with an initial explosion energy in the range $(0.02-1.32)\times10^{51}$~erg and age of 7--21~kyr (see Table~\ref{table:snr_properties}). 

\subsection{Evolved Fe-rich SNRs}
\label{Fe_disc}
We have identified two of our sample as new members of the class of evolved SNRs with Fe-rich interiors, bringing the total number to 13. The members of this class are listed in Table~\ref{table:evolved_fe_snrs}.

\begin{table}
\caption{Evolved Fe-rich SNRs in the LMC.
}
\begin{center}
\label{table:evolved_fe_snrs}
\begin{tabular}{ll}
\hline\hline\noalign{\smallskip}
Names (MCSNR) & Reference \\
\noalign{\smallskip}\hline
J0504--6723, J0542--7014      & this work, \citet{Yew2021}  \\
J0506--7025, J0527--7104     & \citet{Kavanagh2016} \\
J0508--6830, J0511--6759  & \citet{Maggi2014} \\
J0508--6902 & \citet{Bozzetto2014} \\
J0454−-6713 & \citet{Seward2006} \\
J0534−7033, J0536−7039 & \citet{Borkowski2006} \\
J0547-6941 & \citet{Williams2005} \\
J0547−-7025, J0534−-6955 & \citet{Hendrick2003} \\
\hline
\end{tabular}
\end{center}
\end{table}%

The X-ray spectral models for MCSNR~J0504--6723 and MCSNR~J0542--7104 contained pure Fe and O components representing hot ejecta. We estimated various properties of the ejecta following the method described in detail in, e.g., \citet{Kosenko2010,Bozzetto2014,Maggi2014, Kavanagh2016}.
For a Type Ia explosion, we would expect that about half of the ejecta mass is Fe, $\sim0.7$~M$\sun$ \citep[e.g.][]{Iwamoto1999} and can be determined from the emission measure of the ejecta components and the volume they occupy. The Fe mass given by the equation

\begin{equation}
\label{Fe-calc}
M_{\rm{Fe}}  = (V_{\rm{Fe}} EM_{\rm{Fe}})^{0.5} (n_{\rm{e}}/n_{\rm{Fe}})^{-0.5} m_{\rm{U}} A_{\rm{Fe}},
\end{equation}

\noindent where $V_{\rm{Fe}}$ is the volume of the Fe, $EM_{\rm{Fe}}=n_{e}n_{Fe}V$ is the emission measure of the Fe gas, $n_{\rm{e}}/n_{\rm{Fe}}$ is the electron to Fe-ion ratio, $m_{\rm{U}}$ is the atomic mass unit, and $A_{\rm{Fe}}$ is the atomic mass of Fe. $EM_{\rm{Fe}}$ is obtained from the normalisation of the Fe component and is listed in Table~\ref{Fe_fits}. The $n_{\rm{Fe}}/n_{\rm{H}}$ value is calculated from the Fe abundance parameter of the model. There are two limiting cases for the value of the $n_{\rm{e}}/n_{\rm{Fe}}$ ratio as the number of free electrons $n_{\rm{e}}$ is dependent on the amount of H in the ejecta \citep{Hughes2003}. If there is no H then $n_{\rm{e}}/n_{\rm{Fe}}$ only depends on the average ionisation state of the Fe, taken as 18.3 \citep{Shull1982}. Conversely, if a similar mass of H is mixed into the Fe-rich ejecta, the number density of Fe over H is 1/56, and $n_{\rm{e}}/n_{\rm{Fe}}$ is 74.3.  

We estimated the volume range occupied by the Fe-rich plasmas in our objects by assuming that they are ellipsoidal. We fitted ellipses to the emission regions, measured the semi-major and semi-minor axes, and determined a volume range assuming that the third semi-axis was equal to either the semi-major or semi-minor axes. We must also consider whether or not the ejecta are clumpy. In this case, the emitting volume must be modified by a filling factor of $\sim0.4$ \citep[see][and references therein]{Kosenko2010}, in which case the Fe content reduces by a factor of $\sqrt{0.4}\approx0.63$.

\subsubsection{MCSNR~J0504--6723}
\label{0504_disc}
The multi-wavelength images of MCSNR~J0504--6723 showed an SNR which is elongated in a southwest-northeast direction. The X-ray morphology was notable because of the bright and central 0.7--1.1~keV emission surrounded by the softer 0.3--0.7~keV, suggesting that MCSNR~J0504--6723 belongs the class of evolved SNRs with enhanced Fe emission. The associated optical emission has an unusual morphology, appearing more rectangular that circular, most likely because of ambient density variations revealed in the $24~\mu$m image. The size was found to be $58.00\times75.62~(\pm7.71)$~pc, which is consistent with evolved SNRs in the LMC. The shell component was best fitted with a soft thermal plasma (k$T\sim$0.2~keV) in CIE (Sect.~\ref{0504-res}). From the spectral fit results and size estimate, we determined that the SNR shell is consistent with the Sedov dynamical model with an initial explosion energy in the range $(0.04-1.35)\times10^{51}$~erg and age of 15--35~kyr (see Table~\ref{table:snr_properties}).

For the central Fe-rich plasma, the determination of $V_{\rm{Fe}}$ presents a similar problem to MCSNR~J0506--7025 \citep{Kavanagh2016}, namely that the emission in the 0.7--1.1~keV range, where the Fe-L shell lines dominate, is not confined to a well-defined core. While there does appear to be an aggregate of Fe-L shell emission near the SNR centre, there are extensions protruding to the north east and south. It is unclear if these extensions are due to the Fe-rich ejecta or to enhancements in the shell emission. 

\begin{figure}
\begin{center}
\resizebox{\hsize}{!}{\includegraphics[trim= 0cm 0cm 0cm 0cm, clip=true, angle=0]{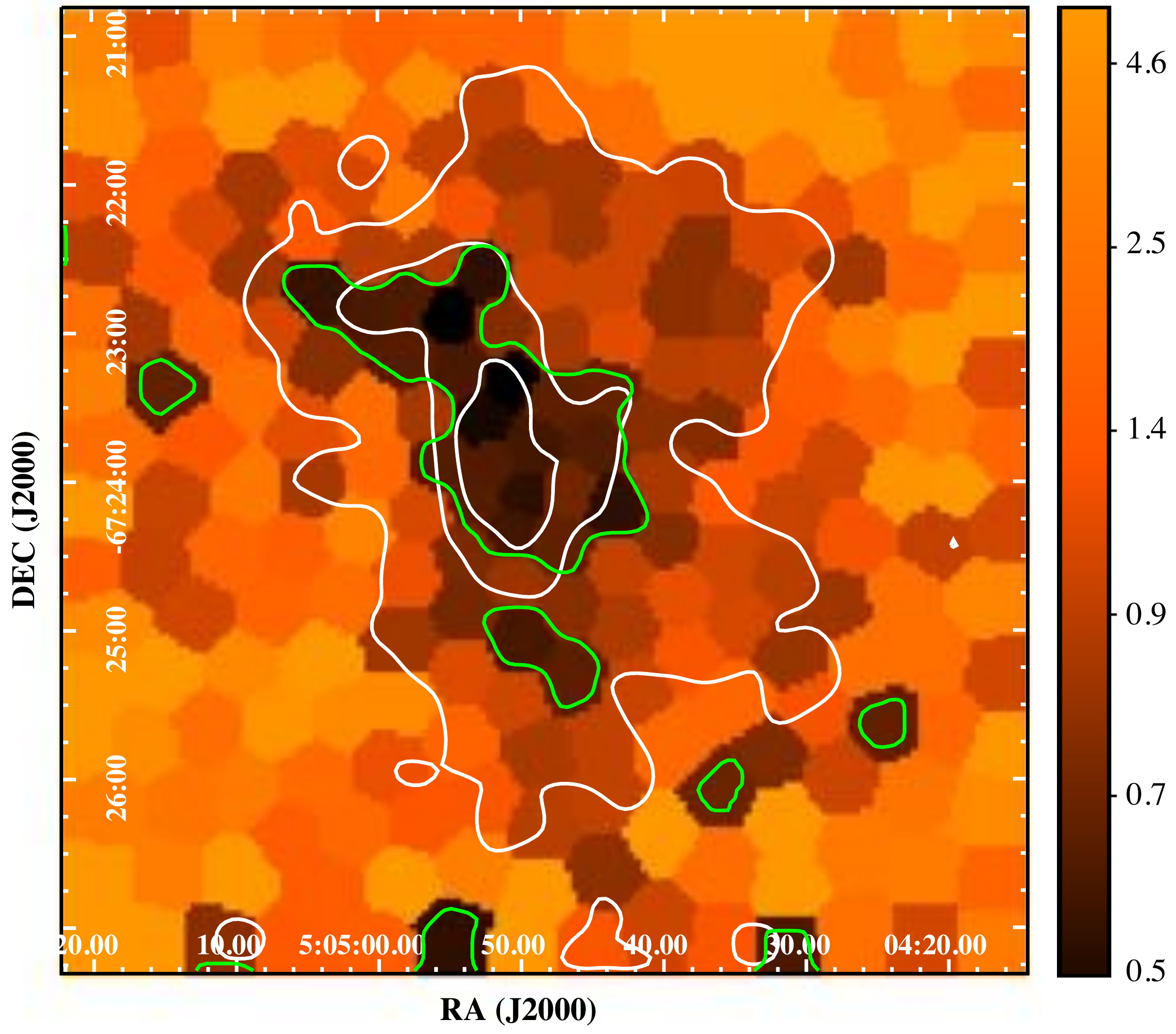}}
\caption{Adaptively binned X-ray colour map of MCSNR~J0504--6723 using Voronoi tessellation. The tessellates are defined so that each has a S/N of 10. Lower values of $S/H$ indicate higher contributions from Fe L-shell emission. The white contours show the 0.3--0.7 keV contours from Fig.~\ref{figures:0504-6723_4panel}-top right and the green contour outlines the Fe L-shell emission regions, set to $S/H$ $<$0.7 (see text).}
\label{0504_wvt}
\end{center}
\end{figure}

As for MCSNR~J0506--7025 in \citet{Kavanagh2016}, we created an X-ray colour map to gauge the distribution of Fe-rich plasma in the core of MCSNR~J0504--6723. We used the same two energy bands, a soft band 
$S$~=~0.3--0.7~keV (where the shell dominates) and a hard band 
$H$~=~0.7--1.1~keV (where the Fe L-shell bump peaks), and the colour $S/H$ is then given by 
$N_{0.3-0.7~\mathrm{keV}}/N_{0.7-1.0~\mathrm{keV}}$. As in \citet{Kavanagh2016}, we binned the combined 0.3--1.1~keV image using Voronoi tessellation to achieve a minimum signal-to-noise ratio per bin but in this case used the VorBin python package which is an implementation of the \citet{Cappellari2003} Voronoi binning algorithm, with the weighted Voronoi tessellation option \citep{Diehl2006} set. We found that a signal-to-noise of 10 gave a useful estimate of $S/H$ in each bin while producing small enough bins to adequately trace the Fe distribution in MCSNR~J0504--6723. We then used the tessellates defined for the 0.3--1.1~keV image to bin the $S$ and $H$ images, and calculated $S/H$, which is shown in Fig.~\ref{0504_wvt}. A threshold level to separate Fe-L shell dominated regions from the shell regions was still needed. Again we employed the method of \citet{Kavanagh2016} of performing spectral simulations assuming that the best-fit core and shell component parameters are constant over the remnant, allowing only the relative normalisation between the shell and core components to vary. These simulations suggested a threshold of $\sim0.7$, with values below this corresponding to Fe L-shell dominated regions. We defined a contour at this threshold (see Fig.~\ref{0504_wvt}) which showed that the Fe emission is non-uniformly distributed, with a narrow extension towards the north east, and a non-contiguous morphology in the south. Given such an inhomogeneous morphology, it is difficult to obtain an estimate of $V_{\rm{Fe}}$. Nevertheless, we attempted to determine the volume of the Fe-rich core by fitting an ellipse by eye around the inner two contours in Fig.~\ref{0504_wvt}. We estimate this core region to have a size of $26.76\times51.78$~pc which results in a volume estimate of $V_{\rm{Fe}}=0.57-1.10\times10^{60}$~cm$^{3}$. We used this in Eq.~\ref{Fe-calc} to determine an Fe mass of 2.3--3.8~M$_{\sun}$ for a pure metal plasma with no H and a filling factor of 1, and 1.1--1.8~M$_{\sun}$ in the case of an equivalent mass of H is mixed into the Fe-rich ejecta. Assuming a clumpy ejecta leads to Fe masses of 1.4--2.4~M$_{\sun}$ and 0.7--1.1~M$_{\sun}$ for a pure metal plasma with no H and with equivalent amount of H mixed into the Fe-rich ejecta, respectively. These results indicate that a clumpy ejecta with an admixture of H of similar mass to the Fe ejecta is required for agreement between our determined Fe mass and that expected from Type Ia explosive nucleosynthesis yields \citep[e.g.][]{Iwamoto1999}, though note that this conclusion is still reliant on our coarse $V_{\rm{Fe}}$ estimate.

\subsubsection{MCSNR~J0542--7104}
Both the X-ray morphology and spectral composition of MCSNR~J0542--7104 indicated that the emission is primarily from Fe-rich ejecta. Unlike MCSNR~J0504--6723, the morphology of the core is much more regular, being approximately circular. We used the size of the core region determined in Sect.~\ref{0542-res} to estimate the ejecta volume to be $V_{\rm{Fe}}=4.30-4.92\times10^{59}$~cm$^{3}$. We used this in Eq.~\ref{Fe-calc} and a filling factor of 1 to determine Fe masses of 1.9--2.5~M$_{\sun}$ and 0.9--1.3~M$_{\sun}$ for the cases of no H and an equivalent mass of H mixed into the Fe-rich ejecta, respectively. Assuming a clumpy ejecta leads to corresponding Fe masses of 1.2--1.6~M$_{\sun}$ for the no H and 0.6--0.8~M$_{\sun}$ for the equivalent amount of H. These results again indicate that a clumpy ejecta with an admixture of H of similar mass to the Fe ejecta is required for agreement between our determined Fe mass and that expected from Type Ia explosive nucleosynthesis yields \citep[e.g.][]{Iwamoto1999}.

\subsection{Other SNRs}
\label{atypsnrs}
Three of our SNR sample can not be considered either typical Sedov-phase or Fe-rich SNRs because of their unusual morphologies and/or spectra. These are discussed in the following sub-sections.

\subsubsection{MCSNR~J0447--6919}
The multi-wavelength images of MCSNR~J0447--6919 showed a large optical shell with faint X-ray emission distributed between the eastern and western shells, elongated in a southeast-northwest direction. Morphologically, this is similar to MCSNR~J0527--7104 in which the X-ray emission is primarily due to Fe ejecta \citep{Kavanagh2016}. Because of the faintness of the X-ray emission in MCSNR~J0447--6919, we could only fit simple models to the spectrum (see Sect.~\ref{0447-res}). Despite the lack of a clear Fe L-shell hump as in the Fe-rich SNRs in our sample, the best fit was provided by a CIE plasma with free Fe abundance, suggesting that the X-ray emission may indeed be due to Fe-rich ejecta. Deeper observations of this object are required to better identify the spectral components. To determine the likely properties of this SNR, we assume the CIE model represents the SNR shell emission. With these results and the size estimate, we used the Sedov dynamical model to determine an initial explosion energy in the range $(0.17-0.95)\times10^{51}$~erg and age of 22--28~kyr. The explosion energy is lower than the canonical $10^{51}$~erg, which could result from our assumed filling factor of unity. If $f<1$, the determined explosion energy would increase and be more consistent with the Sedov model. These, along with other properties, are listed in Table~\ref{table:snr_properties}. 

\subsubsection{MCSNR~J0449--6903}
\label{atypsnrs-0449}
MCSNR~J0449--6903 is perhaps the most puzzling object in our sample. As described in Sect.~\ref{0449-res}, the X-ray emission appears to originate from two distinct regions: a faint, soft shell consistent with an evolved SNR; and a harder core region. The core emission could be fitted equally well with either a thermal plasma or a non-thermal power-law model, though the model complexity and constraints provided by our fits are severely limited by the count statistics. The thermal interpretation yields a k$T$=2.81~(2.06--4.60)~keV and suggests hotter ejecta emission but there is no evidence for enhanced abundances which should also be present if the emission is due to shock-heated ejecta. It is unclear as to what could be the origin of the emission in the non-thermal interpretation. The location of the hard emission at the centre of the remnant and its radius ($1.16~(\pm0.08)\arcmin$, corresponding to $\sim~16.9~(\pm1.1)$~pc at the LMC distance, see Sect.~\ref{0449-res}) could point to a faint pulsar wind nebula (PWN) as has been observed in other LMC SNRs such as MCSNR~J0453--6829 \citep{Haberl2012} and the N~206 SNR \citep{Williams2005}, though the possible PWN in MCSNR~J0449--6903 would be somewhat larger and fainter, e.g. the PWN in MCSNR~J0453--6829 is less than half the size with $R\sim7$~pc \citep{Gaensler2003}. The hard extended source is also larger than other well-known X-ray PWNe, e.g.
the Crab ($R<1$~pc), 0540--69.3 ($R<1$~pc), G21.5--0.9 ($R\sim3.5$~pc), and 3C58 ($R<5$~pc) \citep[][and references therein]{Petre2007}, though these are associated with younger SNRs. The extent of the hard source is even comparable to the size of MCSNR~J0449--6903 itself (see Sect.~\ref{size_fits_sect} and Table~\ref{table:size_fits}). The radii of a PWNe will be closest to the host SNR shell at the end of the initial expansion phase \citep{Gelfand2009}. As described below, we estimated the age of the SNR to be 9--15~kyr and the ambient density to be 1.6--4.4~$f^{-1/2} 10^{-2}$~cm$^{-3}$. From Fig.~2 of \citet{Gelfand2009}, for a PWN-hosting SNR evolving into a similar low density medium, their respective radii will be at their closest point after $\sim5$~kyr, slightly less than the estimated age of MCSNR~J0449--6903. If the hard extended source is a PWN, it is likely in the reverse shock collision and first compression phase. Deeper X-ray observations of this object are required to better constrain the properties of the hard source. In a multi-wavelength context, we would also expect the hard X-ray emission to be correlated with radio emission. \citet{Bozzetto2017} report on the radio emission from MCSNR~J0449--6903. While they do detect radio emission inside the SNR, it is concentrated towards the western shell, offset from the hard X-ray emission (their Fig.~1, top right). Furthermore, the determined radio spectral index of $\alpha$~=~--0.50$\pm$0.01 is steeper than expected from a PWN, which should be flatter. The true origin of the hard X-ray emission remains unclear. 

We assumed the soft component in our models represents emission from the expanding shell. Adopting the soft component parameters from the thermal plasma plus power law fit with the estimated size, we used the Sedov dynamical model to determine an initial explosion energy in the range $(0.01-0.17)\times10^{51}$~erg and age of 9--15~kyr. The explosion energy is much lower than the canonical $10^{51}$~erg, which might be explained by the filling factor being set to unity. If $f<1$, the determined explosion energy would increase and be more consistent with the Sedov model. These, along with other properties, are listed in Table~\ref{table:snr_properties}. 

\subsubsection{MCSNR~J0510--6708}
The interpretation of the morphology of this object is complicated by the large, optical shell. As described in Sect.~\ref{0510-res}, the \ratio\ contours indicative of an SNR are not correlated with the optical shell. Rather they are off-centre, protruding from the shell to the northwest. We identify the optical shell as being the HII region DEM~L90 \citep{Davies1976} which contains the stellar cluster BSDL~777 \citep{Bica1999}. \citet{Glatt2010} estimated the age of the cluster to be $\sim16$~Myr. We suggest that this cluster has created and is currently photo-ionising the optical shell and that the SNR is located in projection. The SN progenitor could have been a member of BSDL~777, though it might also be possible that an older star in the locality underwent a Type~Ia event.

The X-ray supernova remnant is the faintest in our sample. The emission is most prominent in the 0.7--1-1~keV band and fills the region delineated by the shell identified in the \ratio\ image. This could suggest an Fe-rich ejecta origin though the observation is much too shallow to confirm this. We obtained some constraints on the properties of the emitting plasma by assuming it is described by a thermal plasma in CIE of LMC abundance. We used the size of the shell estimated from the \ratio\ image with the Sedov dynamical model to determine an initial explosion energy in the range $(0.01-1.75)\times10^{51}$~erg and age of 12--22~kyr. These, along with other properties, are listed in Table~\ref{table:snr_properties}.

\section{Summary}
\label{sum}
We presented new \xmm\ observations of a sample of X-ray faint SNR candidates and two previously confirmed SNRs. The main findings of our analysis on these objects can be summarised as follows:

\begin{itemize}
\item No X-ray emission was detected from the SNR candidate 0457--6739. Because of the lack of a clear optical shell with enhanced \sii, X-ray emission, or non-thermal radio emission, we cannot confirm this candidate as an SNR.

\item Four of the SNRs, namely MCSNR~J0448--6700, MCSNR~J0456--6950, MCSNR~J0512--6717, MCSNR~J0527--7134, display X-ray morphologies and spectra that are consistent with evolved Sedov-phase remnants.

\item Two SNRs, the newly confirmed MCSNR~J0504--6723 and previously known MCSNR~J0542---7014, were identified as members of the evolved remnants with a bright Fe-rich core \citep[see][for example]{Maggi2014,Kavanagh2016,Maggi2016}. This brings the total number to 13, 18\% of the confirmed LMC SNR population (see Table~\ref{table:evolved_fe_snrs}). The Fe-rich core of MCSNR~J0504--6723 is surrounded by a soft X-ray shell. This remnant is unusually shaped which we suggest has been imposed by the inhomgeneous density environment into which it is evolving. MCSNR~J0542--7104, which was confirmed as an SNR by \citet{Yew2021}, exhibits a distinct Fe-rich core but without the soft Sedov shell. Such a morphology has previous been observed in, e.g. MCSNR~J0508--6830, MCSNR~J0511--6759 \citep{Maggi2014}, and MCSNR~J0527--7104 \citep{Kavanagh2016}, and is interpreted as the shell evolving to the point that it is no longer visible in X-rays.

\item MCSNR~J0447--6919 displayed a clear optical shell with enhanced \sii\ emission, with very faint emission in the interior which is elongated along its major axis.  Morphologically, this is similar to MCSNR~J0527--7104 in which the X-ray emission is primarily due to Fe ejecta \citep{Kavanagh2016}. Because of the faintness of the X-ray emission in MCSNR~J0447--6919, we could only fit simple models to the spectrum, which do indicate a possible Fe enhancement and suggests that this object is a candidate for the evolved Fe-rich class of LMC SNRs.

\item MCSNR~J0449--6903 is the most unusual object in our sample, exhibiting a faint, soft shell consistent with an evolved SNR which surrounds harder emission in the core region. The core emission could be described with either thermal or non-thermal models, though the non-thermal interpretation appears more likely given the lack of abundance enhancements which would be expected from hot ejecta emission. We suggest that the hard emission could be due to a faint PWN, though there does not appear to be a correlated radio counterpart \citep{Bozzetto2017}. The true origin of the hard X-ray emission in the core remains unclear. 

\item MCSNR~J0510--6708 appears to be located in projection with a large, circular optical shell, identified as the HII region DEM~L90 \citep{Davies1976} which is powered by BSDL~777 \citep{Bica1999}. The extent of the SNR is evident by its \ratio\ contour with faint X-ray emission filling this shell. Our X-ray spectral analysis was severely hampered by count statistics, however, the X-ray and \ratio\ characteristics were enough to confirm the SNR nature.\\

\noindent We confirmed seven of our candidates as SNRs for the first time. These add to the 64 previously known SNRs \citep[][and references therein]{Yew2021} for a new total of 71 confirmed SNRs in the LMC. The upcoming new generation radio (ASKAP (\cite{2021MNRAS.506.3540P}, Bozzetto et al. in prep), MeerKAT \& ATCA) and X-ray ({\it eROSITA}) surveys expects to discover even more LMC SNRs such as the LMC Odd Radio Circle J0524--6948 (Filipovi\'c et al., submitted).

\end{itemize}

\section*{Acknowledgements}
This research has made use of the VizieR catalogue access tool, CDS, Strasbourg, France (DOI: 10.26093/cds/vizier).

\section*{Data Availability}
The X-ray data used in this work are available through the \xmm\ Science Archive (XSA): \url{https://www.cosmos.esa.int/web/xmm-newton/xsa}.

\bibliographystyle{mnras}
\bibliography{refs}

\begin{appendix}
\section{Point source analysis}
In this appendix we describe the methodology for the analysis and identification of point sources possibly associated with our SNR sample, followed by a brief description of each. The nature of the point sources could indicate the type of SN explosion that produced the SNRs. We follow the same procedure described in \citet{Kavanagh2015b}. We assigned \xmm\ identifiers to each of the sources following the source naming conventions\footnote{See \url{https://www.cosmos.esa.int/web/xmm-newton/source-naming-convention}}. For a detailed description of the source detection see Sect.~\ref{sect:source_detect}.

We found that almost all sources were too faint for a meaningful spectral analysis. Instead, we performed a hardness ratio (HR) analysis which compares the number of counts in certain energy bands to determine the approximate shape of the spectrum, allowing spectral parameters to be inferred. The source HRs are were determined using Eq.~5 of \citet{Kavanagh2015b} for the same three energy bands, i.e. $i_{1} = 0.3-1$~keV, $i_{2} = 1-2$~keV, and $i_{3} = 2-4.5$~keV, using the BEHR code \citep[ver. 12-12-2013,][]{Park2006}\footnote{See \url{http://hea-www.harvard.edu/astrostat/behr/}} . We then compared our derived source HRs to synthetic HR grids which were produced from either a power law or a CIE model absorbed by material in the Galaxy and LMC (\texttt{phabs*vphabs*pow} and \texttt{phabs*vphabs*apec} in XSPEC, respectively), which allowed us to infer the source spectral parameters implied by the HR values \citep[see][]{Kavanagh2015b}. These were generated by simulating a set of spectra with Galactic absorption ($N_{\rm{H, Gal}}$) fixed as appropriate for each SNR, and varying values of LMC foreground absorption, and photon index ($\Gamma$) or plasma temperature (k$T$). The grids for each model and source HRs are shown for each of our SNRs in Figs.~\ref{hr_comb1}, \ref{hr_comb2}, and \ref{hr_comb3}. 

Source variability can also provide insight into the nature of the point sources. To search for variability, we extracted light curves for each in the 0.3--10~keV range from a barycentric-corrected PN event list, correcting for instrumental effects and background using the SAS task \texttt{epiclccorr}. We only considered EPIC-pn data in our variability analysis because of the higher count statistics compared to the EPIC-MOS data. We found no evidence of variability in any of our sources.

Multi-wavelength counterparts to the X-ray sources were identified by querying all catalogues on the CDS Vizier service using the Astroquery\footnote{See \url{https://astroquery.readthedocs.io/en/latest/}} affiliated package of astropy. The identified counterparts are listed in Table~\ref{table:point_sources}.

\begin{table*}
\caption{Detected point source counterparts.}
\begin{center}
\label{table:point_sources}
\begin{tabular}{lllllll}
\hline
\hline
No. & Name & RA & Dec & Counterpart & Waveband & Source type \\
 & & (J2000) & (J2000) & & &   \\
\hline
\multicolumn{7}{c}{MCSNR~J0447--6919} \\
\hline
1 & XMMU~J044700.7--691721 & 04:47:00.7 & --69:17:21 & -- & -- & -- \\
2 & XMMU~J044702.7--691807 & 04:47:02.7 & --69:18:07 & -- & -- & -- \\
3 & XMMU~J044718.1--691900 & 04:47:18.1 & --69:19:00 & WISEA J044717.89--691859.4 & IR & -- \\
\hline
\multicolumn{7}{c}{MCSNR~J0448--6700} \\
\hline
1 & XMMU~J044804.9--670040 & 04:48:04.9 & --67:00:40 & WISEA J044804.82--670042.1 & IR & -- \\
2 & XMMU~J044808.1--665857 & 04:48:08.1 & --66:58:57 & WISEA J044807.96--665859.2 & IR & AGN \\
3 & XMMU~J044814.0--670039 & 04:48:14.0 & --67:00:39 & WISEA J044814.67--670037.0 & IR & -- \\
4 & XMMU~J044839.2--670159 & 04:48:39.2 & --67:01:59 & WISEA J044839.16--670201.8 & IR & -- \\
\hline
\multicolumn{7}{c}{MCSNR~J0449--6903} \\
\hline
1 & XMMU~J044922.4--690336 & 04:49:22.4 & -69:03:36 & SSTSL2 J044922.14--690339.0 & IR & -- \\
\hline
\multicolumn{7}{c}{MCSNR~J0456--6950} \\
\hline
1 & XMMU~J045622.3--695230 & 04:56:22.3 & -69:52:30 & SSTSL2 J044922.14--690339.0 & IR & -- \\
2 & XMMU~J045623.0--695142 & 04:56:23.0 & -69:51:42 & -- & -- & -- \\
3 & XMMU~J045626.8--694945 & 04:56:26.8 & -69:49:45 & 2MASS J04562647--6949415 & IR & -- \\
4 & XMMU~J045639.3--695026 & 04:56:39.3 & -69:50:26 & -- & -- & -- \\
5 & XMMU~J045639.7--695232 & 04:56:39.7 & -69:52:32 & MACHO  17.2587.1178 & opt,IR & quasar cand. \\
6 & XMMU~J045640.4--695132 & 04:56:40.4 & -69:51:32 & WISEA J045640.58--695134.0 & IR & -- \\
7 & XMMU~J045651.8--695041 & 04:56:51.8 & -69:50:41 & -- & -- & -- \\
\hline
\multicolumn{7}{c}{MCSNR~J0504--6723} \\
\hline
1 & XMMU~J050500.1--672250 & 05:05:00.1 & -67:22:50 & SSTSL2 J050500.52--672252.4 & IR & -- \\
\hline
\multicolumn{7}{c}{MCSNR~J0512--6717} \\
\hline
1 & XMMU~J051224.8--671736 & 05:12:24.8 & -67:17:36 & WISEA J051225.10--671737.1 & IR & -- \\
2 & XMMU~J051231.3--671754 & 05:12:31.3 & -67:17:54 & -- & -- & -- \\
3 & XMMU~J051240.8--671720 & 05:12:40.8 & -67:17:20 & 2MASS J05124058--6717225 & IR,X? & HMXB? \\
\hline
\multicolumn{7}{c}{MCSNR~J0527--7134} \\
\hline
1 & XMMU~J052744.7--713354 & 05:27:44.7 & -71:33:54 & -- & -- & -- \\
2 & XMMU~J052746.5--713305 & 05:27:46.5 & -71:33:05 & WISEA J052746.38--713303.8 & opt,IR & C \\
\hline
\multicolumn{7}{c}{MCSNR~J0542--7104} \\
\hline
1 & XMMU~J054231.9--710603 & 05:42:31.9 & -71:06:03 & WISEA J054231.87--710558.6 & IR & -- \\
2 & XMMU~J054307.4--710409 & 05:43:07.4 & -71:04:09 & PMN J0543--7104 & -- & -- \\
\hline
\end{tabular}
\end{center}
\end{table*}%

\subsection{MCSNR~J0447--6919}
\begin{figure*}
\begin{center}
\resizebox{6.3in}{!}{\includegraphics[trim= 0cm 0cm 0cm 0cm, clip=true, angle=0]{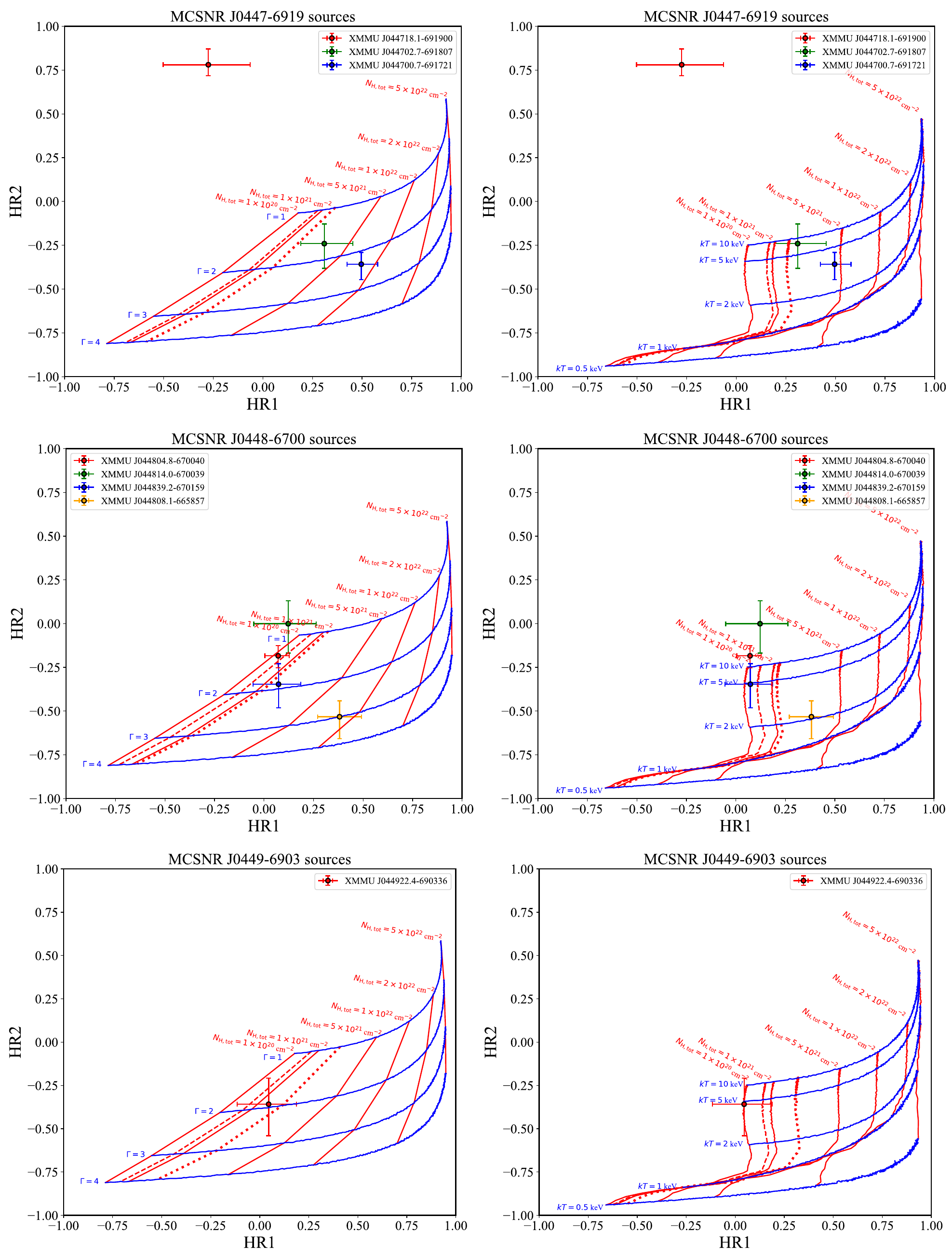}}
\caption{HR plots derived from grids of simulated absorbed power-law spectra (left) and absorbed thermal plasma (right) with the HRs from sources in MCSNR~J0447--6919 (top), MCSNR~J0448--6700 (middle), and MCSNR~J0449--6903 (bottom) overlaid. The blue lines indicate models on the grid of equal $\Gamma$ (left) and k$T$ (right), whereas the red lines mark the models on the grid of equal equivalent hydrogen column $N_{\rm{H,tot}} = N_{\rm{H,Gal}} + N_{\rm{H,LMC}}$. The dashed red line gives the total $N_{\rm{H}}$ measured through the Galaxy and the LMC from \citep{Kalberla2005} and the dotted red line gives the combined $N_{\rm{H}}$ measured through the Galaxy and the LMC by summing the Galactic \nh\ from \citet{Dickey1990} to the LMC \nh\ values from \citet{Kim2003}.
}
\label{hr_comb1}
\end{center}
\end{figure*}

Three point sources were detected in MCSNR~J0447--6919, which are listed in Table~\ref{table:point_sources}. XMMU~J044718.1--691900 was the most centrally located source with XMMU~J044700.7--691721 and XMMU~J044702.7--691807 located near the northern shell (Fig.~\ref{figures:0447-6918_4panel}, top-left). The HRs for each of the sources are plotted on HR diagrams in Fig.~\ref{hr_comb1}, along with the HR grids.

XMMU~J044718.1--691900 is located quite far from either of the grids. This is likely because its spectrum is not well described by either model used to produce the HR grids. The position of XMMU~J044700.7--691721 signifies a high absorption and suggests the source is located beyond the LMC, assuming there is no local or intrinsic absorption of the source emission. A photon index $\Gamma\sim2-3$ is inferred from the power law grid while a plasma temperature k$T\sim2-5$~keV is inferred from the thermal grid. The spectral parameters implied by the power law grid could be consistent with an AGN. XMMU~J044702.7--691807 is less affected by absorption than XMMU~J044700.7-691721, and is consistent with LMC absorption in both power law and thermal plasma derived grids. A photon index $\Gamma\sim1-2$ is inferred from the power law grid while a plasma temperature k$T\gtrsim3$~keV is inferred from the thermal plasma grid.

An IR counterpart was found for XMMU~J044718.1--691900, namely SSTISAGEMA~J044717.97--691859.7 from the \spitzer\ SAGE Survey \citep{Meixner2006}, but is of unknown source type. No counterparts were found for XMMU~J044700.7--691721 or XMMU~J044702.7-691807. The nature of each of these sources remains unclear.

\subsection{MCSNR~J0448--6700}
Four point sources were detected in MCSNR~J0448--6700, which are listed in Table~\ref{table:point_sources}. All four of the sources were located close to the SNR shell (Fig.~\ref{figures:0448-6700_4panel}, top-left) and were too faint for spectral analysis. The HRs for each of the sources are plotted on HR diagrams in Fig.~\ref{hr_comb1}, along with the HR grids.

The location of the three sources XMMU~J044804.9-670040, XMMU~J044814.0--670039, XMMU~J044839.2--670159 on the HR grids indicate low foreground absorptions, consistent with the LMC. A value of $\Gamma\sim1.5$ is inferred from the power law grid while k$T\gtrsim10$~keV is inferred from the thermal grid for XMMU J044804.9--670040. Similar values are seen for XMMU~J044814.0--670039 with $\Gamma\lesssim1.5$ and k$T\gtrsim10$~keV inferred from the power law and thermal grids, respectively. The HRs of XMMU~J044839.2--670159 suggest a softer spectrum with $\Gamma\sim1.5-2.5$ and k$T\gtrsim3$~keV inferred. The position of XMMU~J044808.1--665857 in the HR diagram suggests the source is located beyond the LMC, assuming there is no local or intrinsic absorption of the source emission. A photon index $\Gamma\sim2.5-4$ is inferred from the power law grid while k$T\sim1-3$~keV is inferred from the thermal grid. The implied spectral parameters from the power law grid could be consistent with an AGN.

Counterparts were found for all of these sources. Only the counterpart of XMMU~J044808.10--665857, WISEA~J044807.96--665859.2, had a confirmed source type - an AGN. The nature of all sources except XMMU~J044808.10--665857 remains unclear. The location of all four sources projected near the shell of the SNR suggests that they are unlikely to be the compact remnant of the explosion.

\begin{figure*}
\begin{center}
\resizebox{6.3in}{!}{\includegraphics[trim= 0cm 0cm 0cm 0cm, clip=true, angle=0]{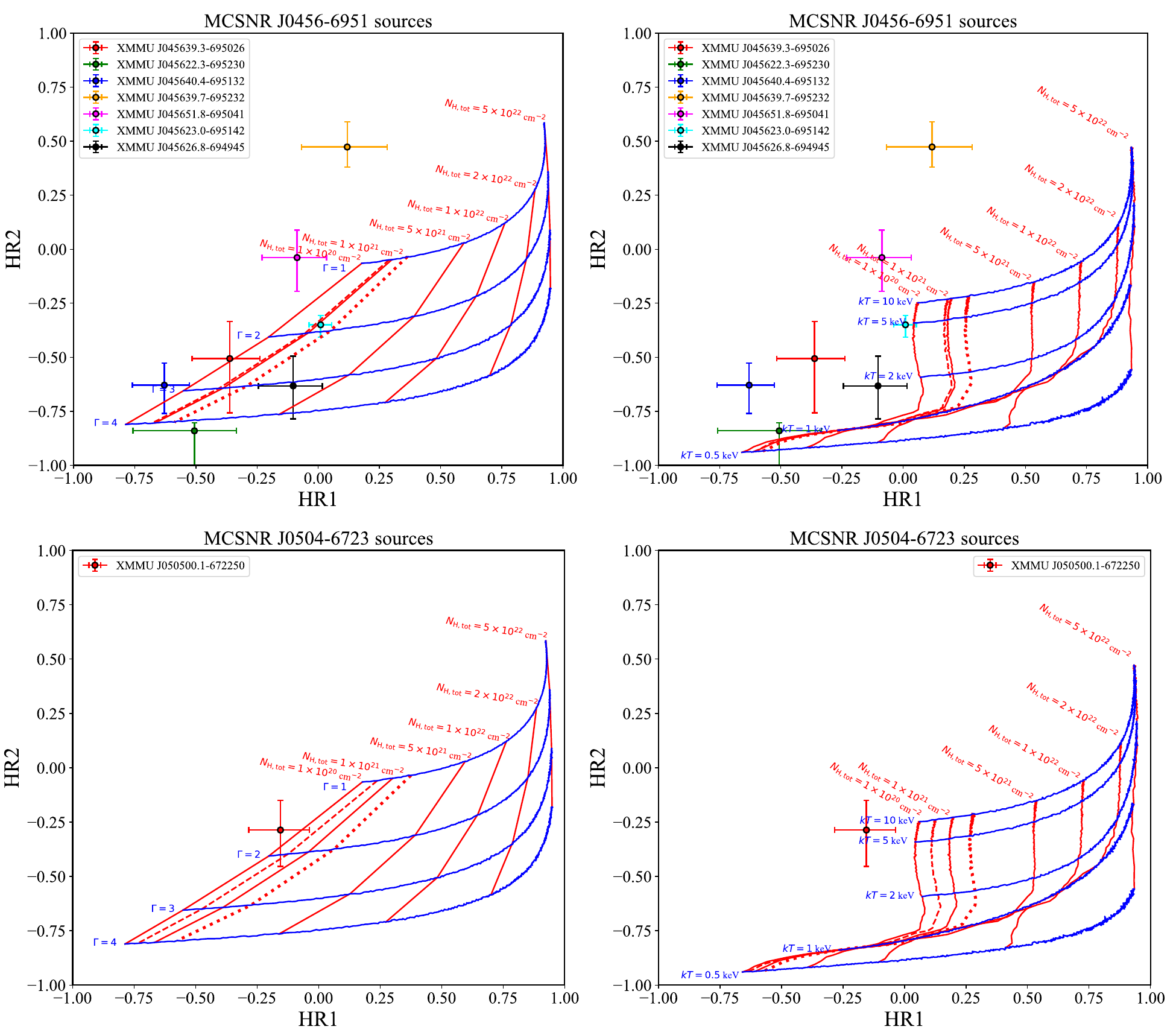}}
\caption{Same as Fig.~\ref{hr_comb1} but for MCSNR~J0456--6950 (top), MCSNR~J0504--6723 (bottom).}
\label{hr_comb2}
\end{center}
\end{figure*}

\begin{figure*}
\begin{center}
\resizebox{6.3in}{!}{\includegraphics[trim= 0cm 0cm 0cm 0cm, clip=true, angle=0]{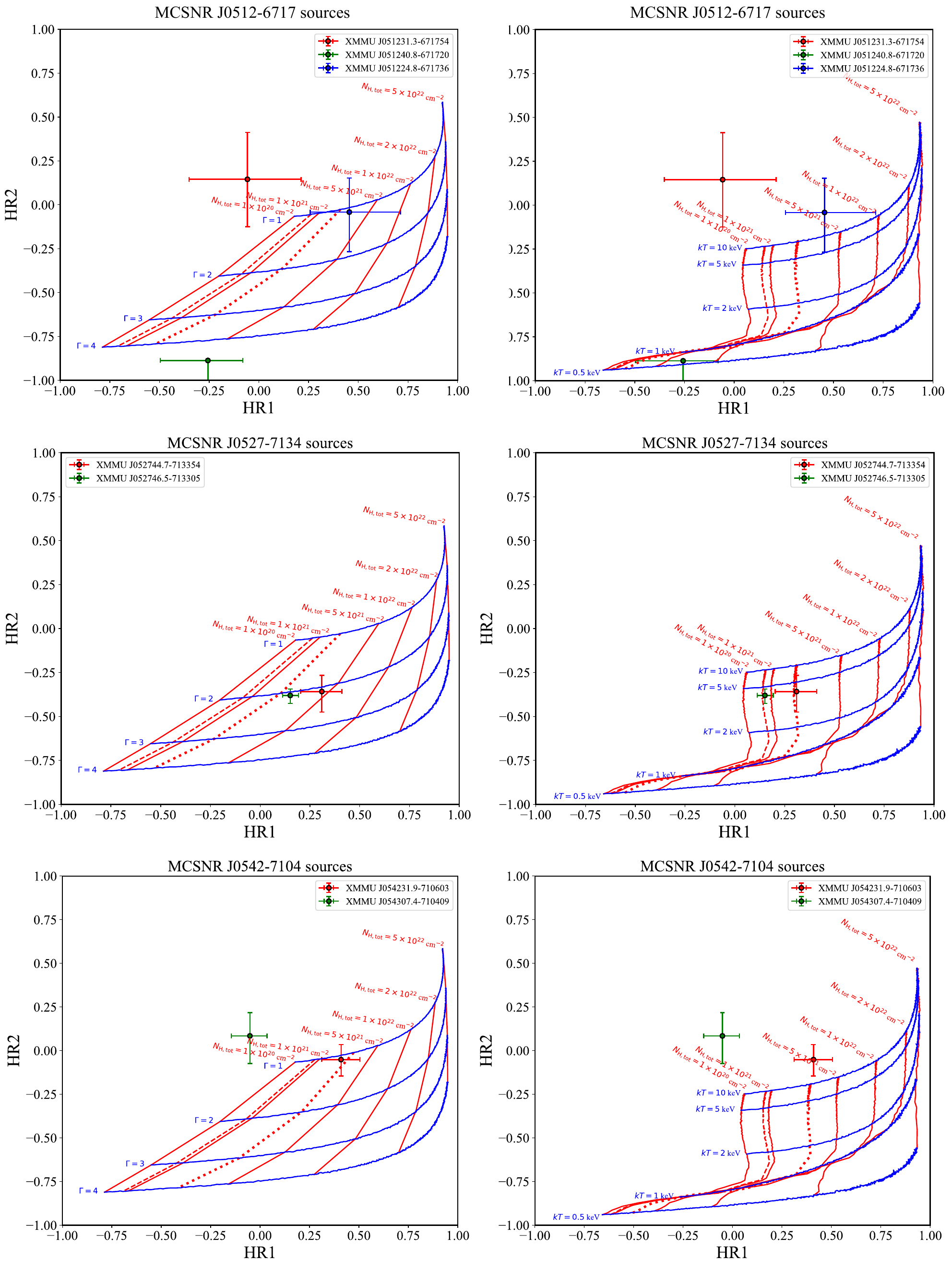}}
\caption{Same as Fig.~\ref{hr_comb1} but for MCSNR~J0512--6717 (top), MCSNR~J0527--7134 (middle), and MCSNR~J0542--7104 (bottom).}
\label{hr_comb3}
\end{center}
\end{figure*}

\subsection{MCSNR~J0449--6903}
Only one point source was detected in MCSNR~J0449--6903, which is listed in Table~\ref{table:point_sources}. The source, XMMU~J044922.4--690336, is located towards the western edge of the remnant (Fig.~\ref{figures:0449-6903_4panel}, top-left). Its HRs are plotted on HR diagrams in Fig.~\ref{hr_comb1} along with the HR grids.

The location of the source on the HR grids indicates a low foreground absorption, consistent with the LMC. A value of $\Gamma\sim2$ is inferred from the power law grid while k$T\gtrsim5$~keV is inferred from the thermal grid. An IR counterpart was identified but there is insufficient information to assign a source class. Therefore its nature remains unclear.

\subsection{MCSNR~J0456--6950}
Seven point sources were detected in MCSNR~J0456--6950, which are listed in Table~\ref{table:point_sources}. The seven sources are distributed around this large and faint SNR (Fig.~\ref{figures:0456-6951_4panel}, top-left).

The HRs for each of the sources are plotted on HR diagrams in Fig.~\ref{hr_comb2} along with the HR grids. Two of the sources, XMMU~J045639.7--695232 and XMMU~J045651.8--695041, are located outside both the power law and thermal plasma HR grids suggesting that these models do not accurately describe their spectra. The sources XMMU~J045640.4--695132, XMMU~J045639.3--695026, and XMMU~J045626.8--694945 are located on the power law HR grid but not on the thermal plasma grid. They are all consistent with lower absorption as expected for the LMC, with power law indices in the range $\Gamma\approx2-4$. The location of XMMU~J045622.3--695230 on both grids also indicates a low absorption and either a high photon index ($\Gamma\gtrsim4$) or a low plasma temperature (k$T\lesssim1$~keV).

Multi-wavelength counterparts were found for four of the seven sources (Table~\ref{table:point_sources}). Only the counterpart for XMMU~J045639.7--695232, MACHO~17.2587.1178, had a source type classification as possible quasar. The nature of the remaining sources remains unclear.

\subsection{MCSNR~J0504--6723}
One point source was detected in MCSNR~J0504--6723, which is listed in Table~\ref{table:point_sources}.  XMMU~J050500.1--672250 is located close to the north-eastern shell of the remnant (Fig.~\ref{figures:0504-6723_4panel}, top-left). The source HRs are plotted on the HR diagrams in Fig.~\ref{hr_comb2}, along with the HR grids.

The location of XMMU~J050500.1--672250 relative to the thermal grid suggests that this model is not a good description of the source emission. Therefore, we only discuss the power law parameters. The location of the source on the power law grid indicates a low foreground absorptions, consistent with the LMC. A value of $\Gamma\sim1.5-2.5$ is inferred. 

The IR source SSTSL2~J050500.52--672252.4 was identified as a counterpart to XMMU~J050500.1--672250. However, the source type of the counterpart is unknown so the nature of this source remains unclear. Its location to the extreme north of the SNR suggests that it is unlikely to be the compact remnant of the explosion.

\subsection{MCSNR~J0510--6708}
No point sources were detected in MCSNR~J0510--6708

\subsection{MCSNR~J0512--6717}
Three point sources were detected in MCSNR~J0512--6717 located to the south and southeast of the centre. XMMU~J051224.8--671736 and XMMU~J051231.3--671754 are located interior to the soft X-ray shell while XMMU~J051240.8--671720 is embedded in the eastern shell.

The source HRs are plotted on the HR diagrams in Fig.~\ref{hr_comb3}, along with the HR grids. The position of XMMU~J051224.8--671736 on both grids indicates a high \nh\ with a location beyond the LMC. The HRs point to a $\Gamma<2$ or k$T>5$~keV, both signifying a hard source and a possible AGN, though this cannot be confirmed with the available data. The location of XMMU~J051231.3--671754 off both the thermal and power law grids suggests that neither model is representative of the source spectrum. XMMU~J051240.8--671720 is located off the power law grid but its position on the thermal plasma grid suggests a soft thermal source with the absorption consistent with the LMC.

Two of the sources, XMMU~J051224.8--671736 and XMMU~J051240.8--671720, were correlated with IR counterparts (see Table~\ref{table:point_sources}), however neither counterpart has a confirmed source type. Interestingly, XMMU~J051240.8--671720 is also a possible counterpart to the high-mass X-ray binary RX~J0512.6--6717 \citep{Haberl1999} identified with \rosat, though its location in the shell suggests it is unlikely to be related to the supernova explosion. No counterpart was identified for XMMU~J051231.3--671754. 

\subsection{MCSNR~J0527--7134}
Two point sources were detected in MCSNR~J0527--7134, XMMU~J052744.7--713354 to the north and XMMU~J052746.5--713305 slightly west of the SNR centre. The source HRs are plotted on the HR diagrams in Fig.~\ref{hr_comb3}, along with the HR grids. The position of XMMU~J052744.7--713354 and XMMU~J052746.5--713305 on the power law grid indicates a high \nh\ with a location beyond the LMC and a $\Gamma=1.5-2.52$ However, their location on the thermal plasma grid suggests a lower foreground absorption consistent with the LMC and k$T>2$~keV. 

We searched for multi-wavelength counterparts to these sources which are listed in Table~\ref{table:point_sources}. XMMU~J052746.5--713305 was the only source with a counterpart, identified as MACHO~21.7644.561 and is a quasar candidate \citep{Kim2012}. Therefore, it is likely unrelated to MCSNR~J0527--7134. No counterpart was identified for XMMU~J052744.7--713354. Its location slightly offset from the SNR centre to the west suggests that it could potentially be the compact remnant of the explosion, though its nature remains unclear.

\subsection{MCSNR~J0542--7104}
Two point sources were detected in MCSNR~J0542--7104, XMMU~J054231.9–-710603 to the south and XMMU~J054307.4–-710409 to the east. Both sources are quite faint and located at the extremities of the SNR shell, already suggesting that they are not the compact remnants of the explosion. The source HRs are plotted on the HR diagrams in Fig.~\ref{hr_comb3}, along with the HR grids. Neither source is located on the thermal plasma grid indicating that this model is not a good description of the source emission. Only XMMU~J054231.9–-710603 is located on the power law grid, with its location suggesting an absorption consistent with the LMC and a hard spectrum with $\Gamma~1$. 

We searched for multi-wavelength counterparts to these sources which are listed in Table~\ref{table:point_sources}. XMMU~J054231.9--710603 is correlated  with an IR source and a radio source but no source classification exists for either. 

\end{appendix}

\bsp	
\label{lastpage}
\end{document}